\documentclass[preprint]{aastex6} 

\begin{document}




\slugcomment{Submitted to ApJSS}

\shorttitle{M87 Novae}
\shortauthors{Shara et al}
\title{A {\it Hubble Space Telescope} Survey for Novae in M87. I. Light and Color Curves, Spatial Distributions and the Nova Rate\altaffilmark{1}}

\author{Michael~M.~Shara\altaffilmark{2}, Trisha F. Doyle\altaffilmark{2,3}, Tod R. Lauer\altaffilmark{4}, David~Zurek\altaffilmark{2}, J. D. Neill\altaffilmark{5}, Juan P. Madrid\altaffilmark{6}, Joanna Miko{\l}ajewska\altaffilmark{7}, D. L. Welch\altaffilmark{8}, and Edward A. Baltz\altaffilmark{9}}

\altaffiltext{1}{Based on observations with the NASA/ESA {\it Hubble Space
Telescope}, obtained at the Space Telescope Science Institute, which is
operated by AURA, Inc., under NASA contract NAS 5-26555.}

\altaffiltext{2}{Department of Astrophysics, American Museum of Natural History, Central Park West and 79th Street, New York, NY 10024-5192, USA}

\altaffiltext{3} {Department of Physics and Space Sciences, Florida Institute of Technology, Melbourne, FL 32901, USA}

\altaffiltext{4} {National Optical Astronomy Observatory, P.O. Box 26732, Tucson, AZ 85726, USA}

\altaffiltext{5} {California Institute of Technology, 1200 East California Boulevard, MC 278-17, Pasadena CA 91125, USA}

\altaffiltext{6} {CSIRO, Sydney, Australia}

\altaffiltext{7} {N. Copernicus Astronomical Center, Polish Academy of Sciences, Bartycka 18, PL 00--716 Warsaw, Poland}

\altaffiltext{8} {Department of Physics \& Astronomy, McMaster University, Hamilton, L8S 4M1, Ontario, Canada}

\altaffiltext{9} {KIPAC, SLAC, 2575 Sand Hill Road, M/S 29, Menlo Park, CA 94025, USA}
 
\begin{abstract}
The Hubble Space Telescope has imaged the central part of M87 over a 10 week span, leading to the discovery of 32 classical novae and nine fainter, likely very slow and/or symbiotic novae. In this first in a series of papers we present the M87 nova finder charts, and the light and color curves of the novae. We demonstrate that the rise and decline times, and the colors of M87 novae are uncorrelated with each other and with position in the galaxy. The spatial distribution of the M87 novae follows the light of the galaxy, suggesting that novae accreted by M87 during cannibalistic episodes are well-mixed.  Conservatively using only the 32 brightest classical novae we derive a nova rate for M87: $363_{-45}^{+33}$ novae/yr. We also derive the luminosity-specific classical nova rate for this galaxy, which is $7.88_{-2.6}^{+2.3} /yr/ 10^{10}L_\odot,_{K}$. Both rates are 3-4 times higher higher than those reported for M87 in the past, and similarly higher than those reported for all other galaxies. We suggest that most previous ground-based surveys for novae in external galaxies, including M87, miss most faint, fast novae, and almost all slow novae near the centers of galaxies.  
\end{abstract}


\section{Introduction and Motivation}

Determining the binary stellar fraction throughout the Universe, a fundamental constraint on star formation,
demands the identification and characterization of substantial populations of extragalactic stars that are unquestionably binary.
Beyond the confines of our own Galaxy it becomes increasingly difficult to separate single from binary stellar populations. 
An important exception is the classical novae (CNe), which are both a 100\% binary population and, during outburst, amongst the most luminous stars in any galaxy.

CNe are white dwarfs accreting hydrogen-rich matter from Roche-lobe
filling secondaries \citep{kra59, war95}. The accumulation of order $10^{-5} M_{\odot}$ of hydrogen on a white dwarf leads to a thermonuclear runaway,
Eddington or greater luminosity, and mass ejection at high speed, observed as
a classical nova eruption \citep{sts72, pri78}. Many novae achieve luminosities 
in excess of $10^{5} L_{\odot}$, making them easily detectable in external galaxies \citep{arp56, ros73, cia90, nei04}.
The most luminous CNe reach absolute magnitudes close to M = -10. Thus very large numbers of novae can be detected with 
the {\it Hubble Space Telescope} (HST) to well beyond the Coma Cluster of galaxies, and their near-constant luminosity 15 days after maximum light can be used as a distance indicator \citep{bus55, shb81, fer03}. 

Four examples of what novae can teach us about the genesis and evolution of close binaries in different stellar populations and types of galaxies are the following. First, by demonstrating that nova rates, luminosities and/or spectra \citep{del98} are different in disk and bulge-dominated galaxies, one could directly show that close binary formation rates and histories are also different in these populations \citep{yun97, mat03}. Second, by mapping the locations of novae in cannibalistic massive central galaxies in clusters, or between galaxies, one can check if dynamical stirring has been effective in mixing swallowed populations throughout these galaxies or intracluster space \citep{nei05,sha06}. Third, important binary parameters - white dwarf mass, luminosity and mass transfer rate - can be determined from the properties of nova eruptions \citep{pri95, yar05, hil16} in external galaxies. Fourth, by using novae as proxies for all close binaries in very different kinds of galaxies, we can determine these binaries' spatial distributions relative to other populations \citep{hat97}. 

In addition, basic tests of the theory of novae become possible. Do the rates of decline from maximum brightness vary systematically with position in a given galaxy, or with Hubble type? Theory suggests that this should not occur, \citep{sha81, liv92}, but a definitive observational demonstration is lacking. Observations could also determine the instantaneous luminosity function and the speed class distribution of novae in different kinds of galaxies - strong constraints on nova evolution theory. 

Studies of novae in galaxies of the Local Group and beyond \citep{hub29, hen54, arp56, ros73, pvb87, tom92, nei04, wil04, sha14, cur15} 
have addressed some subset of these questions. Although M87 has been targeted for novae in the past \citep{pvb85, sha00, sha04, bal04, mad07, cur15}, only a handful of novae with reasonably complete light curves have been detected to date in this massive elliptical galaxy, nine in M49 \citep{fer03} and six in other Virgo cluster ellipticals \citep{pvb87}. The  resolution and  sensitivity of  the Advanced  Camera  for Surveys of HST allowed us  to probe the inner  regions of M87, where galaxy light makes  the detection  of  transients an observational challenge  with ground based data. In this paper we report the detection and detailed characterization via HST of 32 erupting classical novae in M87, and nine likely slow and/or symbiotic novae. We are thus able to provide much more definitive answers to many of the above questions concerning novae in a massive elliptical galaxy for the first time. 

The datasets used and our analysis methodology are described in Section 2. The time-lapse images and light and color curves of the M87 novae are presented in Section 3.  Correlations amongst nova properties, and with position in M87 are presented in Section 4. The spatial distribution of the M87 novae and the observed nova luminosity distribution are detailed in Section 5. The nova rate in M87 (and in other galaxies) is discussed in Section 6. We briefly summarize our results in Section 7.

\section{Observations}
 
The data used to find novae were images taken for the HST Cycle-14 program 10543 (PI - E. A. Baltz). The goal of that program was to search for microlensing events in M87 using $I$ band imagery for the primary search, and $V$ band imagery to provide colors. Four observations, separated by five days, were followed by daily imaging for the remainder of the program, with the HST Advanced Camera for Surveys (ACS) in the F606W (hereafter $V$ or $V$ band) and F814W (hereafter $I$ or $I$ band) filters. The F814W exposures were sub-pixel dithered to ensure Nyquist sampling, allowing for the optimal use of the superb HST resolution. The dithers also allowed for the rejection of hot pixels and other defects, as well as cosmic ray rejection.

The ACS detector has a field-of-view (FOV) of 202" x 202"  and a  pixel scale of 0.05" (Mack  et al.  2003). At the 16.4 $\pm 0.5$ Mpc distance of M87 \citep{bir10}, 1" corresponds to 80 pc; WFC imaging thus covers an area of about 16 x 16 kpc centered on the M87 nucleus. A total of 254 ACS exposures were taken in 61 different visits over a period of 72 days from 2005  December 24 to 2006 March  5. The majority of exposures (205) were taken in the $I$ band totaling 73,800 s.  Color information was  obtained with 49 images in the $V$ band  that combined amount to 24,500 s. Four 360 s exposures in the $I$ band filter, yielding a 1440 s total exposure were followed by a single 500 s $V$ exposure during each orbit. These observations provide unmatched temporal coverage of extragalactic transients as well as an unprecedented large FOV at very high resolution.

Archive-given names, observation dates and exposure times for all observations in each of the filters are given in Table 1.  Several observations contain no useful data, but are kept in the table for completeness. 

\subsection{Finding the Novae}

The 32 classical novae reported here were discovered independently by three different collaborators. We discuss their detection first. 
Nine additional, faint variables are likely to be slow and/or symbiotic novae, and are discussed separately at the end of this section.

Using only the $I$ band images, 26 novae were found by TL. The methodology used by TL for careful exclusion of all cosmic ray events and hot pixels, 
choice of a baseline level from which to seek changes, thresholds above which to flag candidates, candidate vetting and candidate photometry are discussed in detail in \citet{bal04}. The images were reduced to preserve and optimize the angular resolution.  The four F814W images at a given epoch were interlaced to produce Nyquist image by the Fourier method of \citet{lau99}.  This method recovers a Nyquist image (2x finer pixels) from non-optimal dithers without interpolation kernals that can degrade the angular resolution. The Nyquist images were then rectified with sinc-function interpolation, which preserves the resolution of the images.  Note that these procedures offer superior resolution to drizzle, which blurs the images at the resolution scale. The nightly reduced images were convolved with an optimal filter to highlight point-source variables. The search was done on difference frames, using the entire image set with rejection of statistical outliers as a template for the M87 background. A handful of very faint candidates, beyond the 26 noted above, were rejected in the interests of retaining only candidates that were clear and obvious variables.

The data were also independently searched for novae by TFD and by DLW. Both authors retrieved the flat-fielded science files  (FLT.FITS)  and best reference files from  the HST public archive. Prior to their delivery the raw HST images are  processed through the standard pipeline at the STScI. The  standard ACS calibration pipeline (CALACS) performs basic reductions, viz.,  overscan subtraction, bias subtraction, dark subtraction, and flat fielding \citep{sir05}. Subsequent data reduction  was  carried  out  with the one step task  MULTIDRIZZLE \citep{koe02} of  the Space Telescope Data  Analysis System (STSDAS) run within Pyraf. Multidrizzle performs distortion correction using updated reference files, removes cosmic rays, and combines all exposures taken in the  same epoch into a final drizzled image (DRZ.FITS). For the final drizzled images that were used for this study TFD and DLW preserved the native ACS/WFC pixel size of 0.05".

All of the 26 novae discovered by TL, and five additional slow novae were discovered by TFD using one day and five-day Multidrizzled \citep{koe02} $V$ band images (not difference images). Each five-day image was visually blinked against the first five days' Multidrizzled image to locate even the faintest variables in M87. In addition to the 31 candidates noted above, an additional 29 variables were also found by TFD, but noted to be fainter and redder than the 31 candidates noted above. Visual inspection led us to eliminat 20 candidates as borderline detections - all are too faint for reliable photometry. Nine candidates with measurable brightnesses, all of which are redder than the 31 novae noted above, were retained; we consider them to be likely symbiotic and/or very faint novae, as discussed below.

Most of the novae found by TL and TFD, and one additional nova, too faint to be detected by TL and only marginally seen in the 5 day Multidrizzled images, was discovered by DLW. This nova was only observed while declining in brightness.  Its color and brightness behavior are consistent with those of the other 31 novae. DLW used the Welch-Stetson variability index \citep{wel93} on the Multidrizzled images to find candidates. The driving assumptions of  the Welch-Stetson variability index is that variability of a star is correlated between the two filters in use. That is, for instance, if a truly variable star becomes brighter in the $V$ band it will also become brighter in the coeval $I$ band frame. Random errors in magnitude are, by definition, uncorrelated between filters. Extraction of candidates required that they be detected at more than 8 epochs on multiple images. Background sky and galaxy light was take eliminated via ALLFRAME \citep{ste94}.

Our completeness and detection limits are discussed in section 5.2, but we note here that they differ by four magnitudes between the center of M87 and the periphery of our FOV. Our faintest detections reach $V$ $\sim$ 27 and $I$ $\sim$ 28 for nova 29, near the edge of HST's FOV, described below. Nova 15, near the center of M87, is only detected when it brighter than  $V$ $\sim$ 24.5 and $I$ $\sim$ 23.5 .

The positions of all candidate novae were checked in the HST archival images of M87 taken for program GO-8592 (PI - J. Silk) \citep{bal04}. This dataset consisted of 30 consecutive days of images of M87 taken with the Wide Field and Planetary Camera 2 (WFPC2) of HST through the F814W and F606W filters during 28 May through 25 June 2001, inclusive. The goal of this comparison of the two epochs'  (separated by over 4 years) images was to flag candidates that might be recurrent and/or symbiotic novae, or Mira variables. None of the 32 classical novae were detected in the stacked 2001 frames. Only one of the nine faint variables (nova number 33) was unambiguously detected in the stacked 2001 frames.

Our final decision on whether to eliminate a variable candidate that was unambiguously real (i.e. seen on multiple nights in both filters) from the nova pool was determined by its colors. The evolution of the $(V - I)$ color curves of novae are distinct from those of Miras. Inspection of dozens of Mira variables in the American Association of Variable Star Observers database demonstrates that Miras never display a $(V - I)$ color bluer than +2.0.  Every single one of the 32 classical novae, and 9 fainter variables noted above displays at least one epoch with $(V - I)$ $<$ 0.5 . (The one exception with no well-defined colors, nova 15, reaches absolute magnitude $ \sim -8$ in both filters; it is far too luminous to be anything but a classical nova). Those blue colors unambiguously demonstrate that none of our nova candidates is a Mira variable. 

The 32 classical novae and nine likely faint novae detected in our survey are mapped over an HST image of M87 in Figure 1. The spatial distribution of the novae, and checks for correlations between nova properties and position in the galaxy are described below.

\clearpage

\begin{figure}
\figurenum{1}
\epsscale{1.0}
\plotone{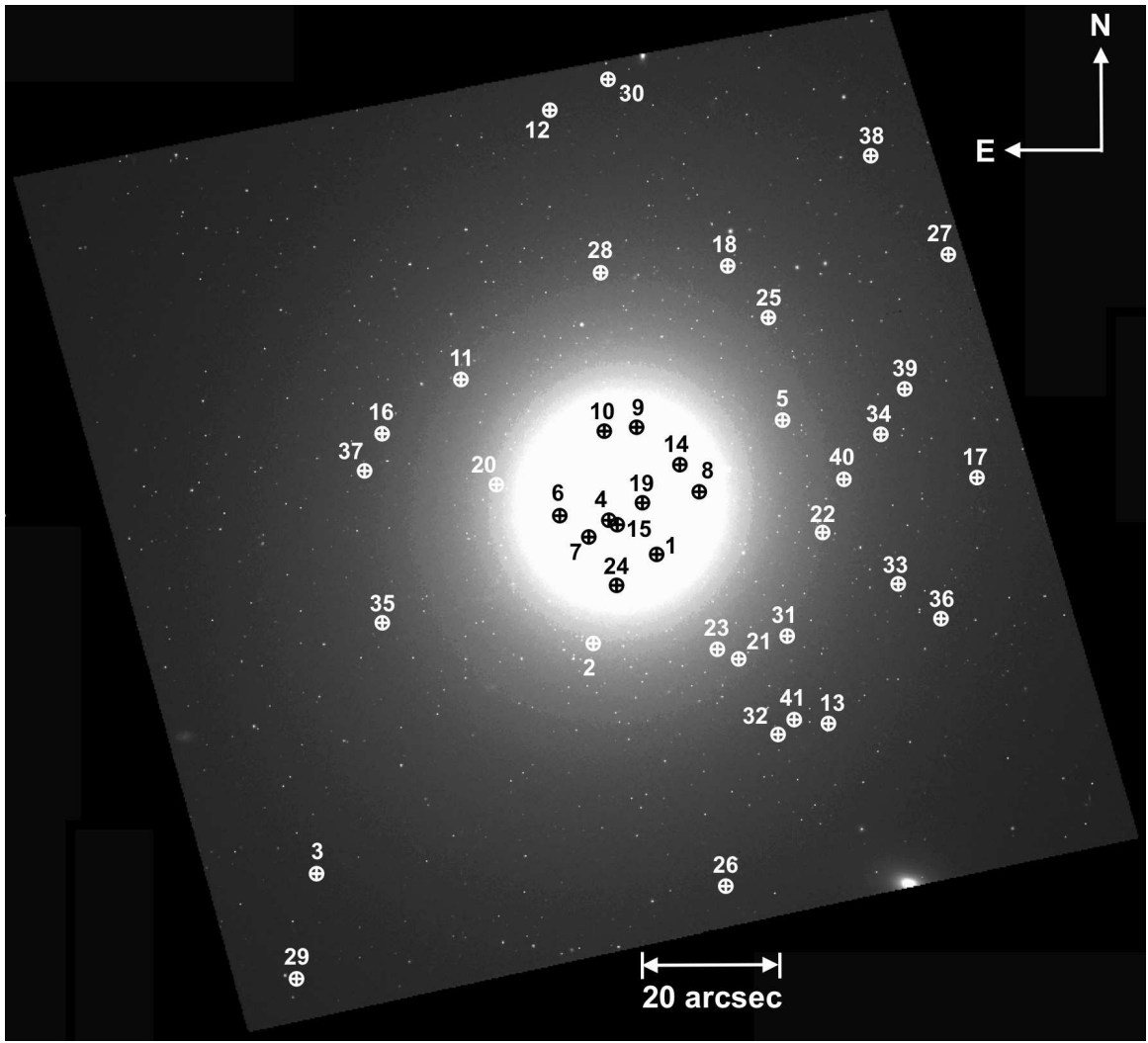}
\caption{A map of the locations of 32 classical novae (numbers 1-32) in M87 overlaid on the full drizzled HST ACS/WFC $V$ band image. Nine fainter variables (numbered 33-41), which are too blue to be Miras, are likely to be very slow and/or symbiotic novae.}
\end{figure}

\clearpage

\section{Results}

\subsection{Time-lapse Images}

Figures 2a through 2u, inclusive, are montages of 1.5 arcsec x 1.5 arcsec ``postage stamp", $I$ band images of each of the 32 classical novae and 9 likely slow novae detected in M87.  We used the $I$ band images for this display because they have better resolution and signal to noise for the single days than the $V$ images, as there were four times as many individual exposures taken in each epoch.  The novae are ordered from brightest to faintest in peak $V$ magnitude. In cases where the nova is only detected on the rising or falling part of its light curve, the peak measured brightness is probably not the brightest that the nova eventually becomes. The sequence of observations starts in the top-left corner, and proceeds horizontally across a row, followed by each of the four succeeding rows. Each nova is marked with a vertical and horizontal tick mark on the day of its maximum brightness.  

The brighter novae are both striking and obvious, usually rising to maximum brightness in just a few days. Inspection of the Figure 2 images suggests, at first glance, that some of the novae are ``out of order", with apparently brighter novae ordered later in the sequence than earlier, ``fainter" novae. An example is Nova 15, detected for only 6 successive days, which appears much fainter than Nova 16 (which is seen on 14 days). The reason for this apparent discrepancy is the very large spatial variation in M87's surface brightness. As is clear from Figure 1, Nova 15 is located considerably closer to M87's bright center than Nova 16, making it much harder to detect against its bright background. We will return to this point when discussing incompleteness corrections in our determination of M87's nova rate. 

\subsection{Light and Color Curves}

We used aperture photometry (in the daily $V$ band and $I$ band original images) to measure the light curves of each nova. We determined that there was no need to subtract a smoothed image of M87 from these frames for accurate photometry as long as we avoided including globular clusters in the apertures.  Aperture photometry was checked against PSF photometry in the $I$ images to ensure consistency, and excellent agreement was found. All photometric measurements are listed in Table 2, which contains the light curve and color curve data for each detected nova: the HST file number, days from maximum light of each observation, date of observation, $V$ band magnitude and error,  $I$ band magnitude and error, $V - I$ color and error in $V - I$. Six of the novae were already in decline when the HST observations began (Novae 9, 23, 25, 27, 30 and 32), while 12 others (Novae 3, 8, 11, 12, 13, 14, 16, 17, 18, 22, 27 and 29) were fading but still visible when the observations ended. Figures 3a through 3u, inclusive, display the classical 32 novae and 9 likely slow novae $V$ and $I$ light curves, and their $(V-I)$ color curves, respectively. Table 2 contains the light curve data. 

Figures 4a, 4b and 5 display 31 nova light curves (since these are $V$ band curves, M87 Nova 32 is not included) over-plotted onto each other to show their $V$ magnitudes, rises, and declines in comparison to each other.  Figures 6a, 6b and 7 complement Figures 4a, 4b and 5 by displaying the $I$ band light curves for all 32 classical novae. Our novae run the gamut from bright, fast novae to faint, slow novae.  

Based on its planetary nebula luminosity function, surface brightness fluctuations method, linear diameters of globular clusters and tip of the red giant branch method, the distance to M87 was determined to be d= (16.4 $\pm 0.5$) Mpc by \citet{bir10}, corresponding to a distance modulus of 31.1, which we adopt. Then the apparent F606W magnitude of our brightest nova (nova 1) at peak brightness (which is V = 21.8), corresponds to an absolute V magnitude of -9.3. The brightest novae, like our nova 1, are an order of magnitude more luminous than the Eddington luminosity of a nearly-Chandrasekhar mass white dwarf \citep{sha81}. The peak brightness of our faintest classical nova (nova 32) is $V$ = 25.3, corresponding to an absolute $V$ magnitude of -5.8. The peak brightness of our faintest slow, redder nova candidates is $I$ = 25.6, corresponding to an absolute $I$ magnitude of -5.5. These correspond to the Eddington luminosities of the lowest mass white dwarfs, roughly 0.4 - 0.5 Msun, that can produce thermonuclear runaways \citep{yar05}. The vast majority of classical novae \citep{sha81,shb81} are found in exactly the brightness range we find for the M87 novae.

The only nova whose light curve displays a somewhat time-symmetric rise and fall is nova number 23. This raises the possibility that the brightening of this one object might be due to a microlensing event. However, the $V - I$ color of object 23 changes by over 1.5 magnitudes during the outburst, ruling out microlensing.

The faintest novae and the most luminous Mira variables overlap in luminosity \citep{dar04}. Thus Miras, near maximum, may be mistaken for slow novae in surveys which only search for variables without measuring their colors. Fortunately, Miras become bluer as they approach maximum light, while novae are reddest at maximum \citep{dar04}. The shape of a variable's color curve is thus a good indicator whether it is a Mira or a nova. We demonstrate this in Figure 8a, where the color curves of 31 M87 classical novae, and our fainter variables are plotted. The smoothed, median classical nova light curve (seen as a thick red line in Figure 8a) is clearly reddest, with $(V - I)$ $\sim 0.4$, near maximum brightness. This happens because nova envelopes' photospheres are at maximum size, and coolest, near maximum light. At first glance, the faintest nova candidates (labelled Symb novae for convenience) might be mistaken for Miras as they display the opposite color evolution - bluest at peak - as seen in Figure 8a. However, as already noted, these variables are 2.5 magnitudes too blue to be Miras. A closeup of the nova color curves is shown in Figure 8b.
 
\section{Rise and Decline Time Correlations}

In Table 3 we list the angular distance (in arcseconds) from the center of M87, the Right Ascension and Declination, the maximum brightnesses in $V$ and $I$ band
magnitudes, and the times to rise to, and fall from maximum brightness by 1 and 2 magnitudes, for each of the classical novae in our sample. 

The novae included in the decline and rise time statistics are only the ones where a clear decline and rise are measurable.  If a nova is only observed to rise or only to decline, then the color at peak brightness $(V-I)_{peak}$ cannot be accurately determined, so these novae were excluded from the plots and statistics.  22/32 novae were used in the decline statistics and 21/32 novae were used in the rise statistics. In Figure 9 we plot the $(V - I$) color, at maximum V light, versus the time to rise 1 or 2 magnitudes to maximum light, or the time to decline 1 or 2 magnitudes from maximum light. It is clear from the plots that there is no correlation between nova color and the rapidity with which it rises to, or declines from peak brightness.

In Figure 10 we plot the time required for a nova to rise either 1 or 2 magnitudes, to reach peak brightness, versus the time required to decline 1 magnitude from peak brightness. There is no correlation between the rise and decline times of the M87 novae.

The absolute magnitude - rate of decline relationship (often referred to as the MMRD relation) \citep{zwi36, mcl45} has been investigated as a distance indicator for nearly a century. We will defer a full discussion of the implications of the results of the present work for MMRD to a subsequent paper. Here, in Figure 11, we simply plot the peak $V$ magnitudes of the 23 novae we observed in M87 (with well-observed peak magnitudes) versus the time to decline 2 magnitudes from peak brightness. The extensive nova grid models of \citet{yar05} first predicted the existence of faint, fast novae. \citet{kas11} first observationally demonstrated the existence of eight of these novae in M31. These faint, fast novae greatly weaken the MMRD correlation. Six of the 23 novae in M87 with well-measured properties are similar to the faint, fast novae of \citet{kas11}, confirming that these objects are both common and ubiquitous. Figure 11 strengthen's \citet{kas11}'s conclusion that the scatter in the MMRD is far too large for it to be anything but a very rough distance indicator.

In Figure 12 we plot the histograms of the observed times required for novae to rise to maximum, or decline from maximum by one or two magnitudes. The decline time histograms are subject to at least two observational biases. First, the interval over which we observed M87 - 72 days - is significantly shorter than the decline times (years) of the most slowly declining novae. Second, the most slowly declining novae are also amongst the faintest such objects. The incompleteness of detection of faint novae, discussed below, rises towards the center of M87.  Thus we are confident that the relative numbers of very rapidly declining novae (t$_{1decline}$ and t$_{2decline}$ less than 5 days and 10 days, respectively) are larger than those of novae taking 10 days and 20 days, respectively. The relative numbers of fainter, more slowly declining novae are subject to significant incompleteness corrections.

It is often said that novae rise to maximum brightness in just a day or two. Figure 12 demonstrates that this statement is an oversimplification. 9 of 22 novae (41\%) take longer than 2 days to complete the last 1 magnitude rise to maximum. Even more striking is that 6 of 22 novae (27 \%) require 5 days or more to rise the last 2 magnitudes to maximum light. This is in accord with theoretical models of nova light curves \citep{hil14}.

In Figure 13 we plot the rise and decline times, and the $(V - I)$ colors at peak brightness, as functions of the radial distance of each nova from the center of M87. This is motivated by the suggestion of \citet{del98} that faster and more luminous novae are more concentrated to the plane of our Galaxy than fainter and slower novae. While there is no way of knowing how many Virgo Cluster galaxies M87 has absorbed, that number is likely to be large. The M87 novae must have been accreted from many galaxies. If we regard the novae as tracers of past accretion events, we can check if dynamical stirring has thoroughly mixed M87's accreted galaxies, and by implication, its novae. It is clear that there are no trends in any of the plots of Figure 13; we do not see faster or bluer novae concentrated in any particular region of M87. 

\clearpage

\section{Spatial distribution}

\subsection{Cumulative novae/M87 light}

We measured the cumulative $V$ band light of M87 in circular, concentric annuli. This is a good approximation as M87 is almost spherical, and its isophotal twists are minor \citep{kor09}. We then plotted the cumulative nova fraction and the cumulative light of M87 in Figure 14. The novae appear to track the light of M87 with remarkable fidelity. This is confirmed by the Kolmogorov-Smirnoff test which finds no evidence for a difference in the M87 light and nova distributions, even at just the 95\% confidence level.
\citet{sha00} first suggested that the novae in M87 track the galaxy's light from a sparse ground-based sample (9 novae); this is now demonstrated to be valid into the central regions of M87 on the basis of our larger sample. 

The 32 novae detected in our survey, as suggested by Figure 1, show no obvious, strong concentration in the inner 10 arcsec of M87, contrary to the suggestion of \citet{mad07}. However, our F606W and F814W filter images are less sensitive than the HST STIS near-ultraviolet observations of \citet{mad07} to novae. This is especially true in the inner core of M87, where the high surface brightness of the light of red giants makes the detection of novae nearly impossible in visible light. Completeness tests are clearly needed to settle this question.

\subsection{Completeness}

Tests were conducted in order to determine our completeness in finding novae in M87.  We added 10 artificial novae in successive radial annuli of width 20 pixels.  The magnitudes of these artificial novae ranged from $\sim$21 to $\sim$27 in increments of 0.2 mags for $V$ five-day drizzled images, and both $I$ single and five-day drizzled images.  We then blinked through images, going to successively fainter magnitudes, to measure when the artificial novae were no longer visible in each radial annulus.  This process yielded limiting magnitudes for nova detection at specific distances from the center of M87.  The magnitude at which our completeness of detection is 50\% as a function of radial distance from the center of M87 is presented in Figure 15. As noted in Section 3.2, the least luminous novae ever detected display absolute $V$ magnitudes M = -5.8 \citep{sha81,shb81}, corresponding to $V$ = 25.3 in M87. Figure 15 warns that we must be missing a majority of the faint novae in the inner 25" of M87, and almost all novae in the inner 10". In the following section we discuss how we correct for this incompleteness in our determination of the nova rate in M87.

\section{The Nova Rate in M87}

Ground-based searches for extragalactic novae are often plagued by gaps in observing due to bad weather and/or the full moon, as well as large nightly variations in seeing and detection limits. None of these effects, of course, occur for HST data, so we were able to gather a highly uniform imaging dataset for M87 that is unique for studies of extragalactic novae. No other galaxy has ever been surveyed by HST as regularly, for as long a period of time, and with the same high cadence of observations, as M87. 

\subsection{A simple estimate} 

An simple estimate of the M87 annual nova rate R is the number of novae observed to erupt during 72 days (26 novae) divided by the fraction of a year over which they were observed. (We exclude the six novae, numbered  9, 23, 26, 28, 30 and 32, because their eruptions were underway before the observing window started. To be extremely conservative in our rate estimate we also exclude the nine faint variables, numbered 33-41, that are likely slow and/or symbiotic novae). We find R = 26/(72/365) = $132 \pm 11 $ novae/yr, where the estimated error is simply due to Poisson statistics. No corrections for incompleteness due to observing gaps \citep{cia90} are needed because of the high cadence - daily - of observations. We can also be confident that we are not missing novae because of dust obscuration, as evidenced by M87's lack of far infrared emission \citep{bae10}, lack of cool molecular gas \citep{tan08, sal08} and non-detection of significant intrinsic absorption in the galaxy's X-ray spectrum \citep{boh02}. 

A first significant correction that we must apply to the rate calculation is the fraction of novae we miss because they fall outside of HST's 202" x 202" FOV. \citet{kor09} measured the $V$ band magnitude of M87 to be 8.30. We measured M87's surface brightness profile from our F814W image and matched it to the \citet{kor09} extended $V$ band profile, allowing us to correct for the sky background of the ACS field. This also gave us the $V$ band zero point, tying our photometry to that of \citet{kor09}. We calculate the total apparent $V$ mag for the portion of M87 in the ACS as $V$ = 9.28, corresponding to 41\% of the galaxy's light. Applying only this areal correction, while ignoring both our incompleteness in detection of faint novae (discussed below), brings our simple rate estimate to $322 \pm 27$ novae/yr.

A second significant correction that we must apply to our rate calculation is the number of novae we miss because of the high surface brightness in the inner part of M87. Inspection of Figure 1 shows that nova 31, the second faintest of the entire sample (and, with nova 32, as faint as any nova has ever been measured at maximum light), is detected just 30 arc sec from the nucleus. This suggests that very few novae are missed because of background light any further out than nova 31. Conversely, 10 of the 15 brightest novae in our sample of 32 novae are located within 10" of M87's nucleus. Even more telling is that none of the 17 faintest novae in our sample are found in the inner 10" of M87. Both of these facts strongly suggest that fainter novae are being missed there. Figure 15 quantifies this suggestion, as discussed in the previous section, and we incorporate it in the following simulation.

\subsection{Simulated novae and the nova rate in M87} 

A more sophisticated simulation of the nova rate, which rigorously allows for the incompleteness quantified in Figure 15, and a realistic distribution of nova peak brightnesses and fade times was extensively described and implemented by \citet{nei04} to determine the nova rate in M81. We have adopted the same general methodology, which involves choosing synthetic novae at random from a representative sample of novae with a distribution of speed classes and maximum brightnesses that match those in the galaxy under observation. Each randomly chosen nova is placed in the 202" x 202" HST FOV of M87, weighted by the local surface brightness. If the nova was bright enough to be detected against the M87 background light of Figure 16 on at least 2 occasions then it was counted as part of the nova rate. Full details of the simulation methodology are given in \citet{nei04}.

The accuracy of our nova rate simulations in M87 is tied to the similarity of the distribution of nova maximum brightnesses and rates of decline in the set of light curves from which we draw novae at random. Until recently the largest sample of classical novae detected in uniform fashion in a single, {\it densely} sampled broadband survey was the 30 novae in M31 found by \citep{arp56} over the course of two observing seasons spanning 18 months. (The five year-long survey of \citet{cia87} detected 35 novae, but in H$\alpha$, and with much sparser time sampling.  \citet{ros73} and \citet{hub29} identified 44 and 85 novae in M31, respectively, but their light curves are relatively sparsely sampled and many of the rates of decline are indeterminate as maximum brightness was missed). The 18 month baseline of Arp's M31 sample is important because slow novae (those declining on timescales longer than 100 -150 days) are difficult to detect in shorter surveys, even those with a cadence as high as our M87 survey. The slowest novae in Arp's sample displayed decline times of 150 days, but slower Galactic novae (e.g. PU Vul and V723 Cas) are known to exist. Are they common? It is important to answer this question as a distribution of light curves that is skewed towards fast, luminous novae and missing very slow, less luminous novae will incorrectly estimate a galaxy's nova rate.

Despite the irregular cadence of observations, and gaps due to lunation and variable seeing, the fraction of very slow novae in M31 (those novae with decline times of years), can be estimated from the work of \citet{hub29}. Three novae out of 85 he identified were visible for {\it at least} 4.5, 5.5 and 11 yrs. They are probably symbiotic novae (see \cite{mik10} for a recent review of symbiotic novae). The next longest observed nova in \citet{hub29}'s sample was seen for 238 days; during this time it was within 1.7 magnitudes of its peak brightness. It was not observed in previous or subsequent observing seasons. Three very slow symbiotic novae from a total sample of 85 novae corresponds to an occurrence rate of about 3.5\%. 

Recently the results of a search for novae during the OGLE campaign have been published by \citet{mro15}. 34 Galactic Bulge novae with well-determined decline times were identified in this and other Galactic Bulge surveys, as were two symbiotic novae. This suggests a somewhat higher incidence of symbiotic novae, of order 5.5\%, than detected by \citet{hub29}, but of course both samples suffer from small-number statistics. Nevertheless, it is clear from both the \citet{hub29} and \citet{mro15} samples that very slow novae, detectable for longer than about two years, are probably not more than about 5\% of the total nova population.

The 15-year long OGLE nova survey has the longest baseline and highest cadences (typically several times weekly) of any published continuum-band nova survey. As already noted, high cadence is important to detect the fastest novae, while long baselines (years) are essential to detect the most slowly declining novae. A comparison of the well-determined decline times for 22 novae in M87 from the survey presented in this paper, 30 novae in M31 from Arp's survey and 34 Galactic Bulge novae from \citet{mro15} is shown in Figure 16. It is clear that only the \citet{mro15} Bulge survey detects the slowest novae, including one with a decay time of 600 days omitted from the figure. We thus chose to use the \citet{mro15} sample, rather than our own M87 or the Arp M31 samples as the source of novae for our simulation. 
The decay times of the Bulge sample are very well determined. We assigned their absolute magnitudes at maximum by matching the decay times to those of M31 novae.  

We carried out 10,000 trials in which we chose novae at random from the \citet{mro15} Bulge sample of novae, and placed them at random in M87, weighted by the local surface brightness. Our simulation determined whether the randomly chosen nova was bright enough to be seen against the light of M87 where it was placed. The results are shown in Figure 17, wherein we derive a most probable rate in the 202" x 202" HST FOV to be $R = 149_{-18.3}^{+13.4}$ novae/year. This rate is consistent with the simple estimate ($132 \pm 11 $ novae/yr) given above. Also as noted above, the ACS FOV covers just 41\% of the light of M87. Correcting for this areal incompleteness we arrive at the global M87 nova rate of $363_{-45}^{+33}$ novae/year in M87. This rate is 4 times higher than the ground-based observations' estimate of 91 $\pm 34$ novae/yr in M87 \citep{sha00}, and 2.4 times higher than the ground-based estimated rate of M87 novae of $ 154_{-19}^{+23}$ of \citep{cur15}. 
 
By adopting an M87 distance of $15.2 \pm 0.2$ Mpc, \citet{sha00} derived a K - band luminosity of M87 of $39.8 \pm 8.2 $ x $10^{10}L_\odot,_{K}$. Correcting that luminosity to our adopted distance of $16.4 \pm 0.4$ Mpc, and combining with our rate of $363_{-45}^{+33}$ novae/year we derive a luminosity-specific nova rate (LSNR) of $7.88_{-2.6}^{+2.3} /yr/ 10^{10}L_\odot,_{K}$, which is also 3.4 times higher than the measurement of \citet{sha00}. This discrepancy is understandable if previous (mostly) ground-based surveys miss both the faint, fast novae, and almost all slow novae near the centers of galaxies. We conclude that the nova rates and LSNR of galaxies are several times larger than the values currently accepted. 

\section{Summary and Conclusions}

During a 72 day-long observing campaign we have located 32 erupting classical novae in the giant elliptical galaxy M87. We also found nine likely slow and/or symbiotic novae.Their light and color curves are presented, and their spatial distribution is shown to follow that of the galaxy light. No correlations are found between the colors, rise or decline times, or position within M87, of the novae. Six of the novae are of the faint and fast variety first described by \citet{kas11}. This detection demonstrates the ubiquitous nature of these novae, and further weakens claims of the utility of the decline rates of classical novae as distance indicators. 

We find twice as many of the most rapidly declining novae (in less than 5 days) compared to novae with 5-10 day decline times. We compare the distribution of speed classes of our 32 classical novae in M87 with those in M31 and in the Galactic bulge. Each of the longer time baseline surveys (in M31 and the Galactic bulge) has a larger fraction of slow novae than our M87 sample. After accounting for incompleteness in our discovery of fainter novae, especially near the center of M87, and for the unobserved outer parts of this galaxy, we derive a annual nova rate of $363_{-45}^{+33}$ novae/yr for M87. This is a conservative estimate because we omitted the fainter slow and/or symbiotic nova candidates. We also report the LSNR for this galaxy, which is $7.88_{-2.6}^{+2.3} /yr/ 10^{10}L_\odot,_{K}$. Both rates are 3.4x times larger than those for M87 and M49 reported by \citet{sha00} and by \citet{fer03}, respectively. We suggest that most previous surveys miss both the faint, fast novae, and almost all slow novae near the centers of galaxies. Thus the nova rates and LSNR of galaxies are probably several times larger than the values currently accepted. 

\acknowledgments
We gratefully acknowledge the support of the STScI team responsible for ensuring timely and accurate implementation of our M87 program. Support for program \#10543 was provided by NASA through a grant from the Space Telescope Science Institute, which is operated by the Association of Universities for Research in Astronomy, Inc., under NASA contract NAS 5-26555. This research has been partly supported by the Polish NCN grant DEC-2013/10/M/ST9/00086. MMS gratefully acknowledges the support of Hilary and Ethel Lipsitz, longtime friends of the AMNH Astrophysics  department. We thank a referee for a careful reading of, and useful suggestions that improved earlier versions of the paper.

\clearpage

\begin{deluxetable}{lccccc}
\tabletypesize{\footnotesize}
\tablecaption{HST Observations of M87}
\label{tab:obstable}
\tablewidth{0pt}
\tablehead{
\colhead{Root Name} & \colhead{Obs Date} & \colhead{Exp Time} & \colhead{Root Name} & \colhead{Obs Date} & \colhead{Exp Time} \\
 & DD-MM-YYY & (sec)  & & DD-MM-YYY & (sec)}       
\startdata                
\multicolumn{6}{c}{HST ACS/WFC (F606W)} \\             
\hline
j9e001zdq & 24-12-2005 & 500.0 & j9e032pxq & 04-02-2006 & 500.0 \\     
j9e002hgq & 29-12-2005 & 500.0 & j9e033smq & 05-02-2006 & 500.0 \\     
j9e003d6q & 03-01-2006 & 500.0 & j9e034w9q & 06-02-2006 & 500.0 \\     
j9e004atq & 08-01-2006 & 500.0 & j9e035bkq & 06-02-2006 & 500.0 \\     
j9e005d7q & 09-01-2006 & 500.0 & j9e036dbq & 07-02-2007 & 500.0 \\     
j9e006j4q & 10-01-2006 & 500.0 & j9e037gnq & 08-02-2006 & 500.0 \\     
j9e007lwq & 11-01-2006 & 500.0 & j9e038k1q & 09-02-2006 & 500.0 \\
\enddata
\tablenotetext{a}{Empty observations. Not used in analysis, but shown here for completeness.}
\label{obstab}
\end{deluxetable}
\clearpage

\centering
\begin{deluxetable}{lccccccccccc}
\floattable
\rotate
\tabletypesize{\scriptsize}
\tablecolumns{12}
\tablewidth{0pt}
\tablehead{
\colhead{HST File} & \colhead{Days from Max} & \colhead{Obs Date} & \colhead{$V$} & \colhead{$V_{err}$} & \colhead{HST File} & \colhead{Days from Max} & \colhead{Obs Date} & \colhead{I} & \colhead{$I_{err}$} & \colhead{$(V-I)$} & \colhead{$(V-I)_{err}$} \\
 & days & DD-MM-YYYY & mag & mag & & days & DD-MM-YYYY & mag & mag & mag & mag}
\startdata
\multicolumn{12}{c}{M87 Nova 1} \\
\hline
j9e001zdq & -17.37 & 24-12-2005 & -- & -- & j9e001z6q & -17.37  & 24-12-2005 & 27.843 & 1.06 & -- & -- \\
j9e002hgq & -12.43 & 29-12-2005 & -- & -- & j9e002h9q & -12.39  & 29-12-2005 & 29.933 & 2.72 & -- & -- \\
j9e003d6q & -7.37 & 03-01-2006 & 27.070 & 1.302 & j9e003czq & -6.33  & 03-01-2006 & -- & -- & -- & -- \\
j9e004atq & -2.12 & 08-01-2006 & 23.333 & 0.040 & j9e004amq & -1.13  & 08-01-2006 & 23.223 & 0.03 & 0.110 & 0.050 \\
j9e005d7q & -1.26 & 09-01-2006 & 22.561 & 0.020 & j9e005d0q & -0.26  & 09-01-2006 & 22.423 & 0.01 & 0.138 & 0.022 \\
j9e006j4q & 0.00 & 10-01-2006 & 21.837 & 0.012 & j9e006ixq & 0.94  & 10-01-2006 & 21.733 & 0.01 & 0.104 & 0.016 \\
j9e007lwq & 1.13 & 11-01-2006 & 21.942 & 0.012 & j9e007lpq & 2.13  & 11-01-2006 & 21.543 & 0.01 & 0.399 & 0.016 \\
\enddata
\end{deluxetable}
\clearpage
\begin{deluxetable}{lccccccccc}
\tabletypesize{\footnotesize}
\tablecaption{Positions, and Rise and Decline Times of M87 Novae}
\label{tab:risedec}
\tablewidth{0pt}
\tablehead{
\colhead{Nova} & \colhead{Radial Distance} & \colhead{RA} & \colhead{DEC} & \colhead{I$_{\mathrm{max}}$} & \colhead{V$_{\mathrm{max}}$} & \colhead{t$_{\mathrm{\mathrm{1rise}}}$} & \colhead{t$_{\mathrm{2rise}}$} & \colhead{t$_{\mathrm{1decline}}$} & \colhead{t$_{\mathrm{2decline}}$} \\
\# & arcsec & HH:MM:SS.ss  & DD:MM:SS.ss & mag & mag & days & days & days & days}       
\startdata                
1 & 14.80 & 12:30:48.85 & +12:23:16.91 & 21.54 & 21.84 & 1.97 & 4.60 & 8.52 & 15.24 \\
2 & 33.32 & 12:30:49.85 & +12:22:56.80 & 21.85 & 22.25 & 1.41 & 2.85 & 0.92 & 11.25 \\
3 & 110.87 & 12:30:54.23 & +12:22:03.05 & 22.71 & 22.64 & 4.86 & 9.72 & 0.94 & 2.01 \\
4 & 4.99 & 12:30:49.60 & +12:23:24.82 & 22.28 & 22.74 & 2.27 & 4.74 & 3.56 & 7.66 \\
5 & 14.20 & 12:30:46.86 & +12:23:47.97 & 22.74 & 22.99 & 1.34 & 2.46 & 7.13 & 17.13 \\
6 & 42.23 & 12:30:50.38 & +12:23:25.86 & 22.70 & 23.18 & 0.93 & 2.61 & 3.70 & 11.17 \\
7 & 10.86 & 12:30:49.92 & +12:23:20.93 & 23.61 & 23.21 & 0.90 & 1.80 & 4.75 & 9.33 \\
8 & 18.67 & 12:30:48.18 & +12:23:31.37 & 22.95 & 23.37 & 1.75 & 2.75 & 1.82 & 3.72 \\
\enddata
\end{deluxetable}



\begin{figure*}
\figurenum{2a}
\includegraphics[width=1\columnwidth]{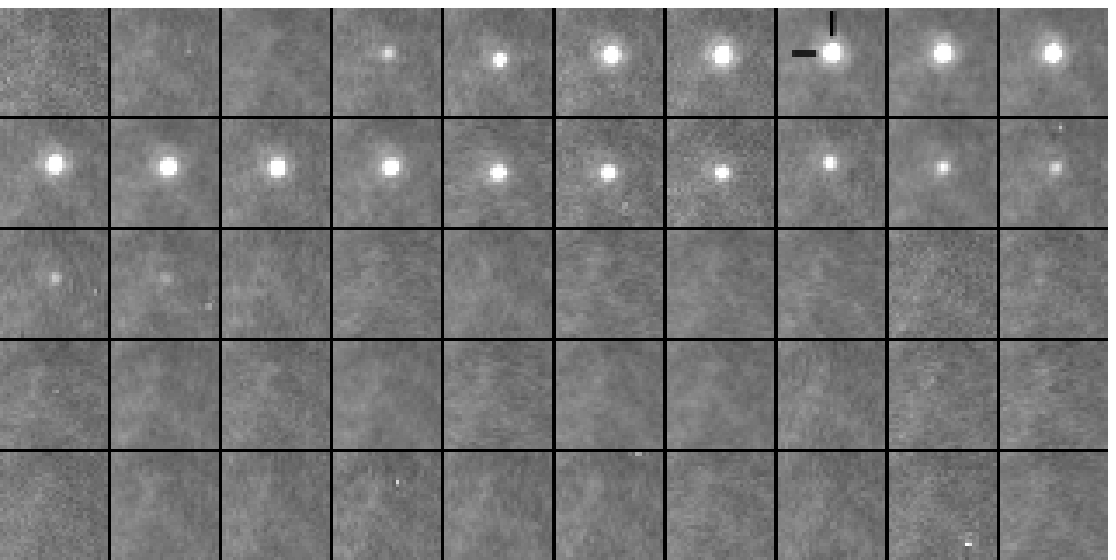}
\includegraphics[width=1\columnwidth]{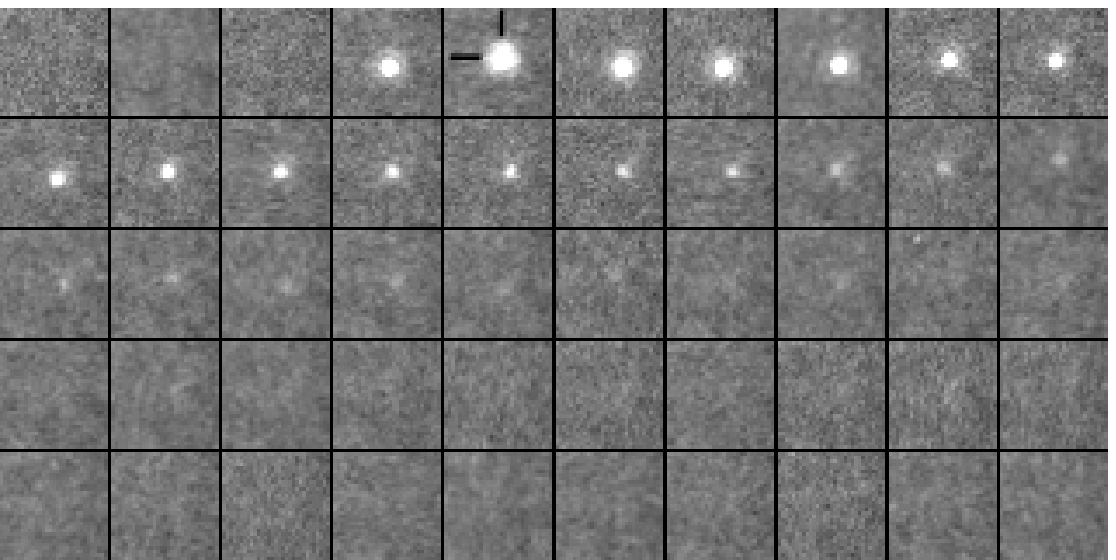}
\caption{Daily $I$ band images of M87 Nova 1 (top) and M87 Nova 2 (bottom).  Novae are ordered by peak brightness in the $V$ band; a vertical and horizontal tic indicate the day of peak brightness. All ``postage stamps" are 1.5 x 1.5 arc sec in size.}
\end{figure*}

\clearpage

\begin{figure*}
\figurenum{2b}
\includegraphics[width=1\columnwidth]{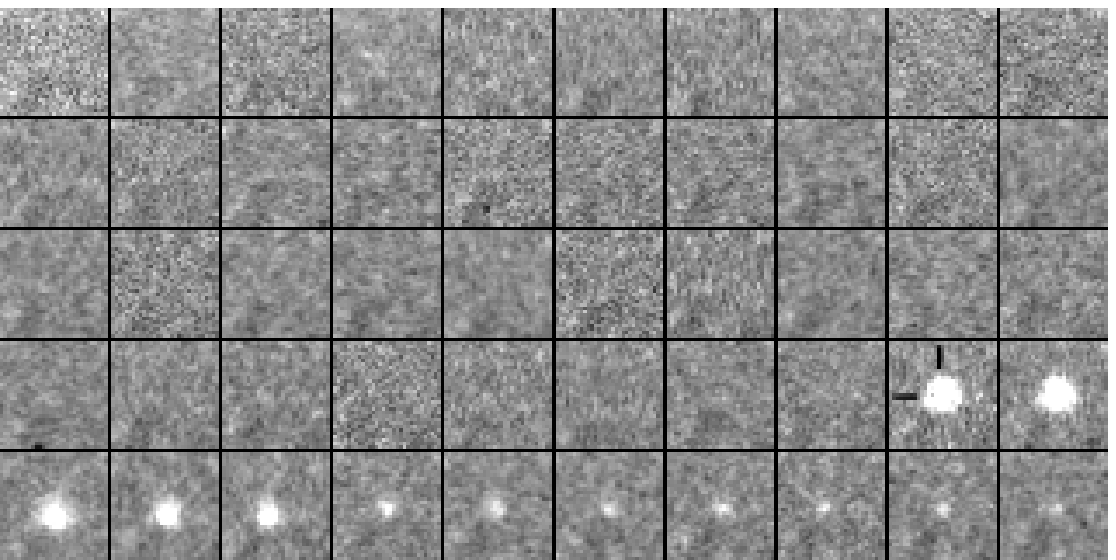}
\includegraphics[width=1\columnwidth]{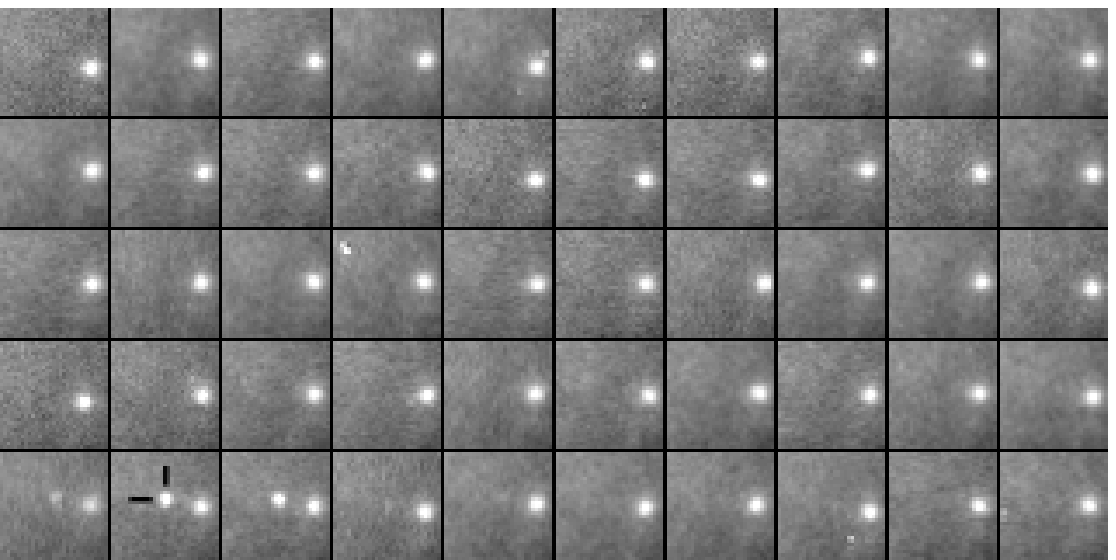}
\caption{Same as Figure 1a but for M87 Nova 3 (top) and M87 Nova 4 (bottom).}
\end{figure*}

\clearpage

\begin{figure*}
\figurenum{2c}
\includegraphics[width=1\columnwidth]{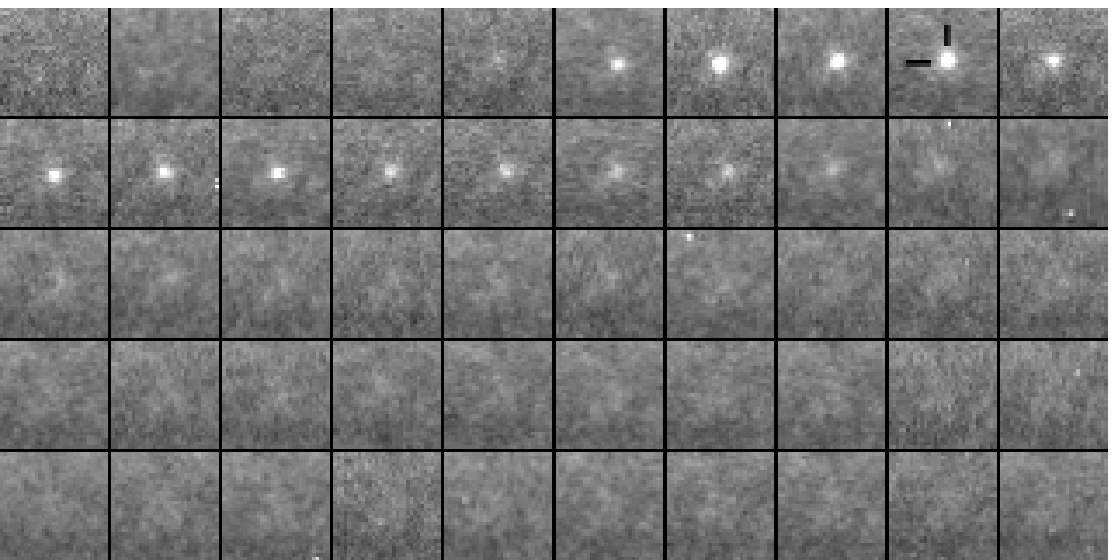}
\includegraphics[width=1\columnwidth]{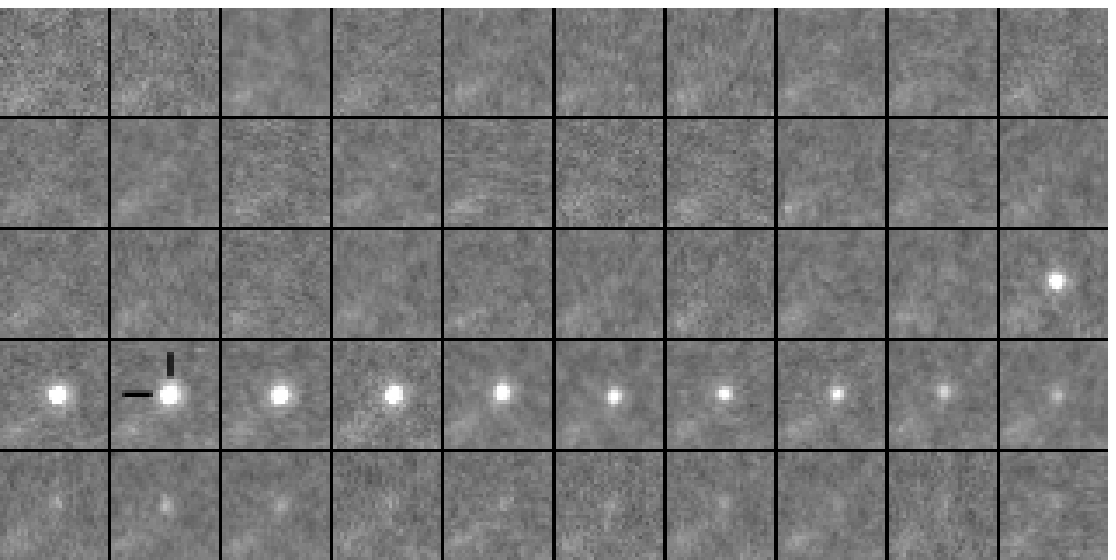}
\caption{Same as Figure 1a but for M87 Nova 5 (top) and M87 Nova 6 (bottom).}
\end{figure*}

\clearpage

\begin{figure*}
\figurenum{2d}
\includegraphics[width=1\columnwidth]{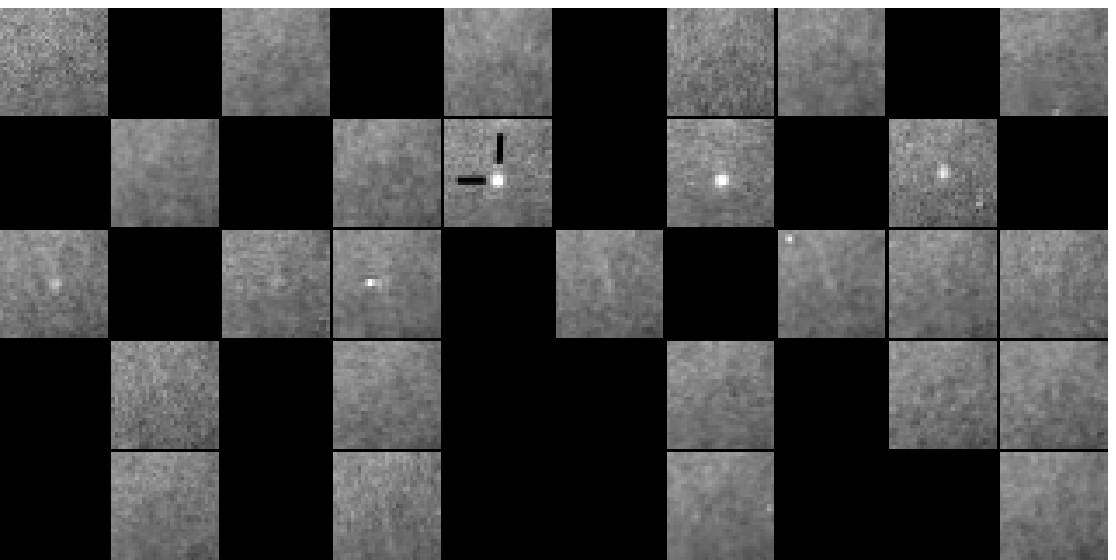}
\includegraphics[width=1\columnwidth]{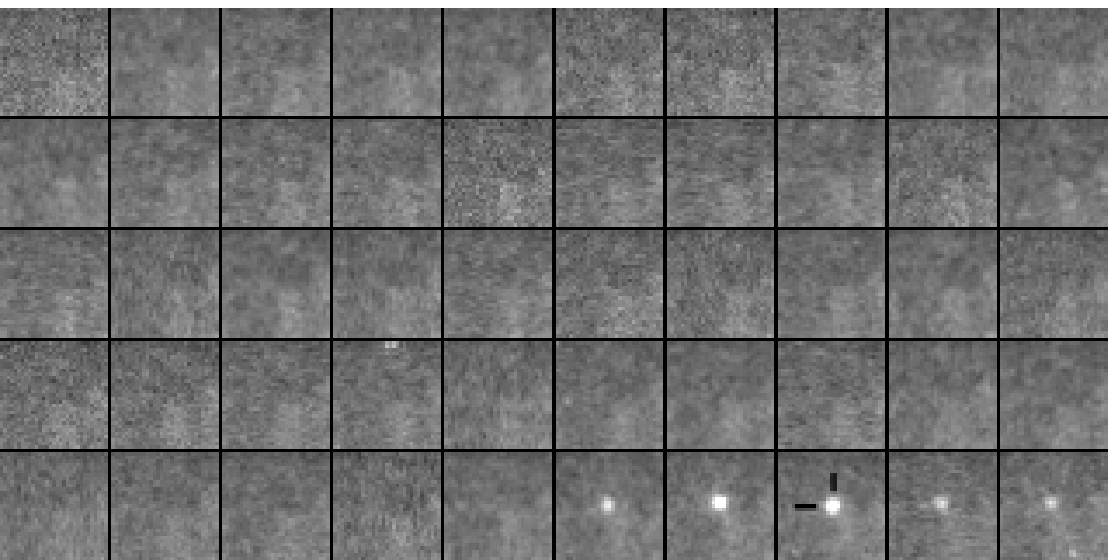}
\caption{Same as Figure 1a but for M87 Nova 7 (top) and M87 Nova 8 (bottom).}
\end{figure*}

\clearpage

\begin{figure*}
\figurenum{2e}
\includegraphics[width=1\columnwidth]{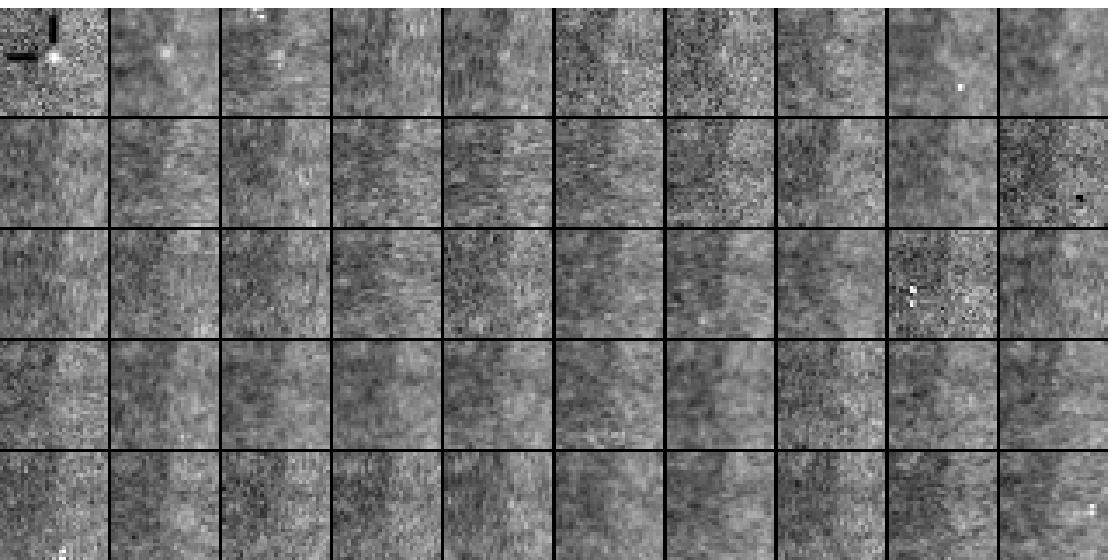}
\includegraphics[width=1\columnwidth]{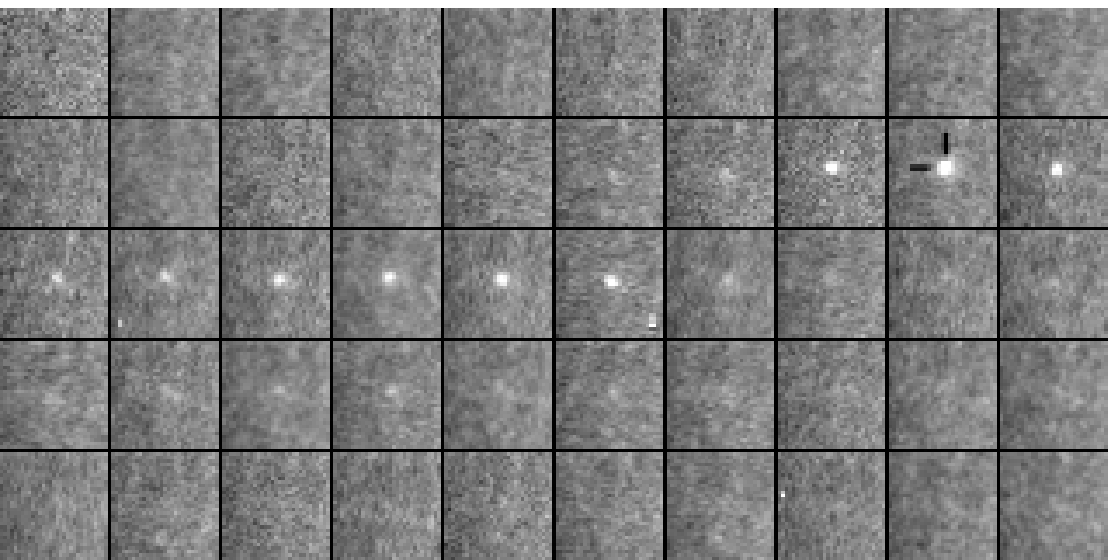}
\caption{Same as Figure 1a but for M87 Nova 9 (top) and M87 Nova 10 (bottom).}
\end{figure*}

\clearpage

\begin{figure*}
\figurenum{2f}
\includegraphics[width=1\columnwidth]{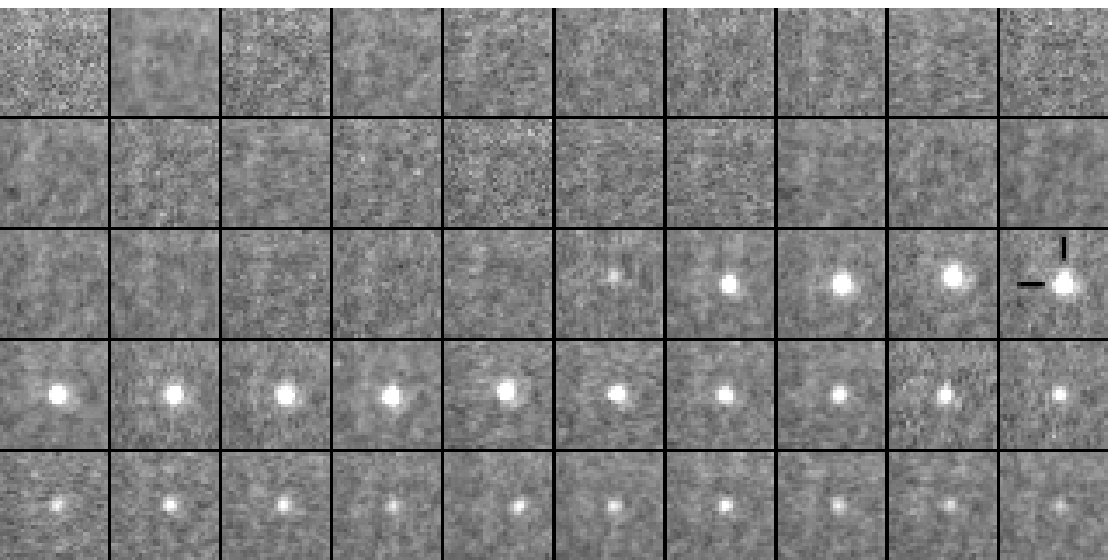}
\includegraphics[width=1\columnwidth]{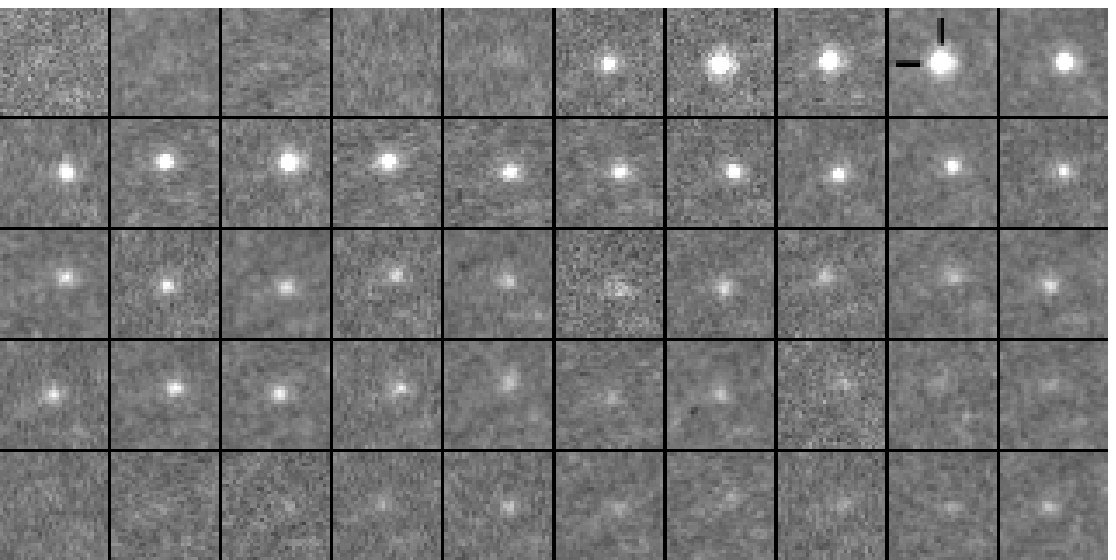}
\caption{Same as Figure 1a but for M87 Nova 11 (top) and M87 Nova 12 (bottom).}
\end{figure*}

\clearpage

\begin{figure*}
\figurenum{2g}
\includegraphics[width=1\columnwidth]{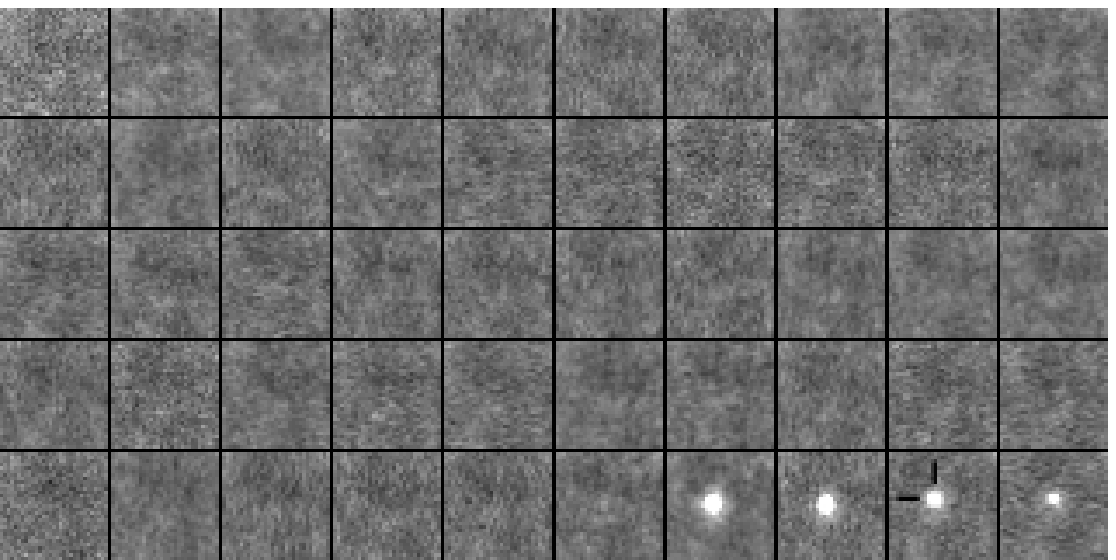}
\includegraphics[width=1\columnwidth]{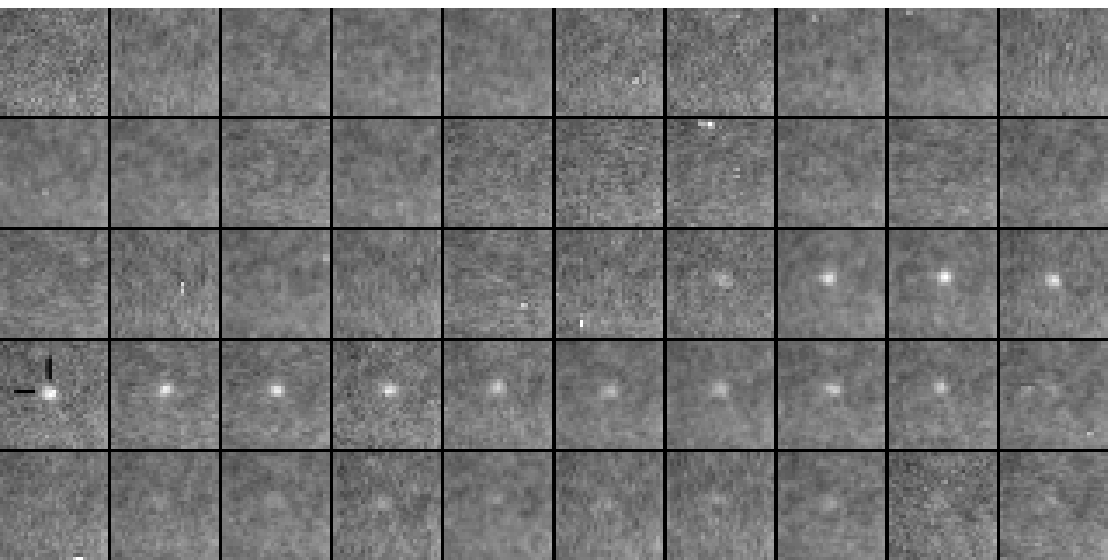}
\caption{Same as Figure 1a but for M87 Nova 13 (top) and M87 Nova 14 (bottom).}
\end{figure*}

\clearpage

\begin{figure*}
\figurenum{2h}
\includegraphics[width=1\columnwidth]{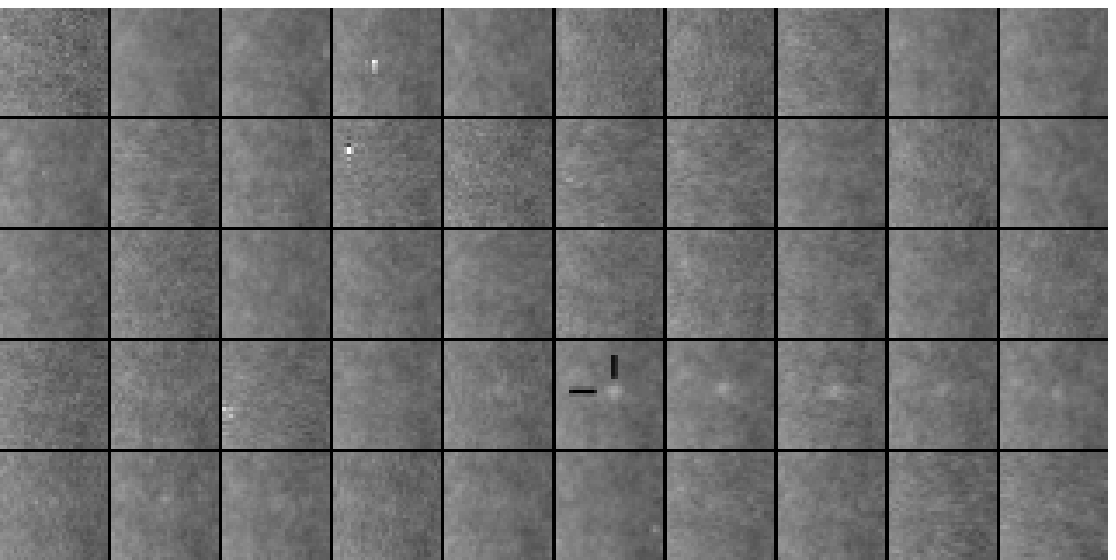}
\includegraphics[width=1\columnwidth]{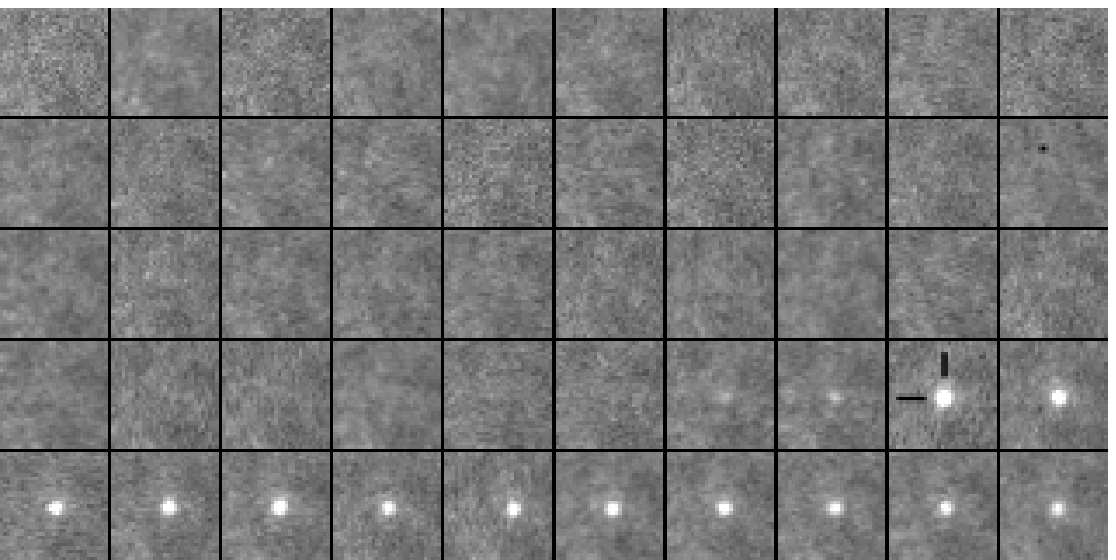}
\caption{Same as Figure 1a but for M87 Nova 15 (top) and M87 Nova 16 (bottom).}
\end{figure*}

\clearpage

\begin{figure*}
\figurenum{2i}
\includegraphics[width=1\columnwidth]{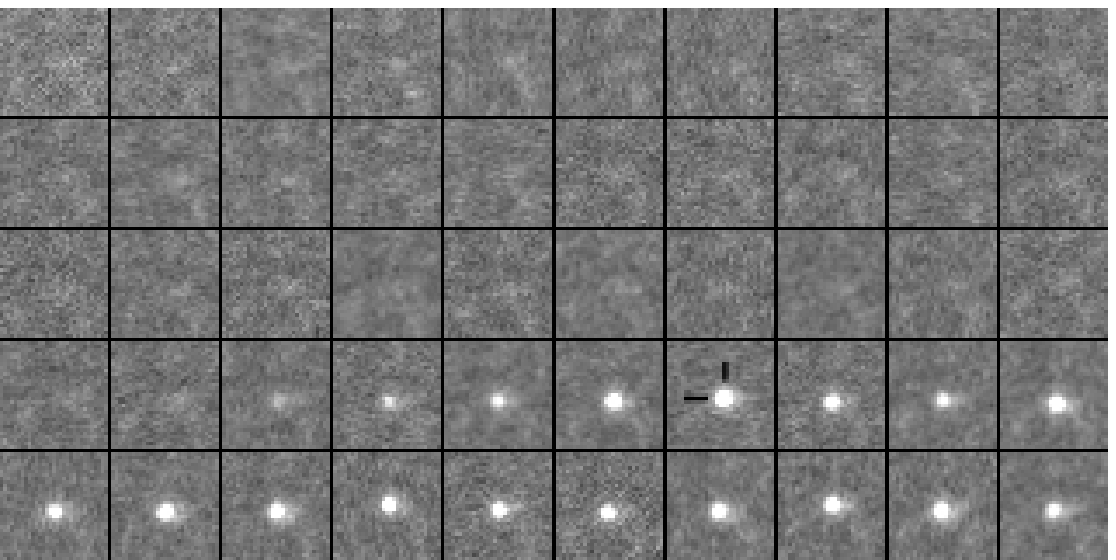}
\includegraphics[width=1\columnwidth]{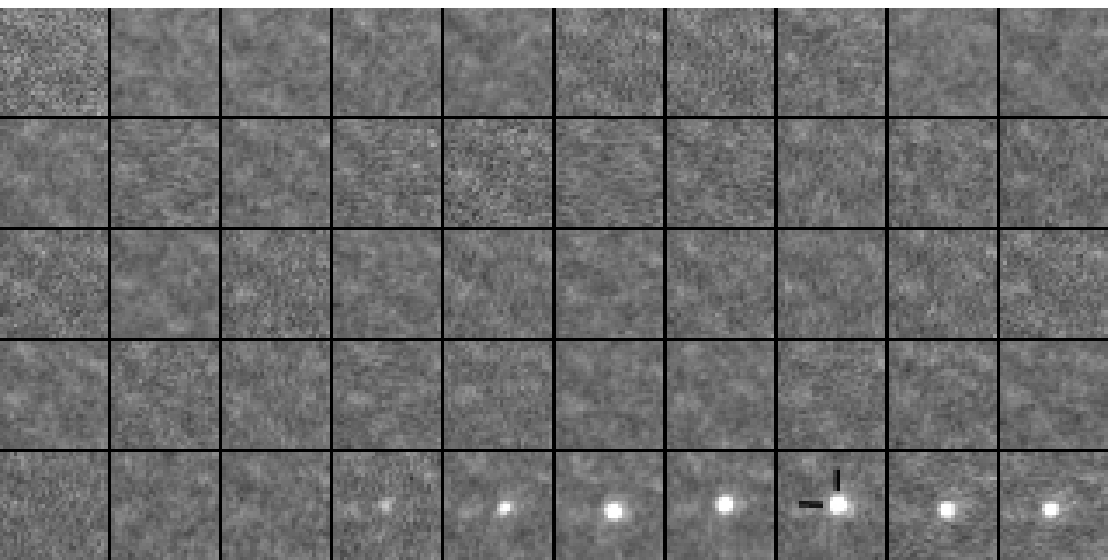}
\caption{Same as Figure 1a but for M87 Nova 17 (top) and M87 Nova 18 (bottom).}
\end{figure*}

\clearpage

\begin{figure*}
\figurenum{2j}
\includegraphics[width=1\columnwidth]{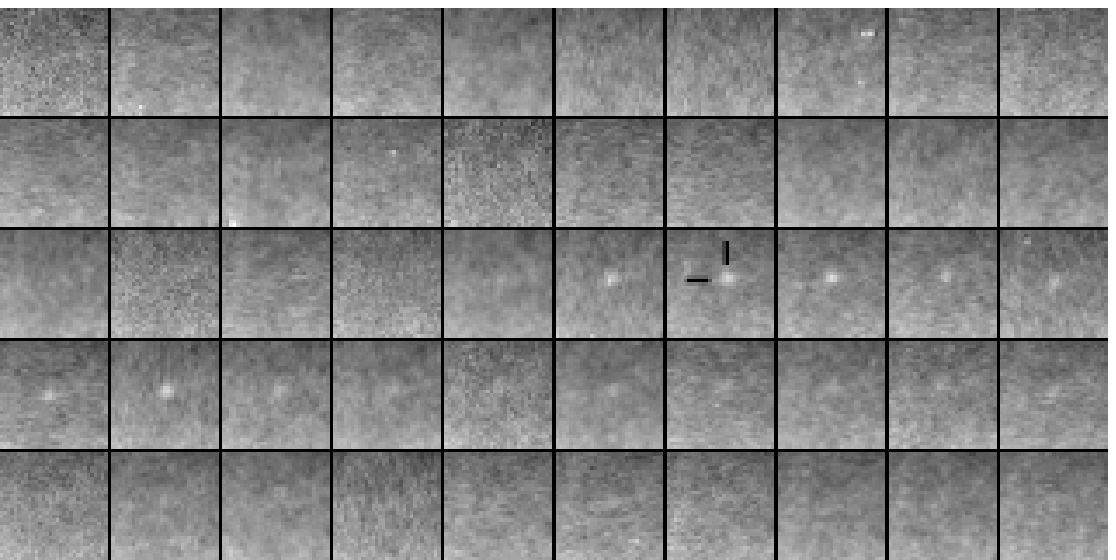}
\includegraphics[width=1\columnwidth]{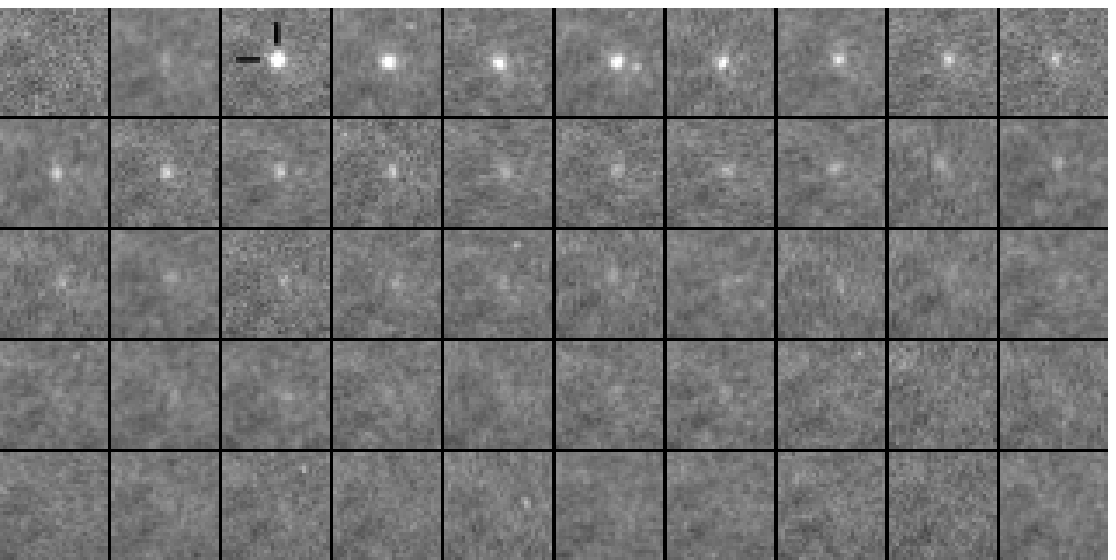}
\caption{Same as Figure 1a but for M87 Nova 19 (top) and M87 Nova 20 (bottom).}
\end{figure*}

\clearpage

\begin{figure*}
\figurenum{2k}
\includegraphics[width=1\columnwidth]{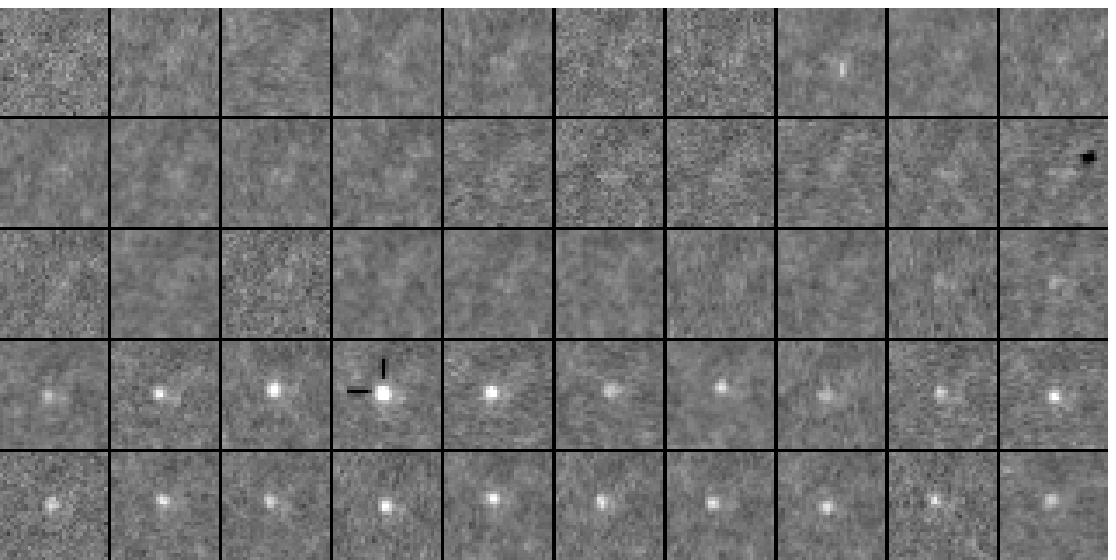}
\includegraphics[width=1\columnwidth]{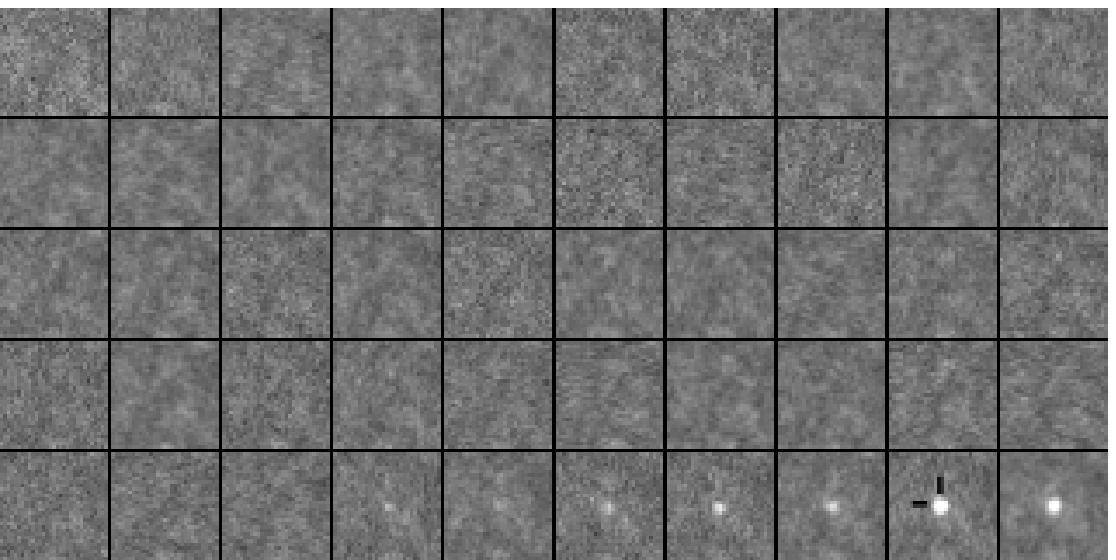}
\caption{Same as Figure 1a but for M87 Nova 21 (top) and M87 Nova 22 (bottom).}
\end{figure*}

\clearpage

\begin{figure*}
\figurenum{2l}
\includegraphics[width=1\columnwidth]{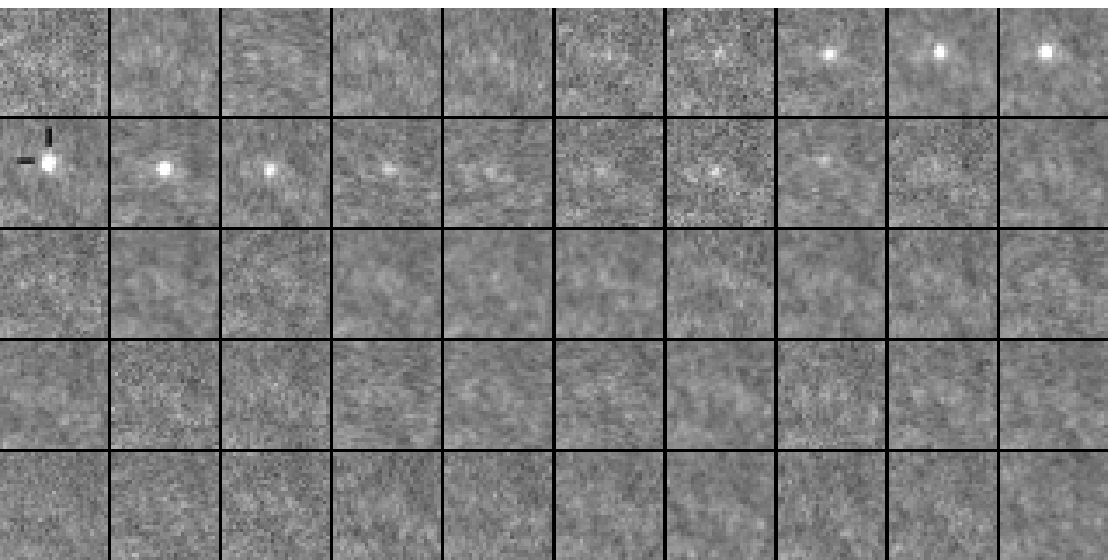}
\includegraphics[width=1\columnwidth]{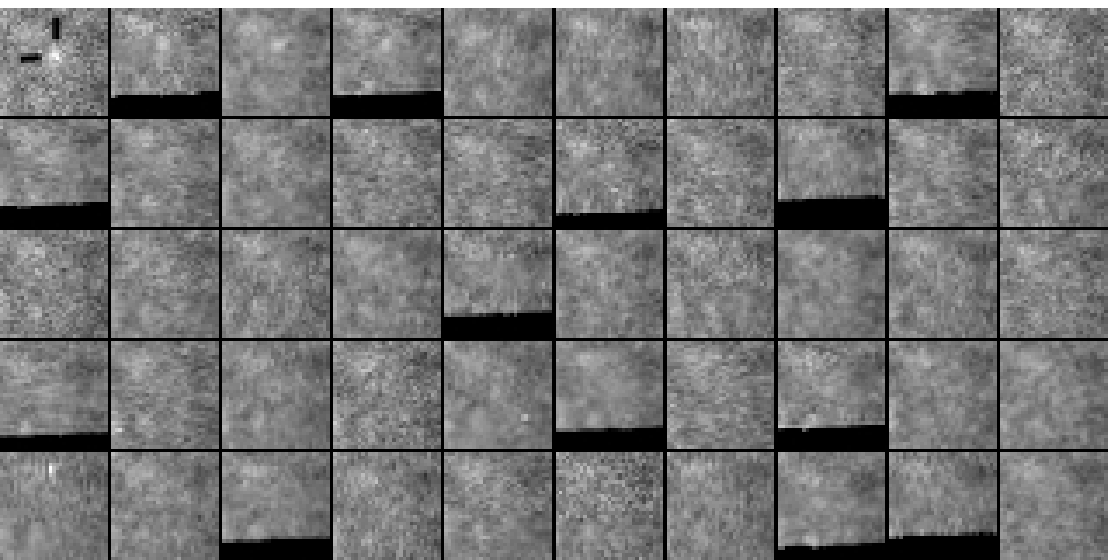}
\caption{Same as Figure 1a but for M87 Nova 23 (top) and M87 Nova 24 (bottom).}
\end{figure*}

\clearpage

\begin{figure*}
\figurenum{2m}
\includegraphics[width=1\columnwidth]{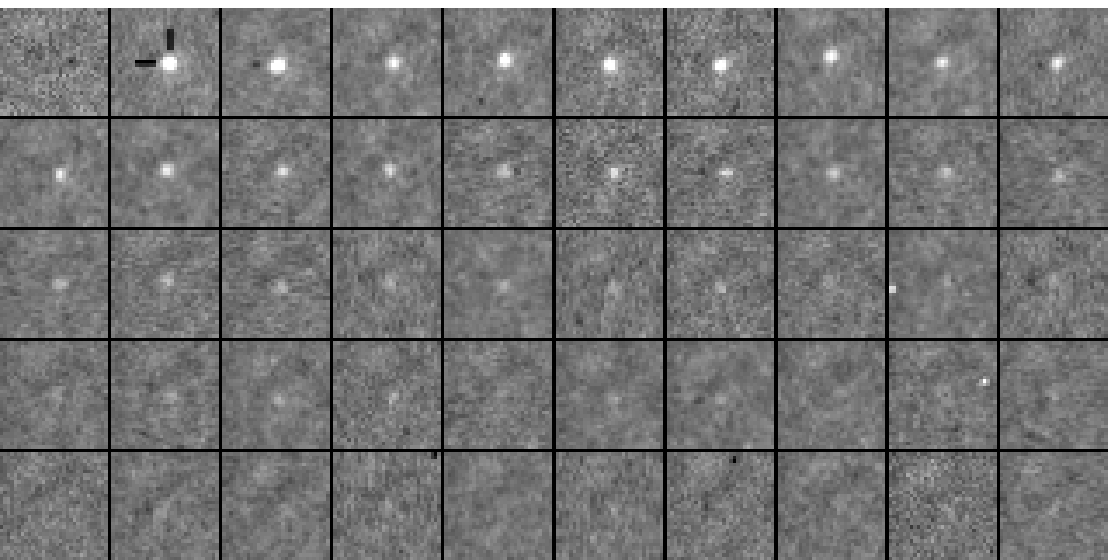}
\includegraphics[width=1\columnwidth]{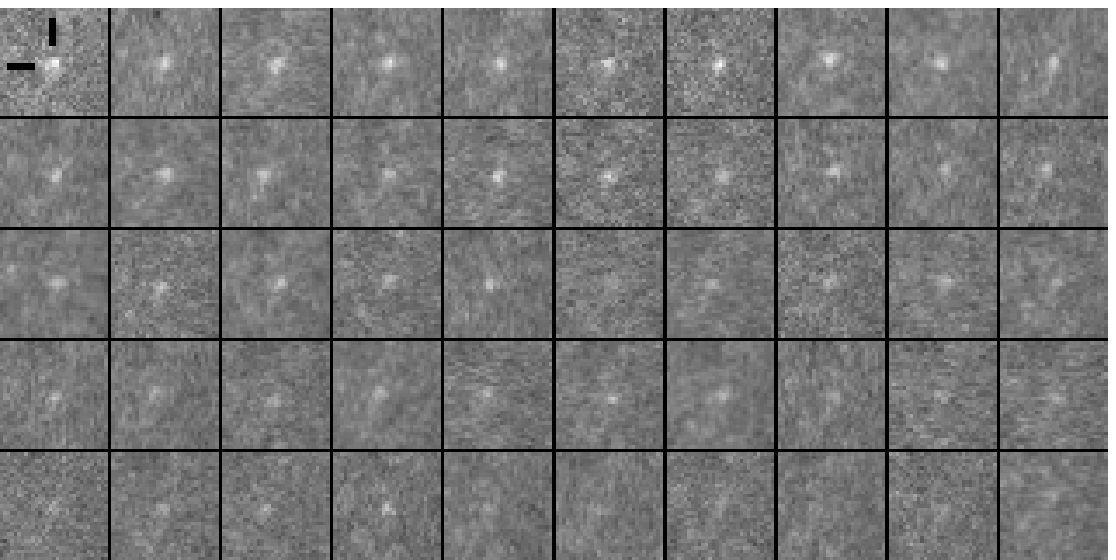}
\caption{Same as Figure 1a but for M87 Nova 25 (top) and M87 Nova 26 (bottom).}
\end{figure*}

\clearpage

\begin{figure*}
\figurenum{2n}
\includegraphics[width=1\columnwidth]{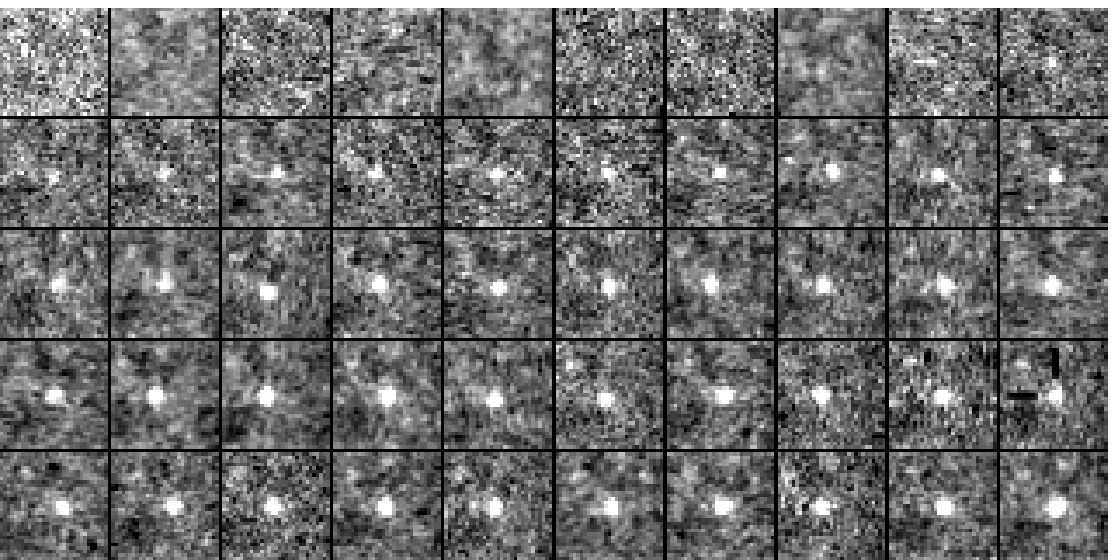}
\includegraphics[width=1\columnwidth]{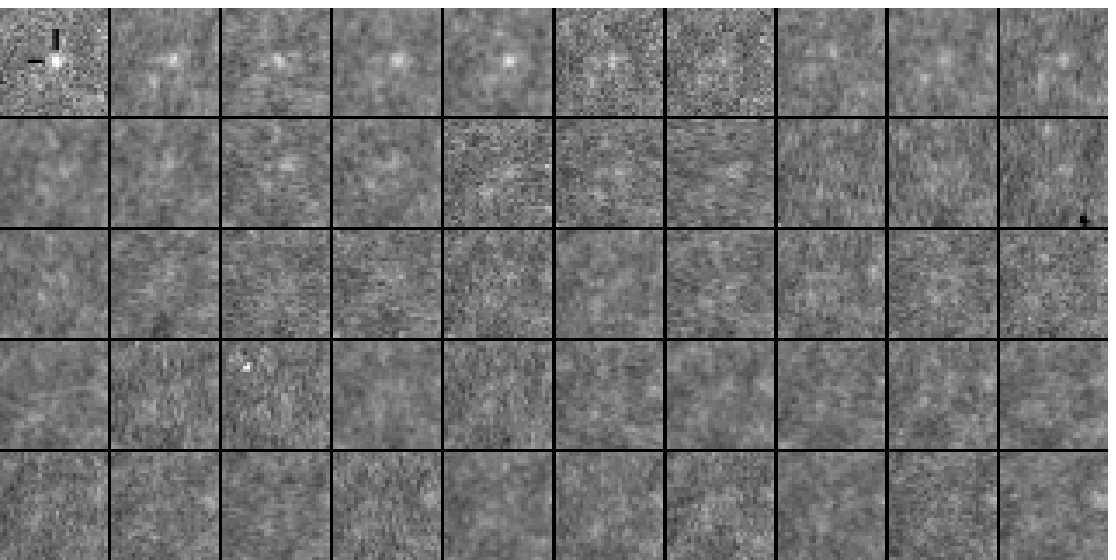}
\caption{Same as Figure 1a but for M87 Nova 27 (top) and M87 Nova 28 (bottom).}
\end{figure*}

\clearpage

\begin{figure*}
\figurenum{2o}
\includegraphics[width=1\columnwidth]{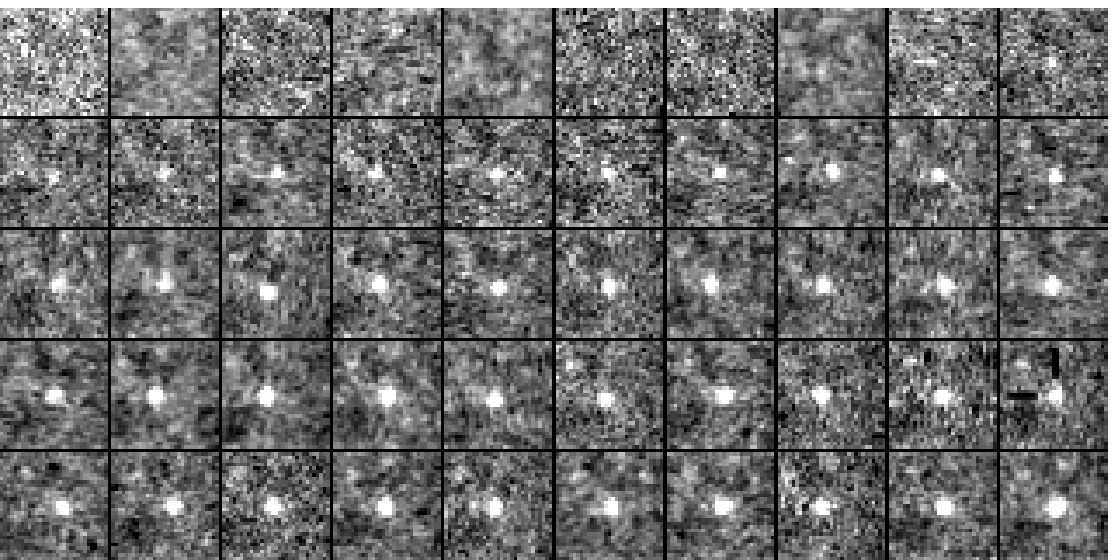}
\includegraphics[width=1\columnwidth]{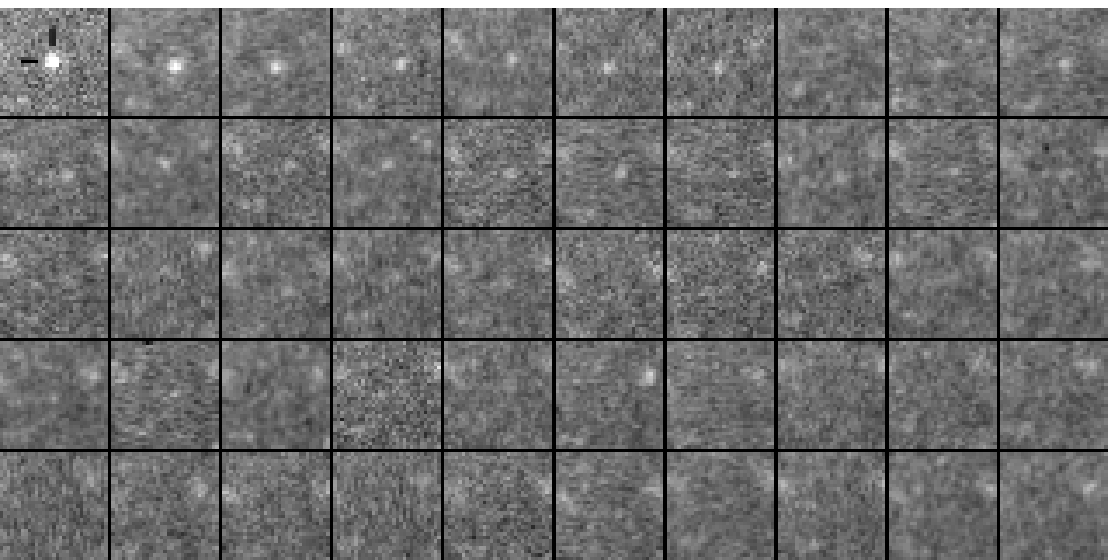}
\caption{Same as Figure 1a but for M87 Nova 29 (top) and M87 Nova 30 (bottom).}
\end{figure*}

\clearpage

\begin{figure*}
\figurenum{2p}
\includegraphics[width=1\columnwidth]{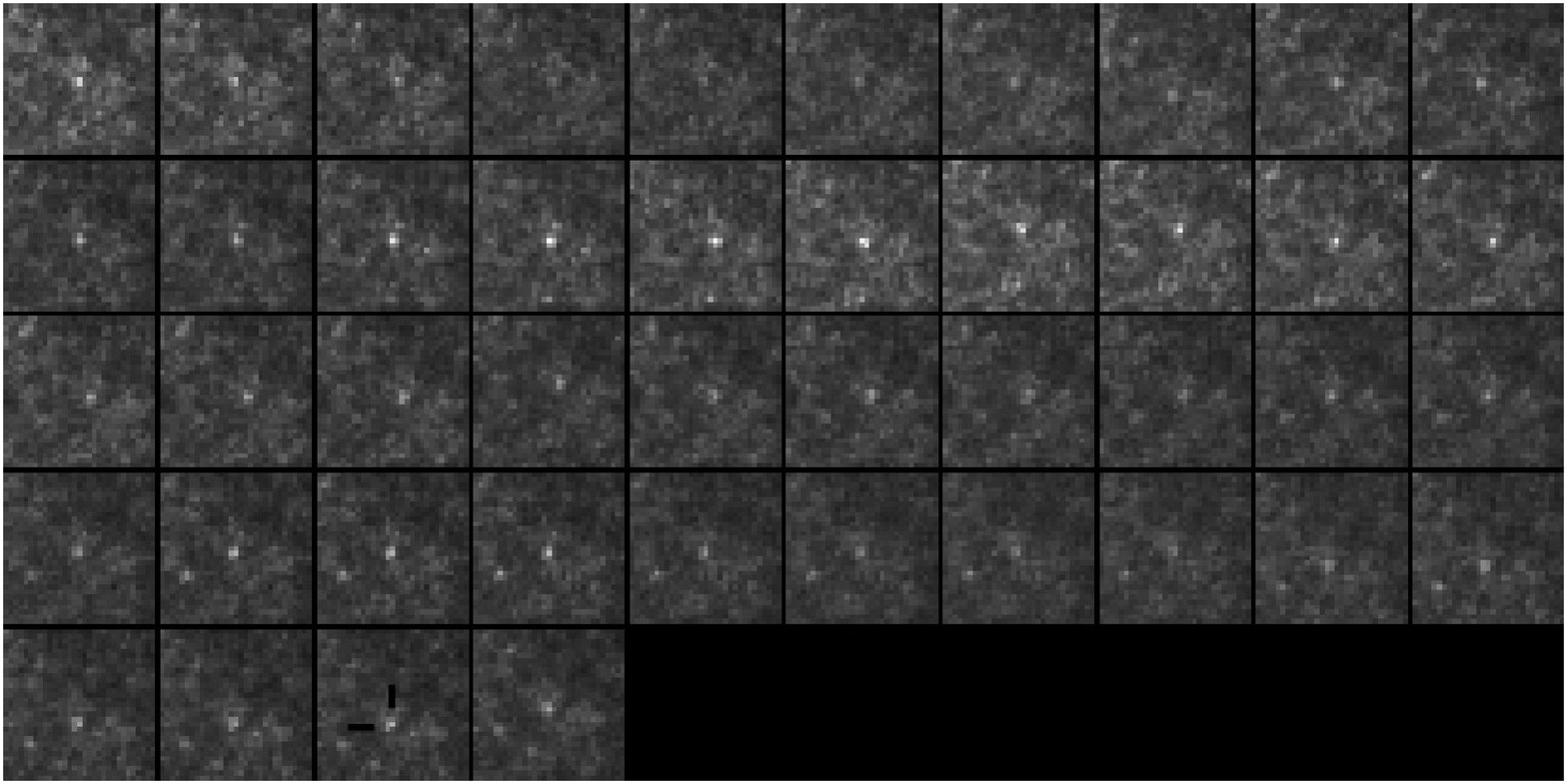}
\includegraphics[width=1\columnwidth]{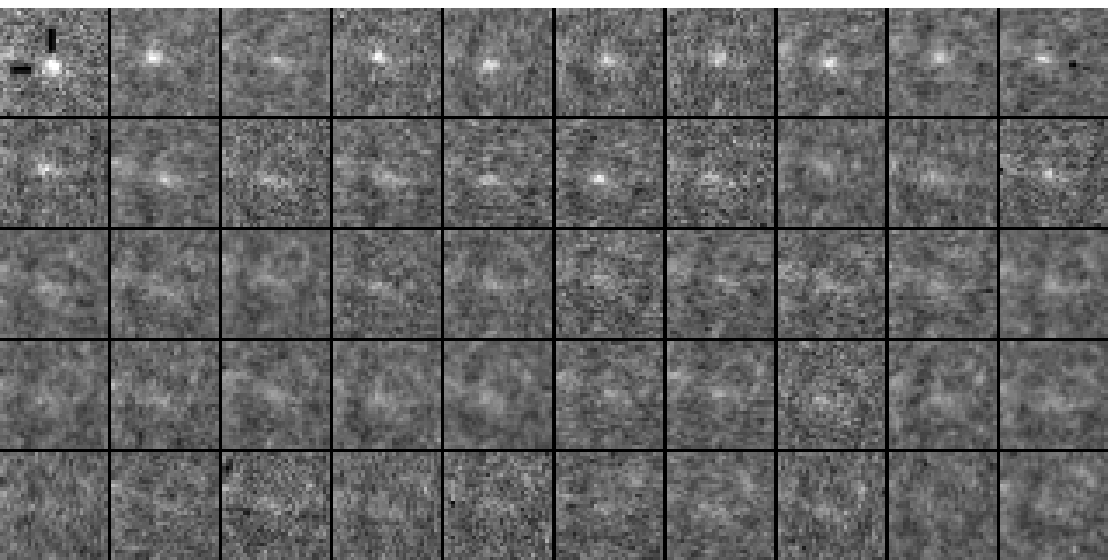}
\caption{Same as Figure 1a but for M87 Nova 31 (top) and M87 Nova 32 (bottom).}
\end{figure*}

\clearpage

\begin{figure*}
\figurenum{2q}
\includegraphics[width=1\columnwidth]{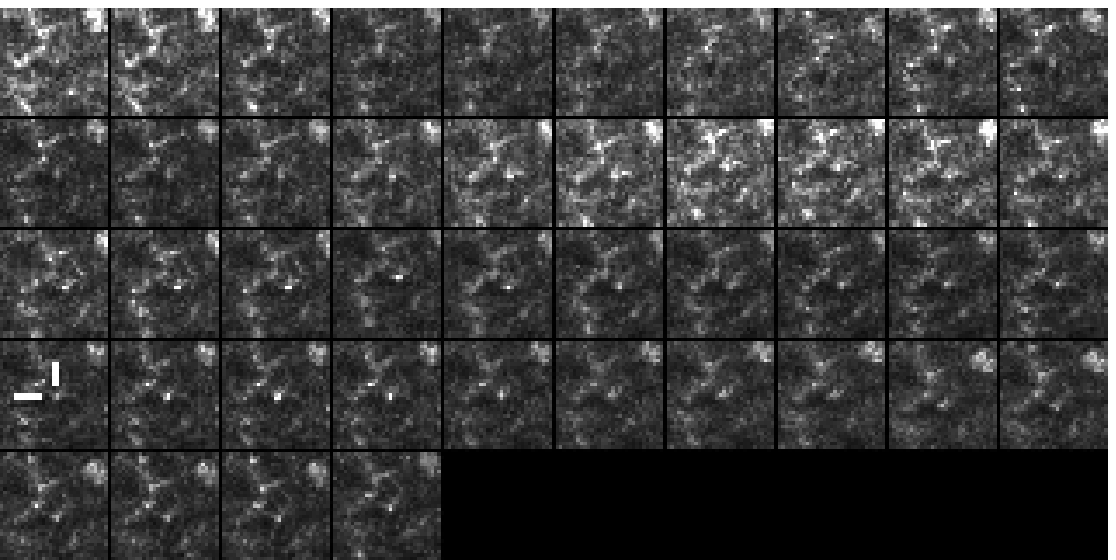}
\includegraphics[width=1\columnwidth]{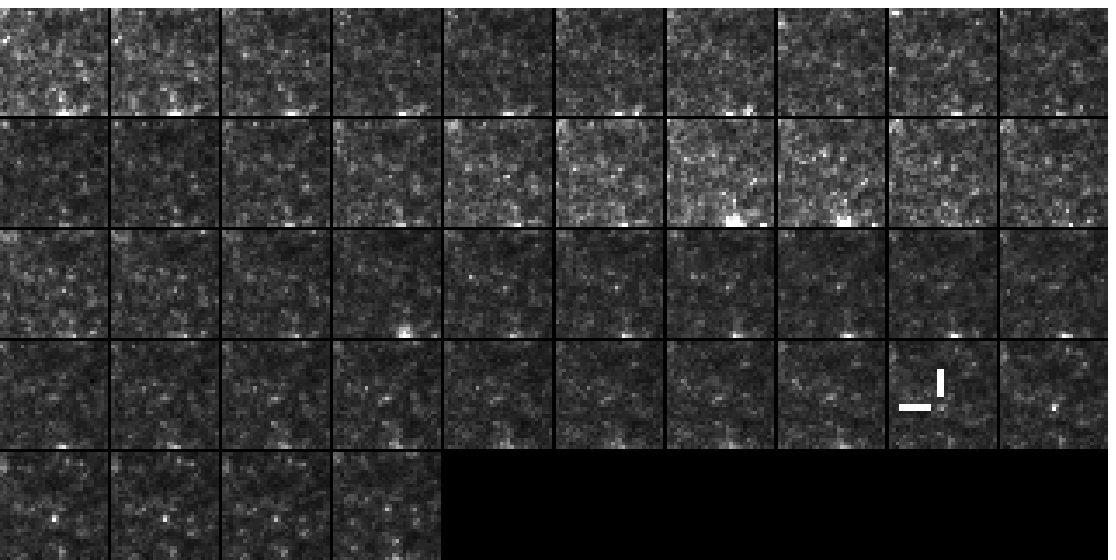}
\caption{Same as Figure 1a but for M87 Nova 33 (top) and M87 Nova 34 (bottom).}
\end{figure*}

\clearpage

\begin{figure*}
\figurenum{2r}
\includegraphics[width=1\columnwidth]{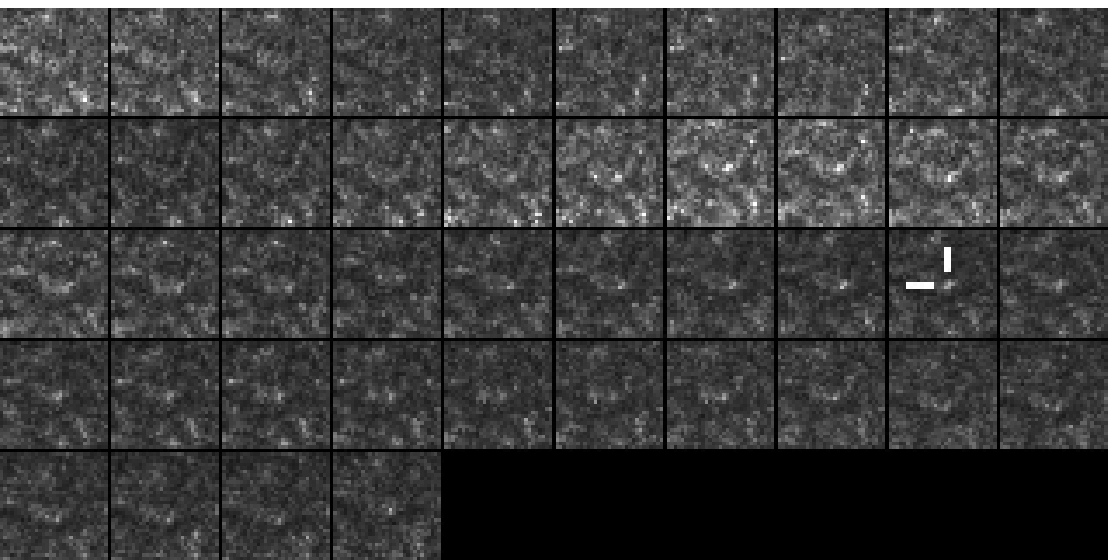}
\includegraphics[width=1\columnwidth]{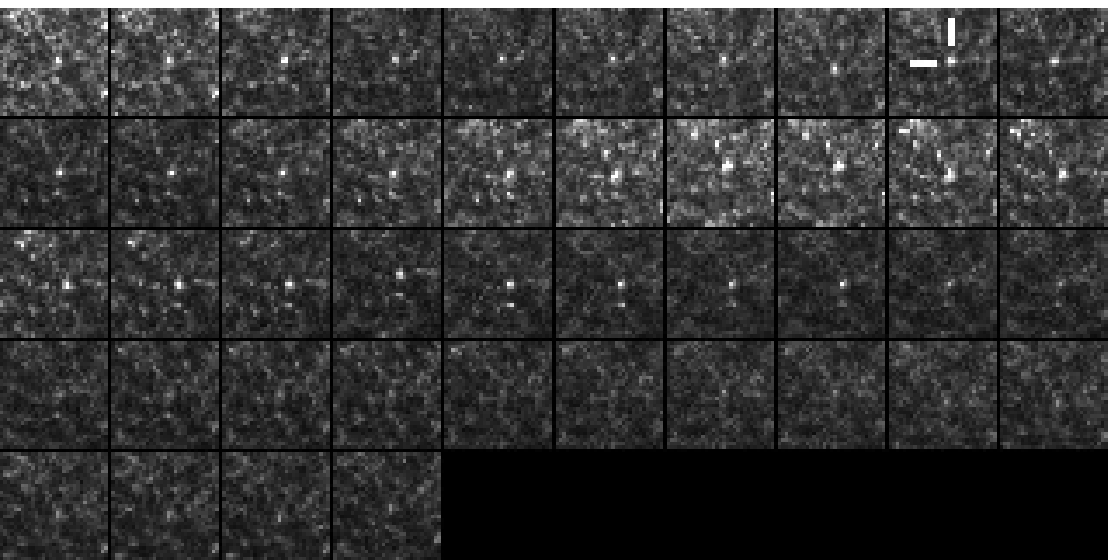}
\caption{Same as Figure 1a but for M87 Nova 35 (top) and M87 Nova 36 (bottom).}
\end{figure*}

\clearpage

\begin{figure*}
\figurenum{2s}
\includegraphics[width=1\columnwidth]{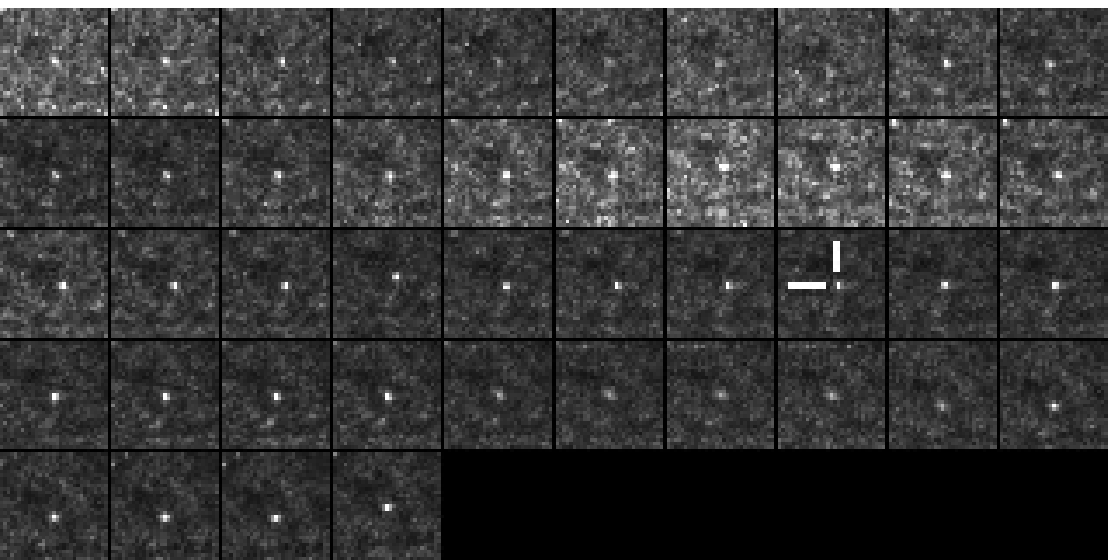}
\includegraphics[width=1\columnwidth]{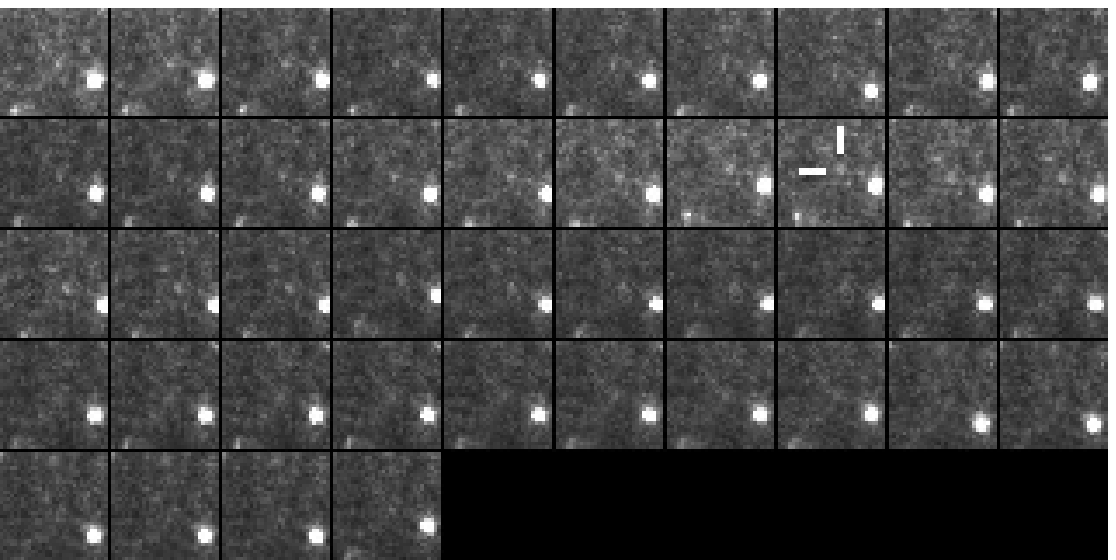}
\caption{Same as Figure 1a but for M87 Nova 37 (top) and M87 Nova 38 (bottom).}
\end{figure*}

\clearpage

\begin{figure*}
\figurenum{2t}
\includegraphics[width=1\columnwidth]{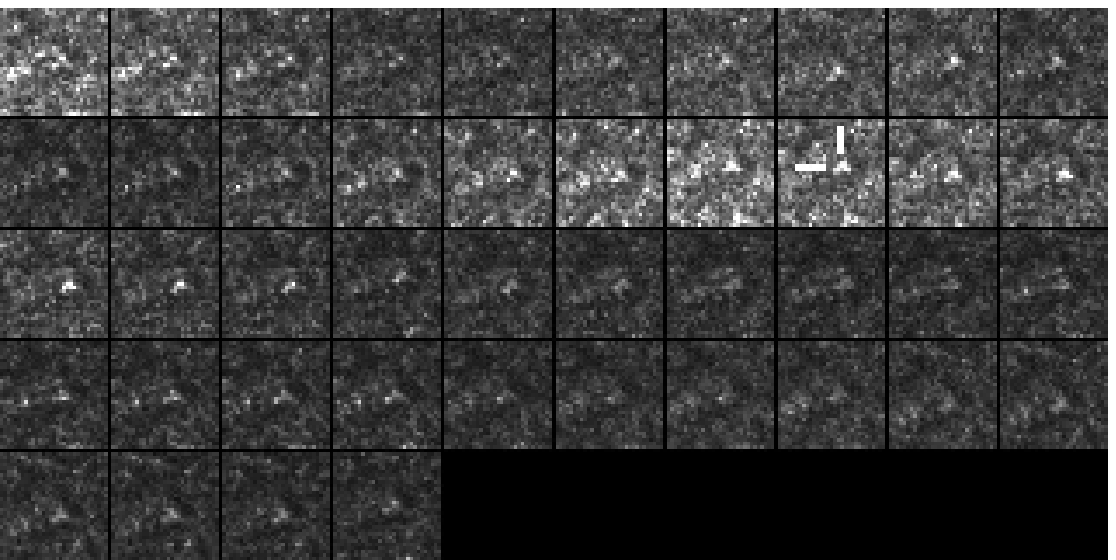}
\includegraphics[width=1\columnwidth]{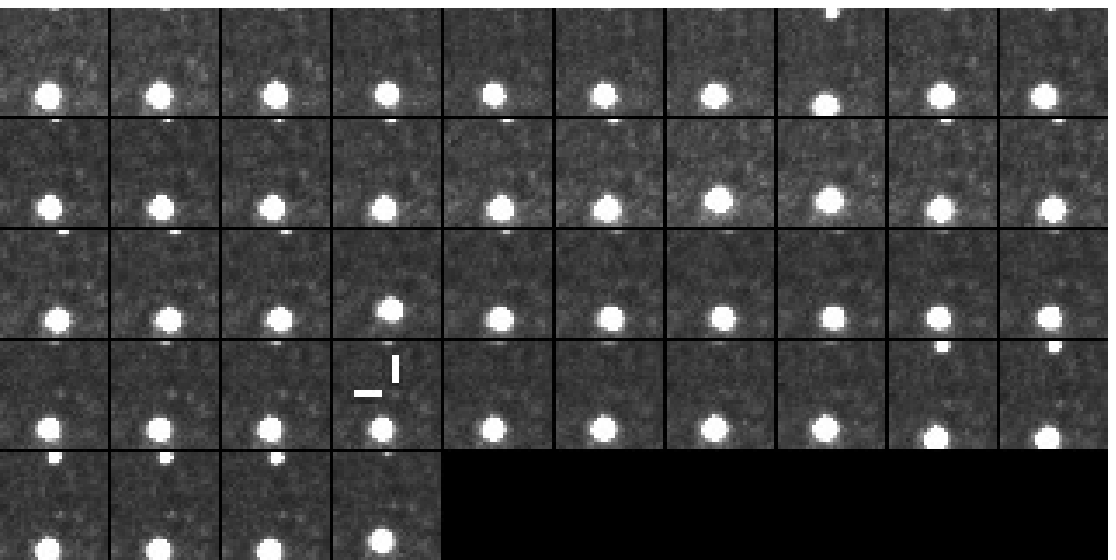}
\caption{Same as Figure 1a but for M87 Nova 39 (top) and M87 Nova 40 (bottom).}
\end{figure*}

\clearpage

\begin{figure*}
\figurenum{2u}
\includegraphics[width=1\columnwidth]{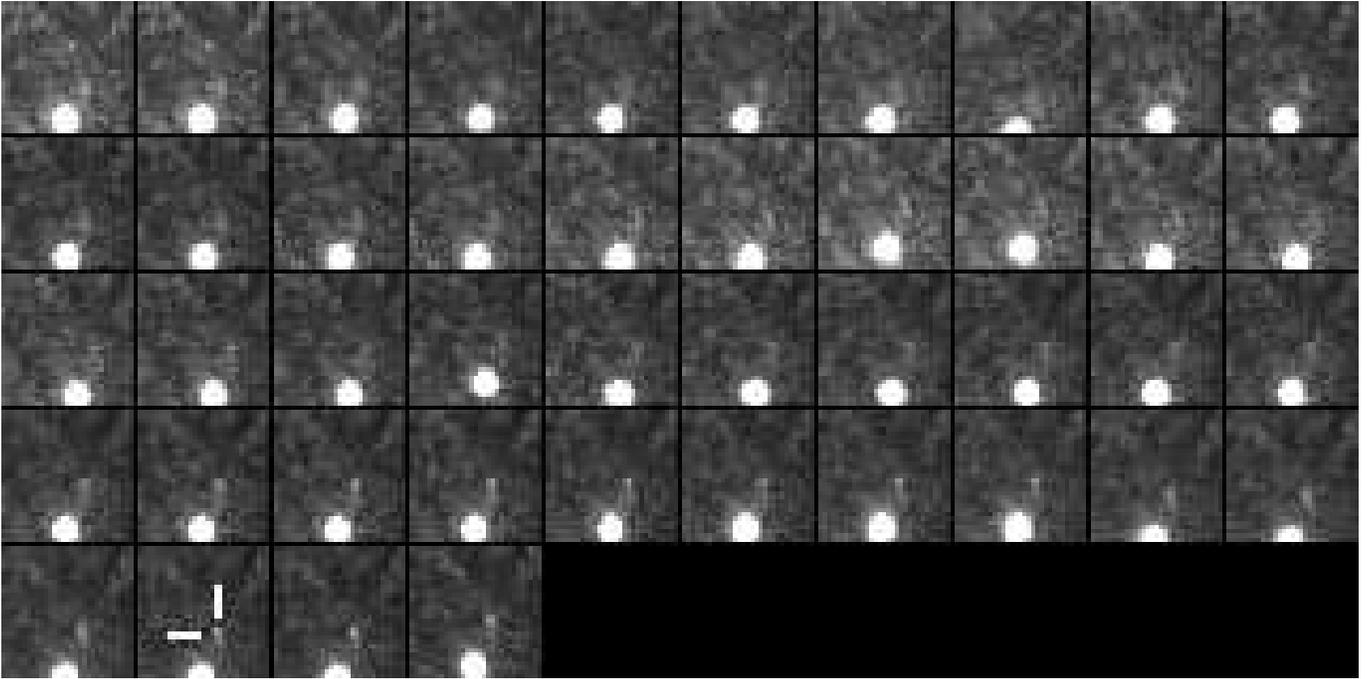}
\caption{Same as Figure 1a but for M87 Nova 41.}
\end{figure*}

\clearpage

\begin{figure}
\centering
\figurenum{3a}
\epsscale{1.0}
\includegraphics[width=180mm]{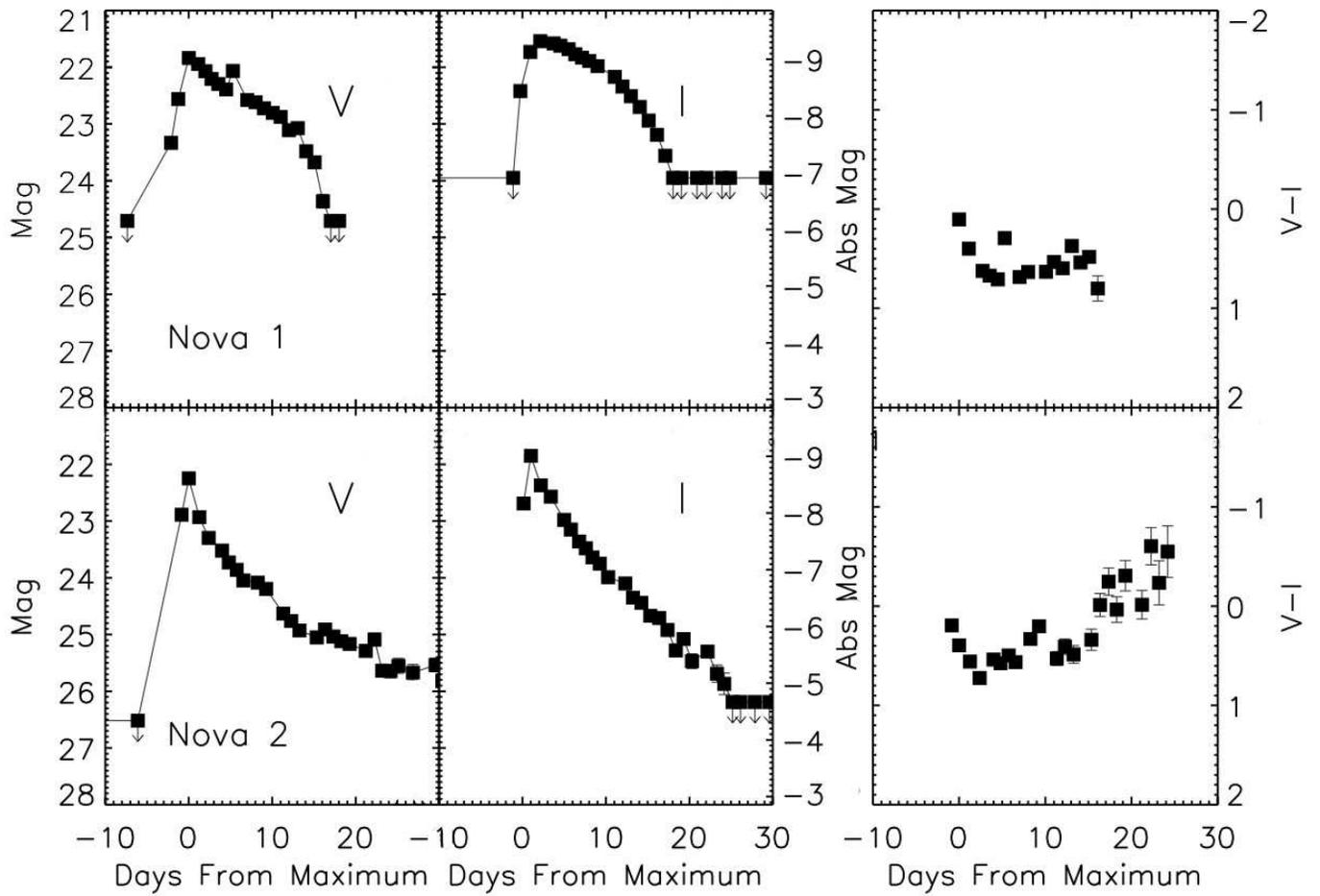}
\caption{Light curves of nova 1 and nova 2 in $V$ (left), $I$ (middle), and $V-I$ (right).  As in Figure 1, novae are again ordered by peak brightness in the $V$ band. Upper limits are indicated by 
square data points with downward pointing arrows. Error bars are displayed for all measurements except for those smaller than the solid squares representing data points.}
\end{figure}

\clearpage

\begin{figure}
\centering
\figurenum{3b}
\epsscale{1.0}
\includegraphics[width=180mm]{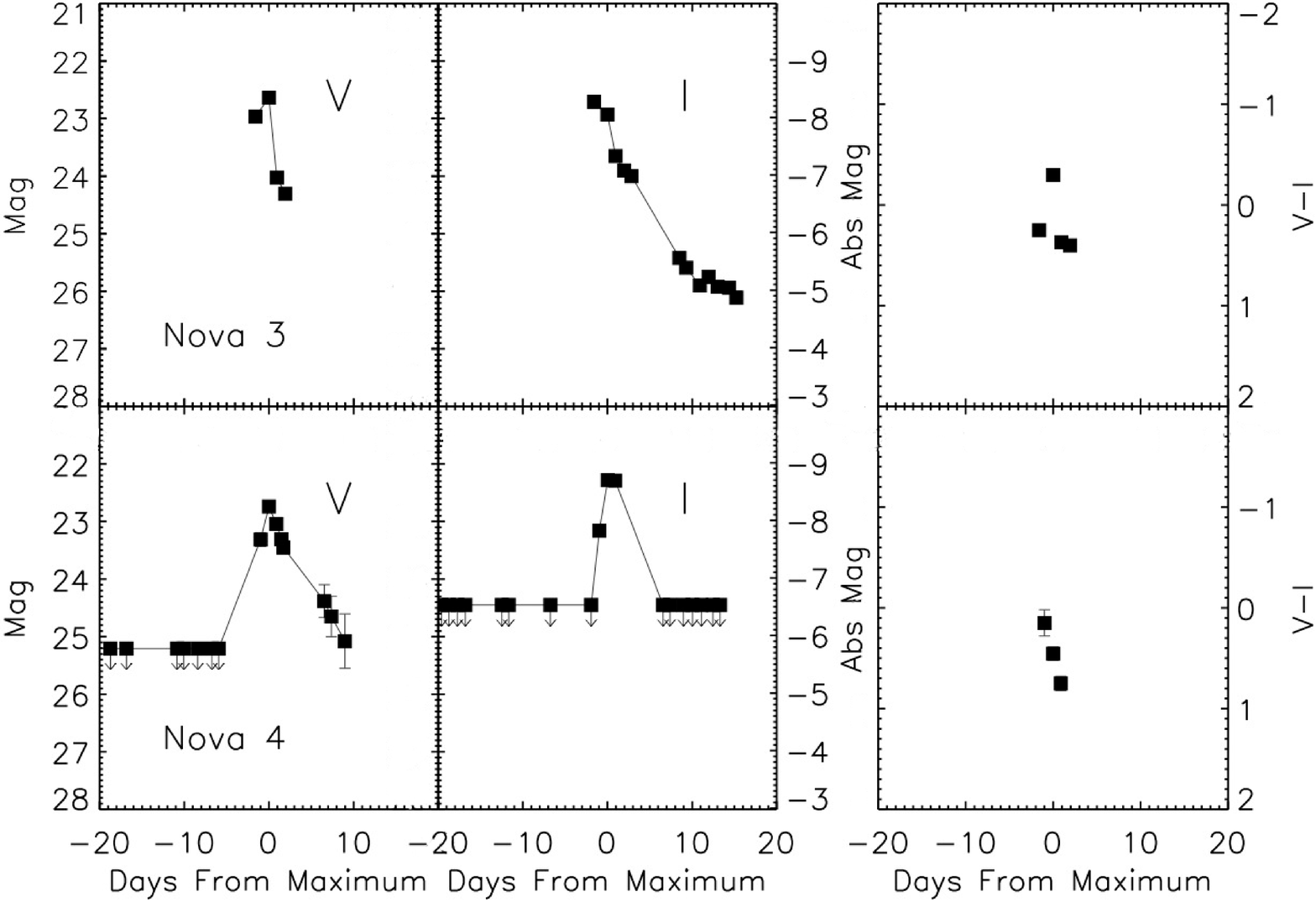}
\caption{Same as Figure 3a, except for novae 3 and 4.}
\end{figure}

\clearpage

\begin{figure}
\centering
\figurenum{3c}
\epsscale{1.0}
\includegraphics[width=180mm]{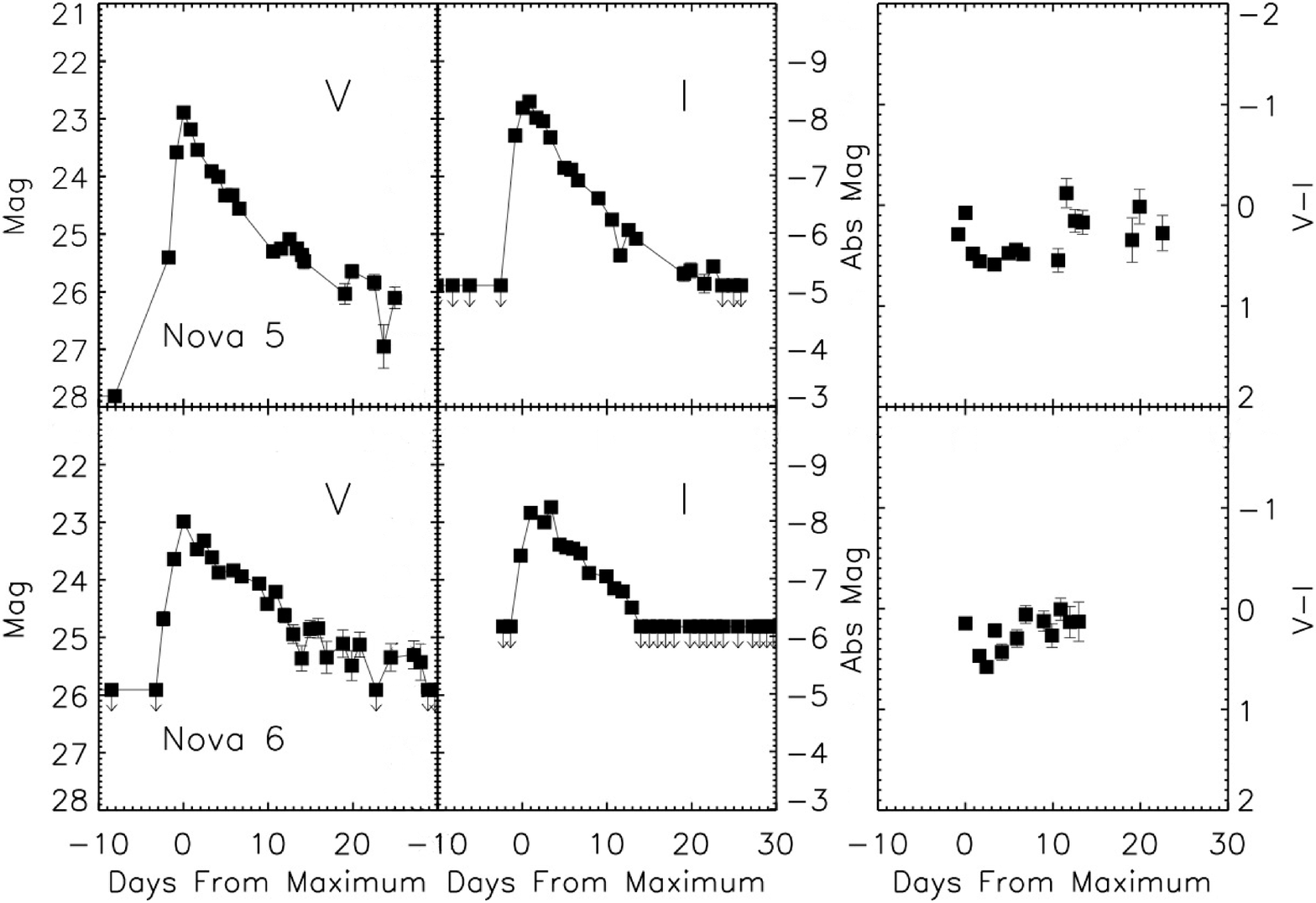}
\caption{Same as Figure 3a, except for novae 5 and 6}
\end{figure}

\clearpage

\begin{figure}
\centering
\figurenum{3d}
\epsscale{1.0}
\includegraphics[width=180mm]{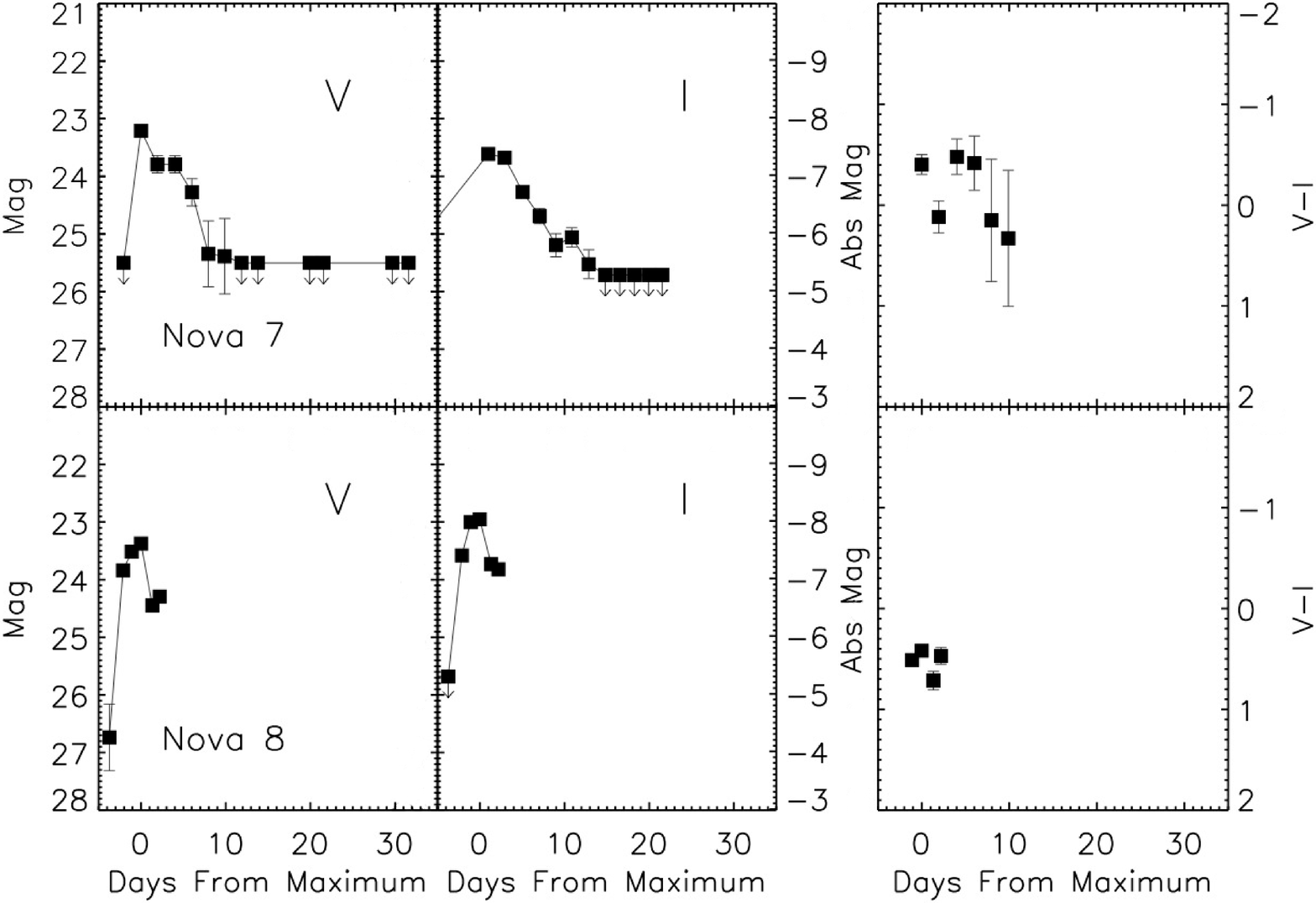}
\caption{Same as Figure 3a, except for novae 7 and 8.}
\end{figure}

\clearpage

\begin{figure}
\centering
\figurenum{3e}
\epsscale{1.0}
\includegraphics[width=180mm]{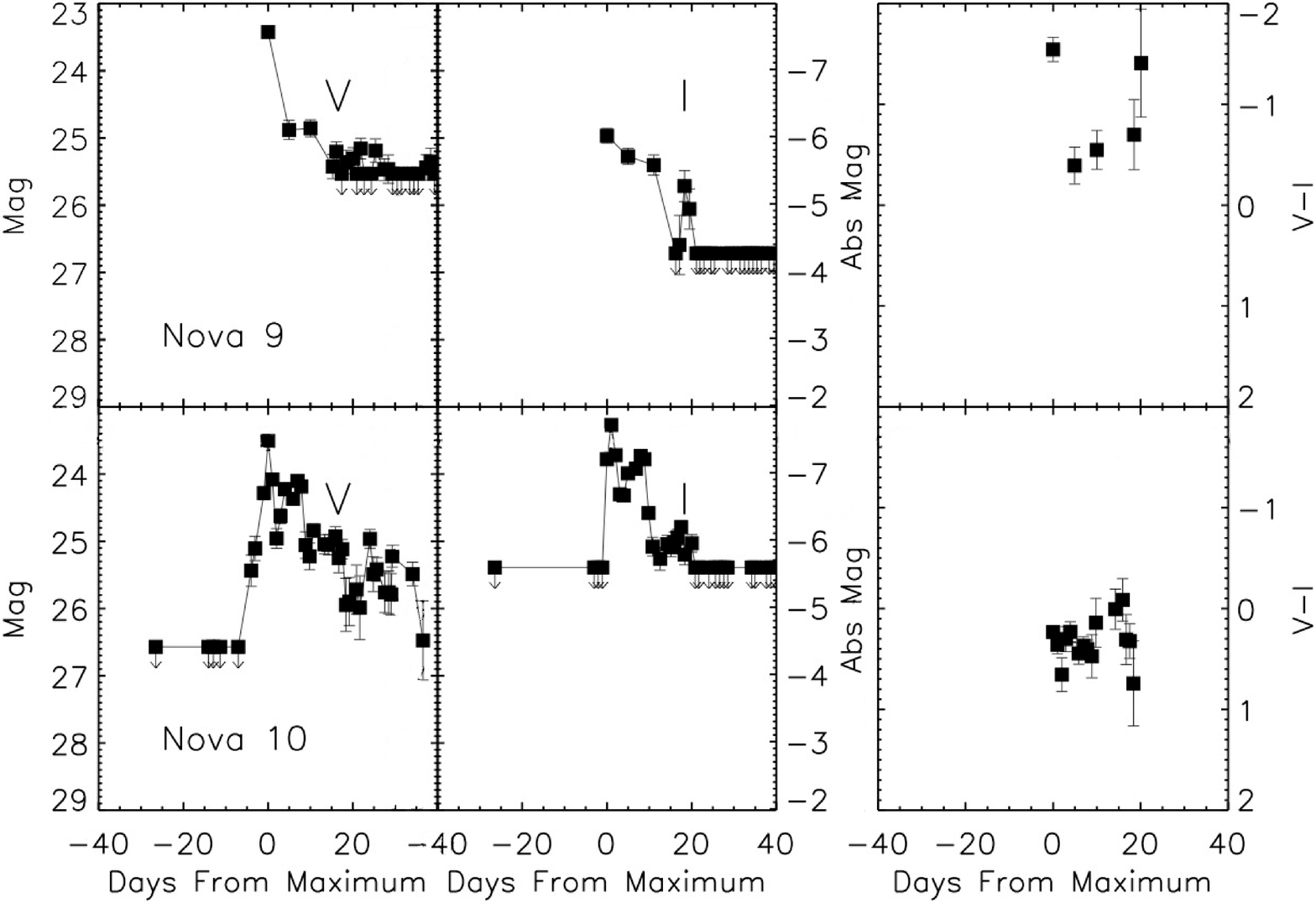}
\caption{Same as Figure 3a, except for novae 9 and 10.}
\end{figure}

\clearpage

\begin{figure}
\centering
\figurenum{3f}
\epsscale{1.0}
\includegraphics[width=180mm]{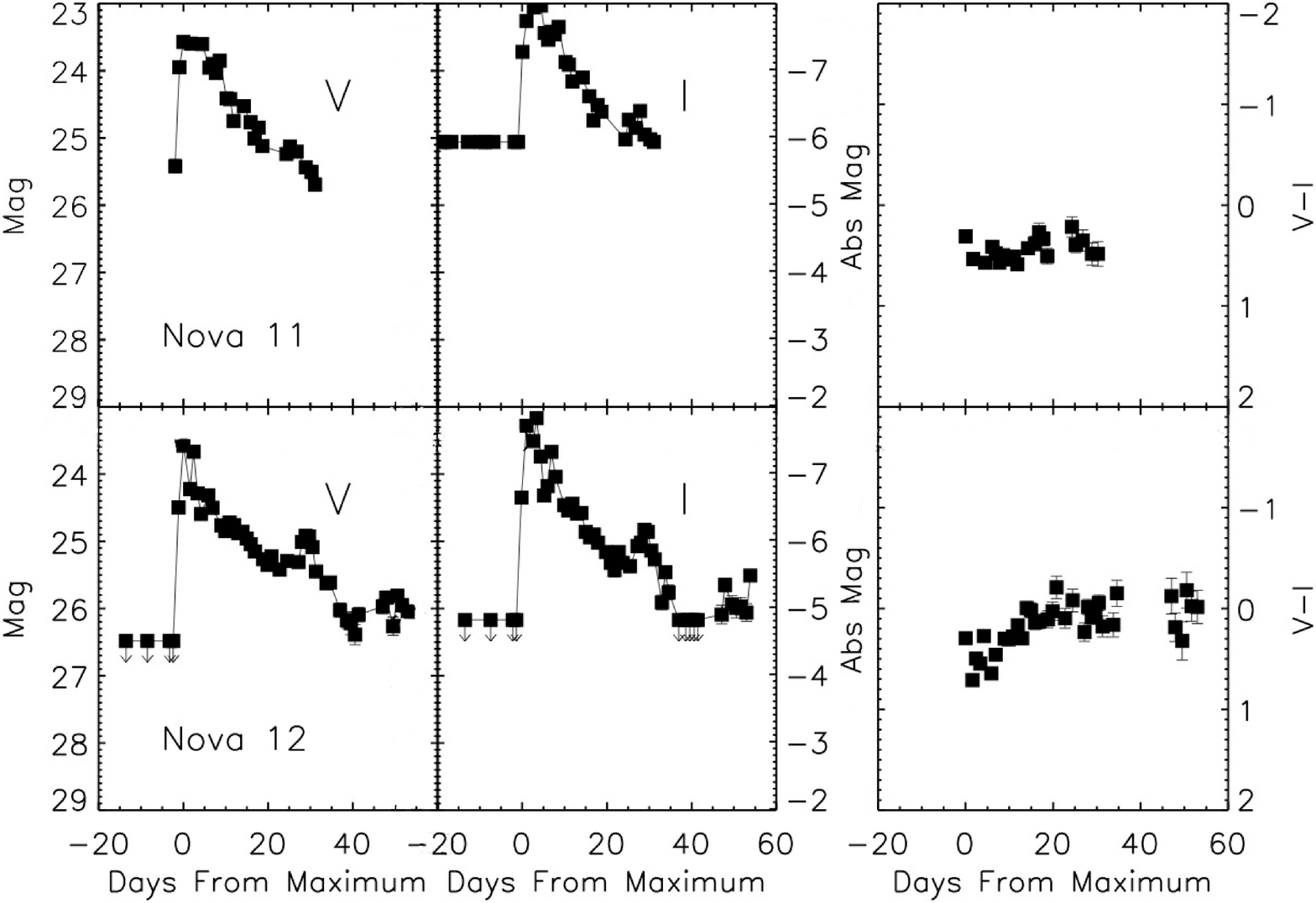}
\caption{Same as Figure 3a, except for novae 11 and 12.}
\end{figure}

\clearpage

\begin{figure}
\centering
\figurenum{3g}
\epsscale{1.0}
\includegraphics[width=180mm]{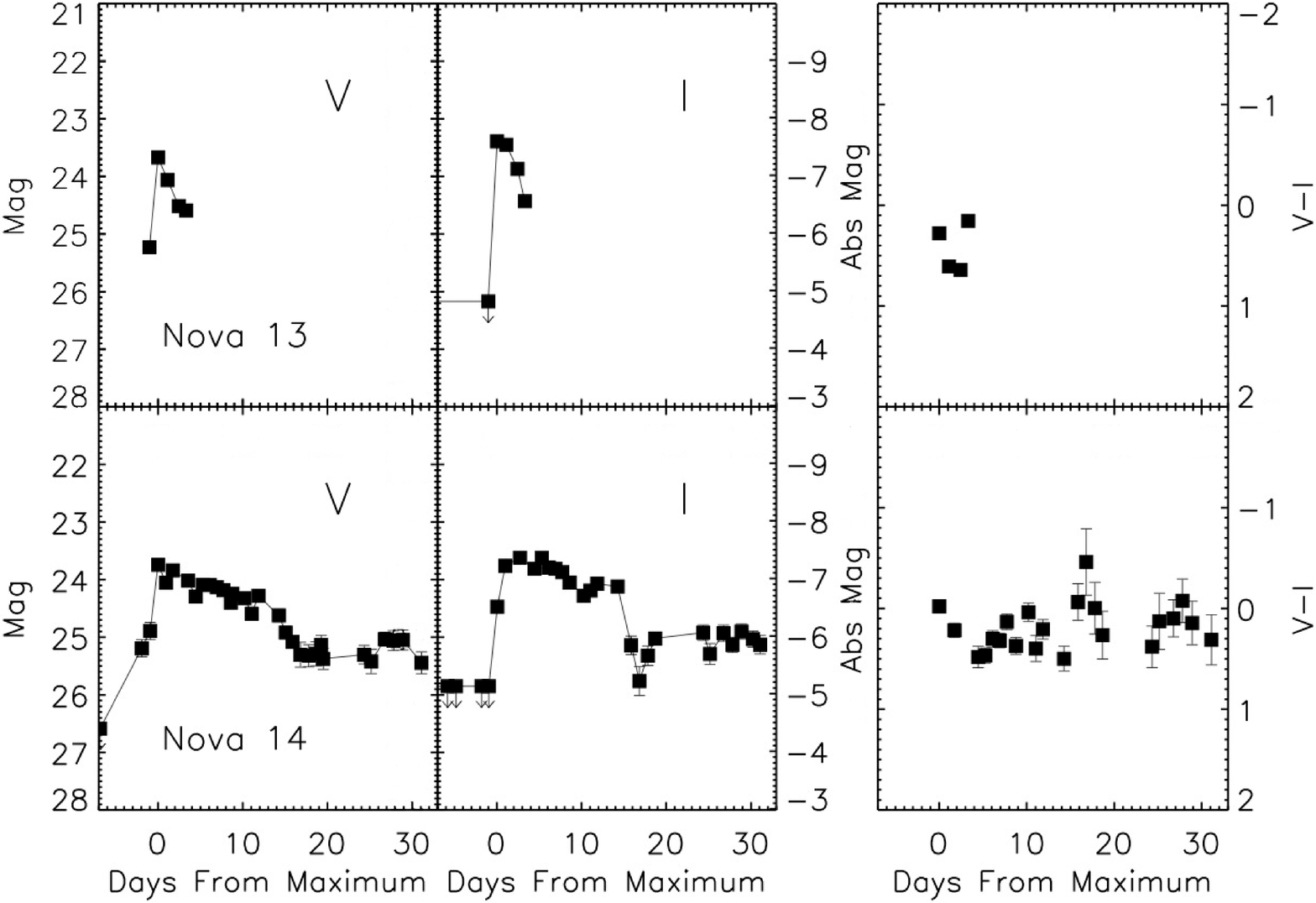}
\caption{Same as Figure 3a, except for novae 13 and 14.}
\end{figure}

\clearpage

\begin{figure}
\centering
\figurenum{3h}
\epsscale{1.0}
\includegraphics[width=180mm]{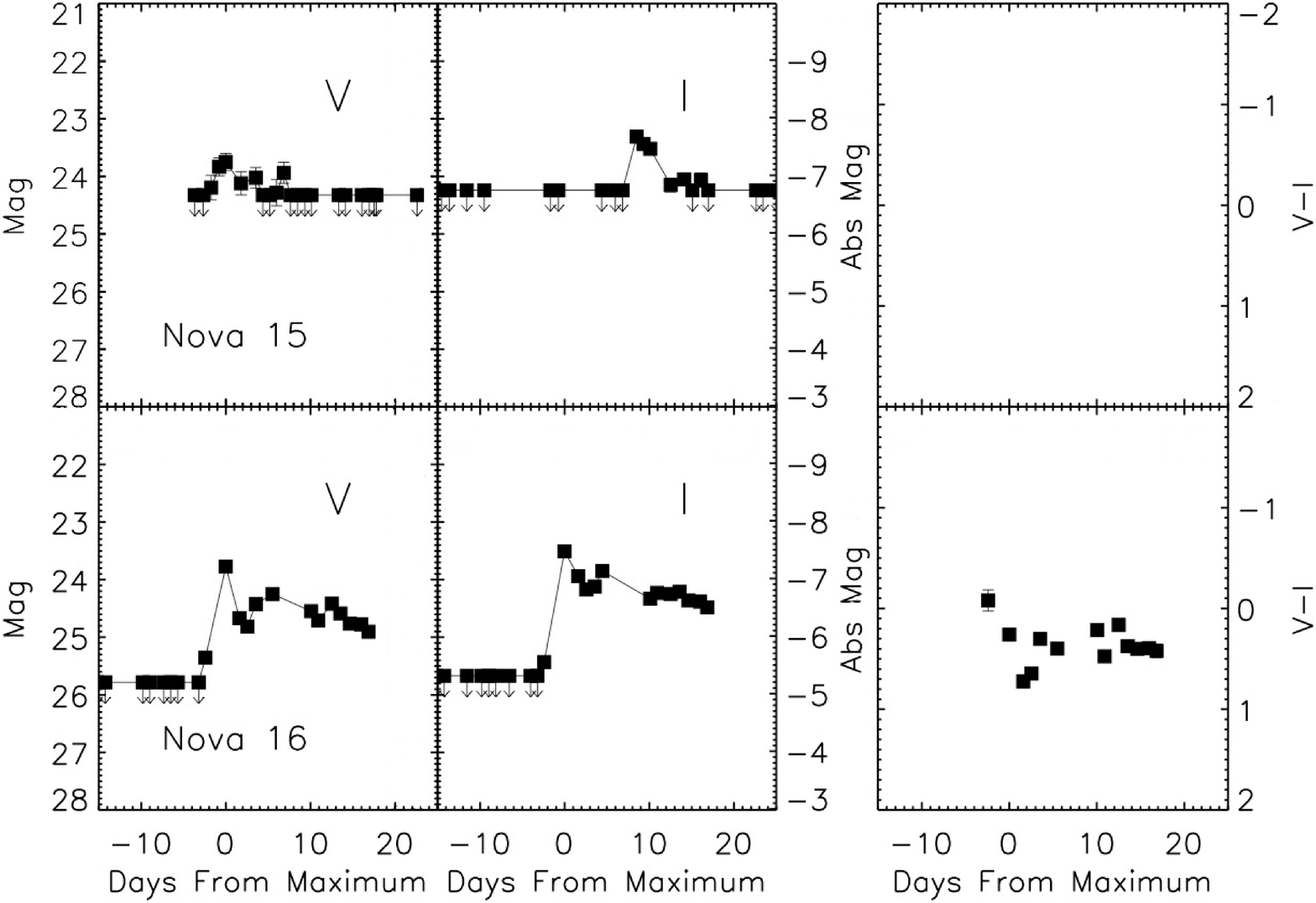}
\caption{Same as Figure 3a, except for novae 15 and 16.}
\end{figure}

\clearpage

\begin{figure}
\centering
\figurenum{3i}
\epsscale{1.0}
\includegraphics[width=180mm]{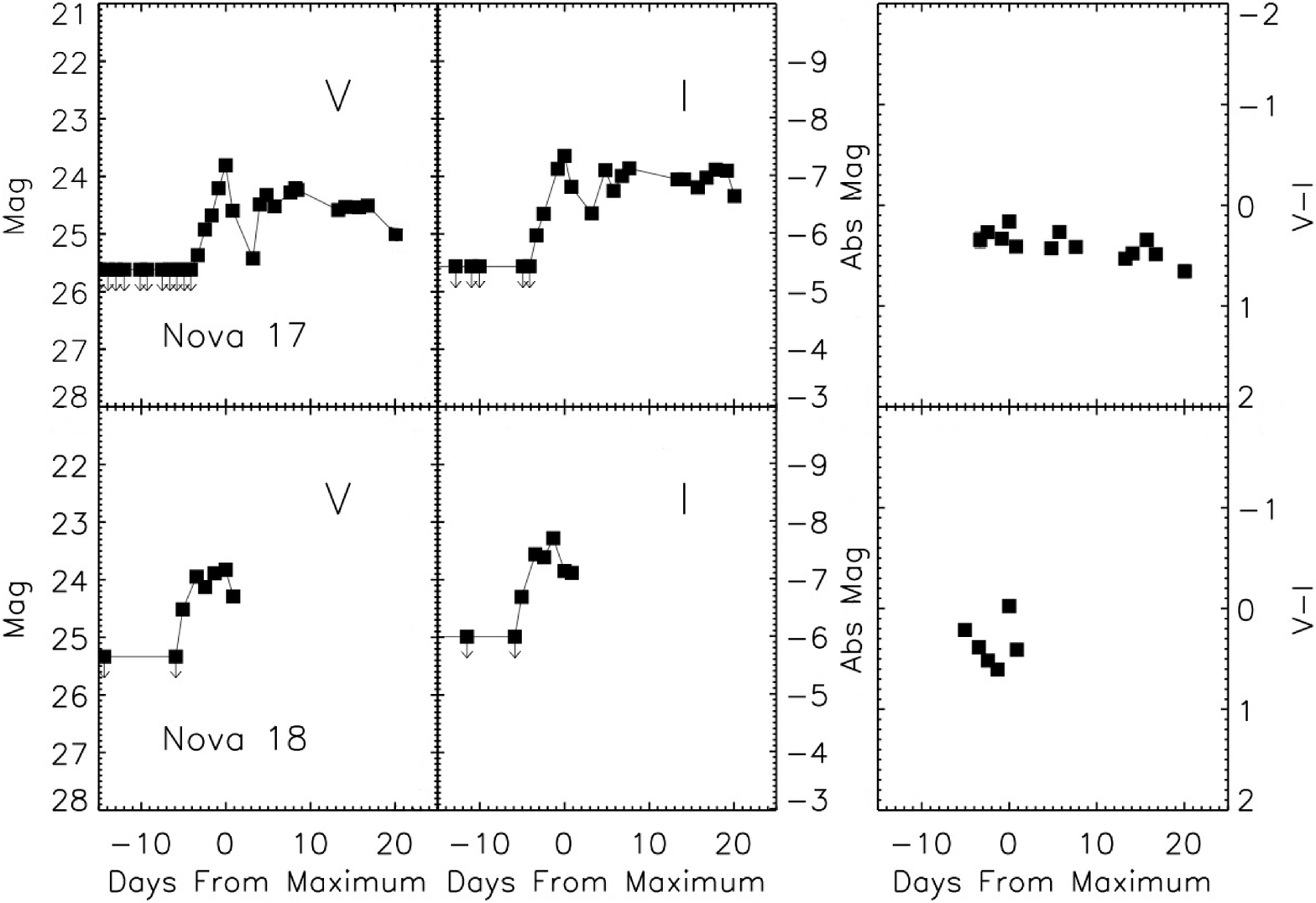}
\caption{Same as Figure 3a, except for novae 17 and 18.}
\end{figure}

\clearpage

\begin{figure}
\centering
\figurenum{3j}
\epsscale{1.0}
\includegraphics[width=180mm]{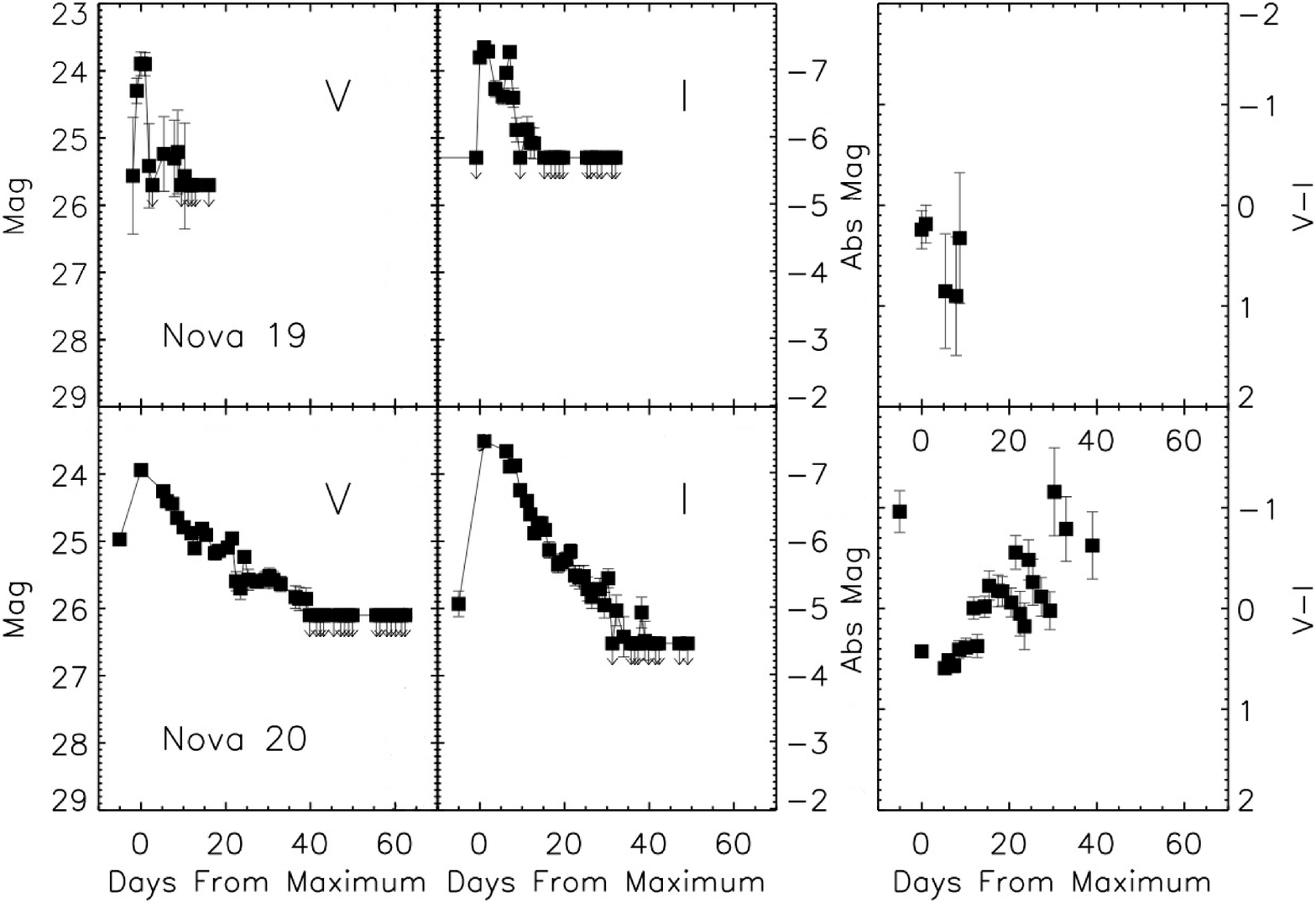}
\caption{Same as Figure 3a, except for novae 19 and 20.}
\end{figure}

\clearpage

\begin{figure}
\centering
\figurenum{3k}
\epsscale{1.0}
\includegraphics[width=180mm]{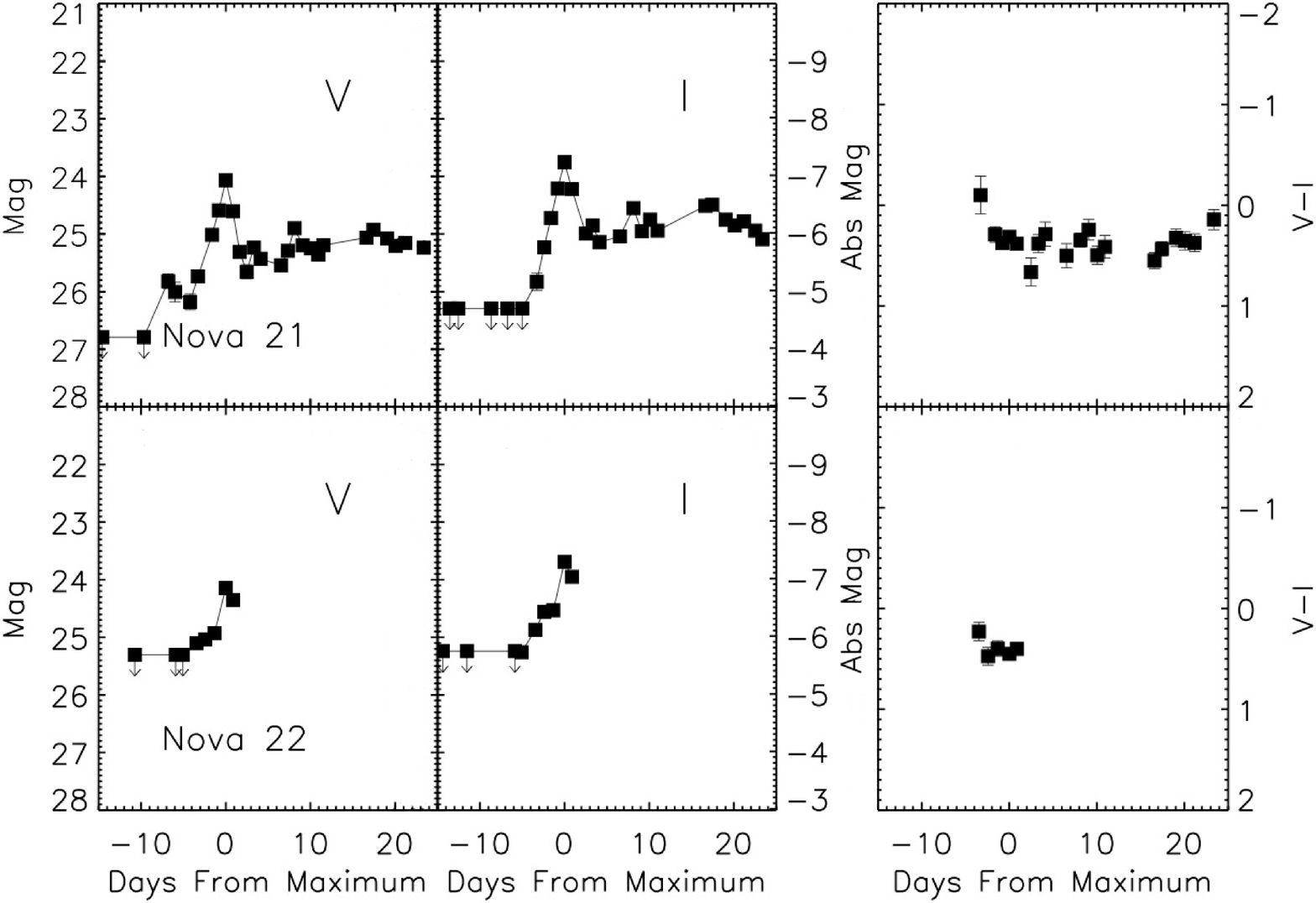}
\caption{Same as Figure 3a, except for novae 21 and 22.}
\end{figure}

\clearpage

\begin{figure}
\centering
\figurenum{3l}
\epsscale{1.0}
\includegraphics[width=180mm]{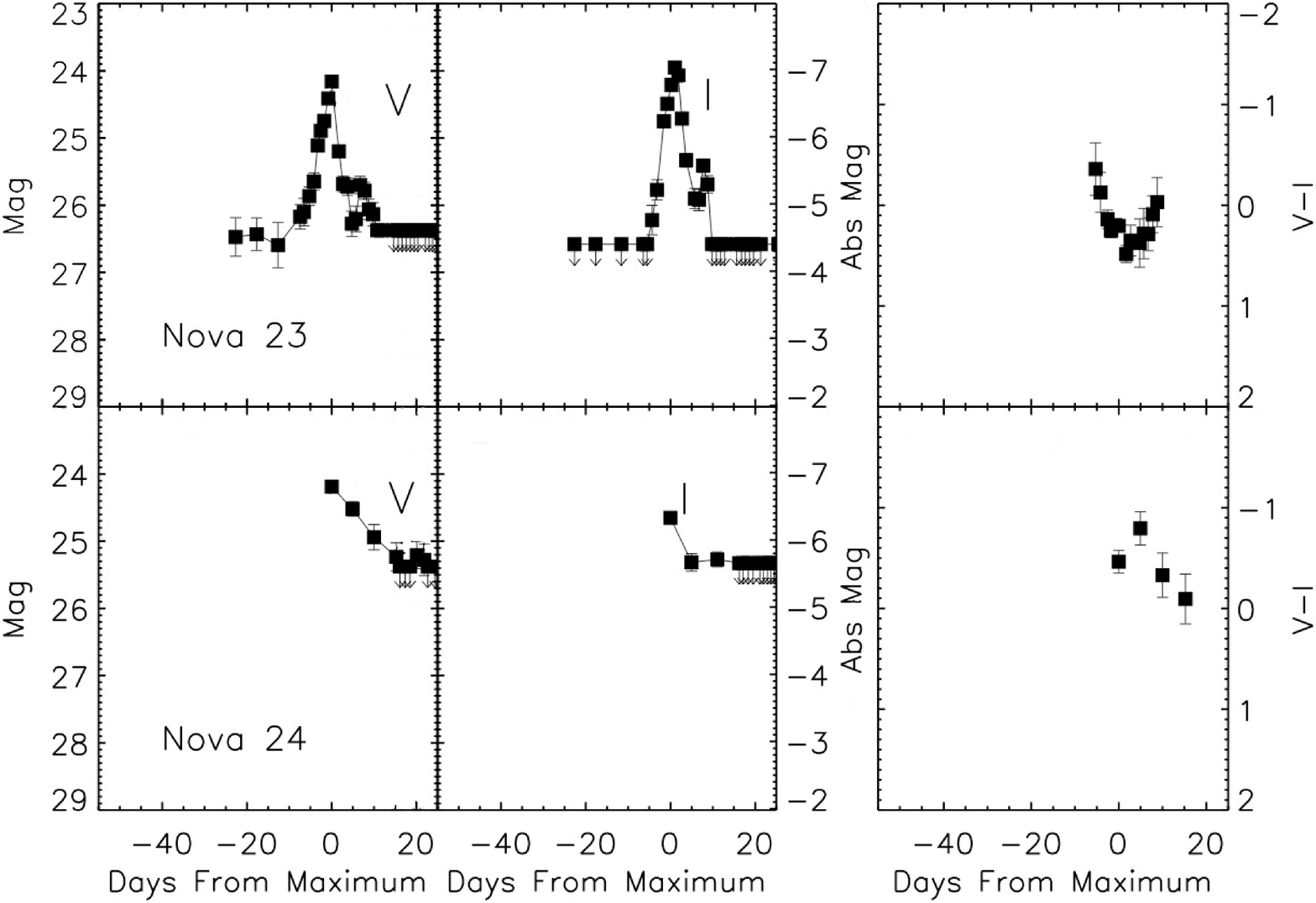}
\caption{Same as Figure 3a, except for novae 23 and 24.}
\end{figure}

\clearpage

\begin{figure}
\centering
\figurenum{3m}
\epsscale{1.0}
\includegraphics[width=180mm]{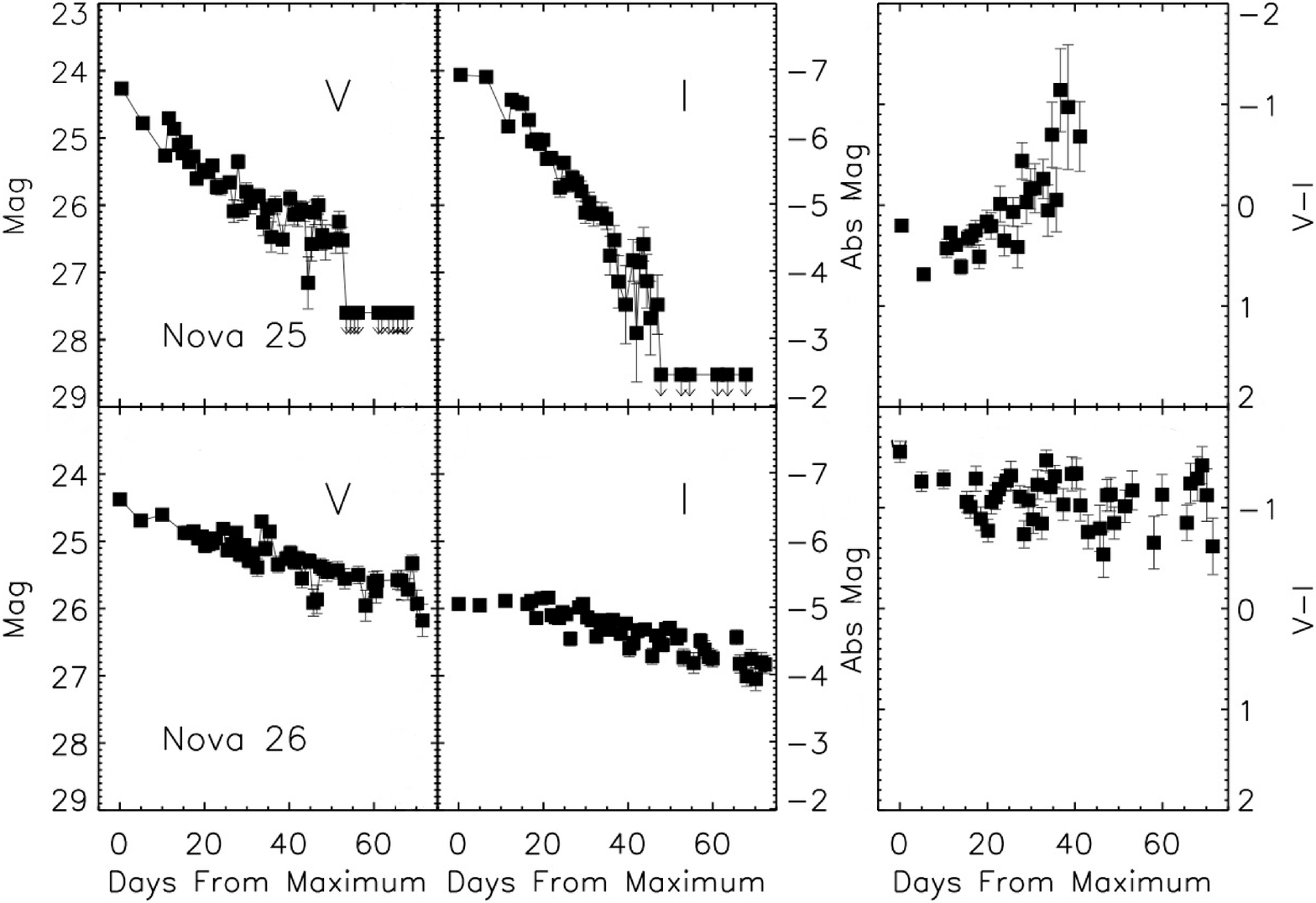}
\caption{Same as Figure 3a, except for novae 25 and 26.}
\end{figure}

\clearpage

\begin{figure}
\centering
\figurenum{3n}
\epsscale{1.0}
\includegraphics[width=180mm]{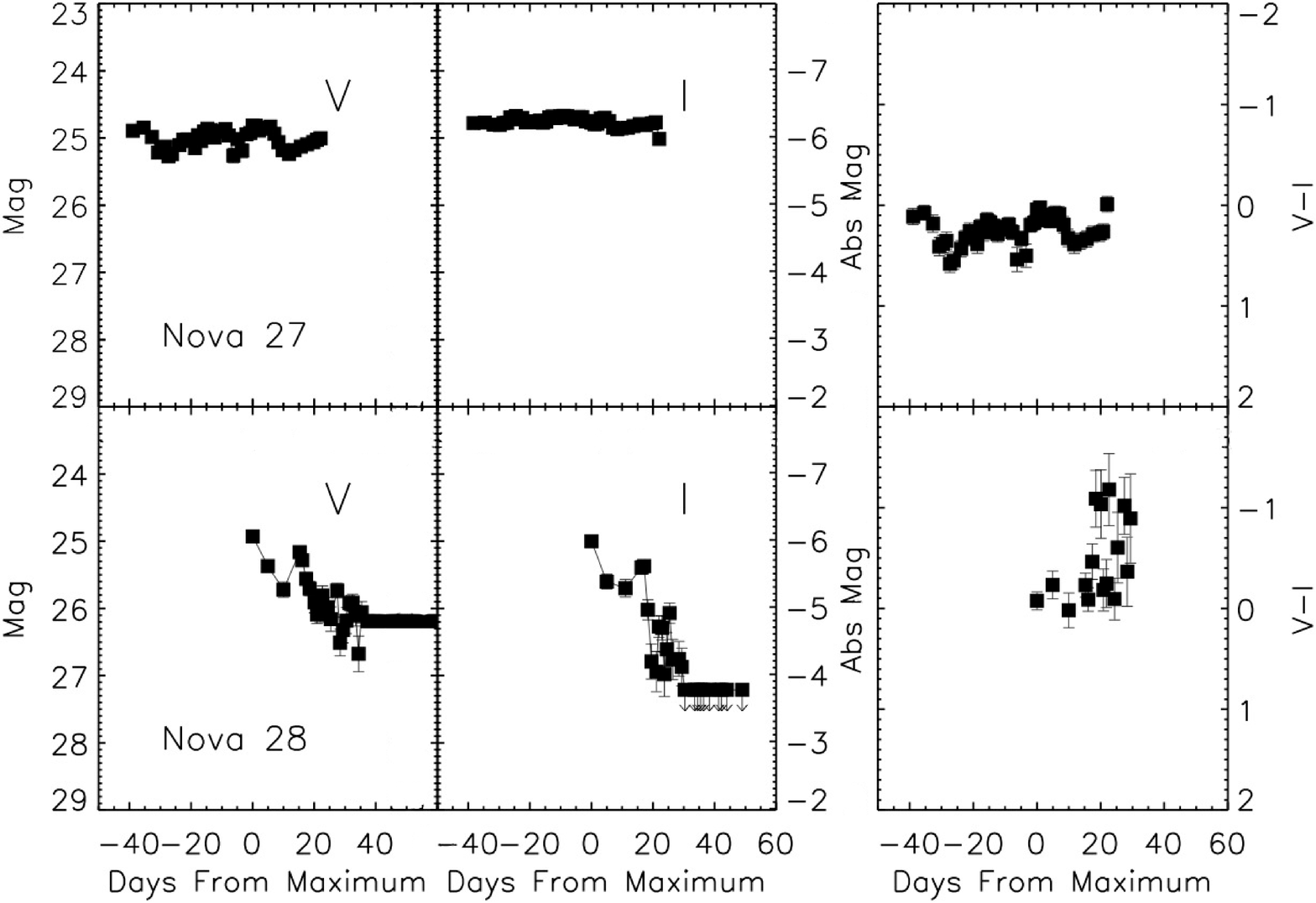}
\caption{Same as Figure 3a, except for novae 27 and 28.}
\end{figure}

\clearpage

\begin{figure}
\centering
\figurenum{3o}
\epsscale{1.0}
\includegraphics[width=180mm]{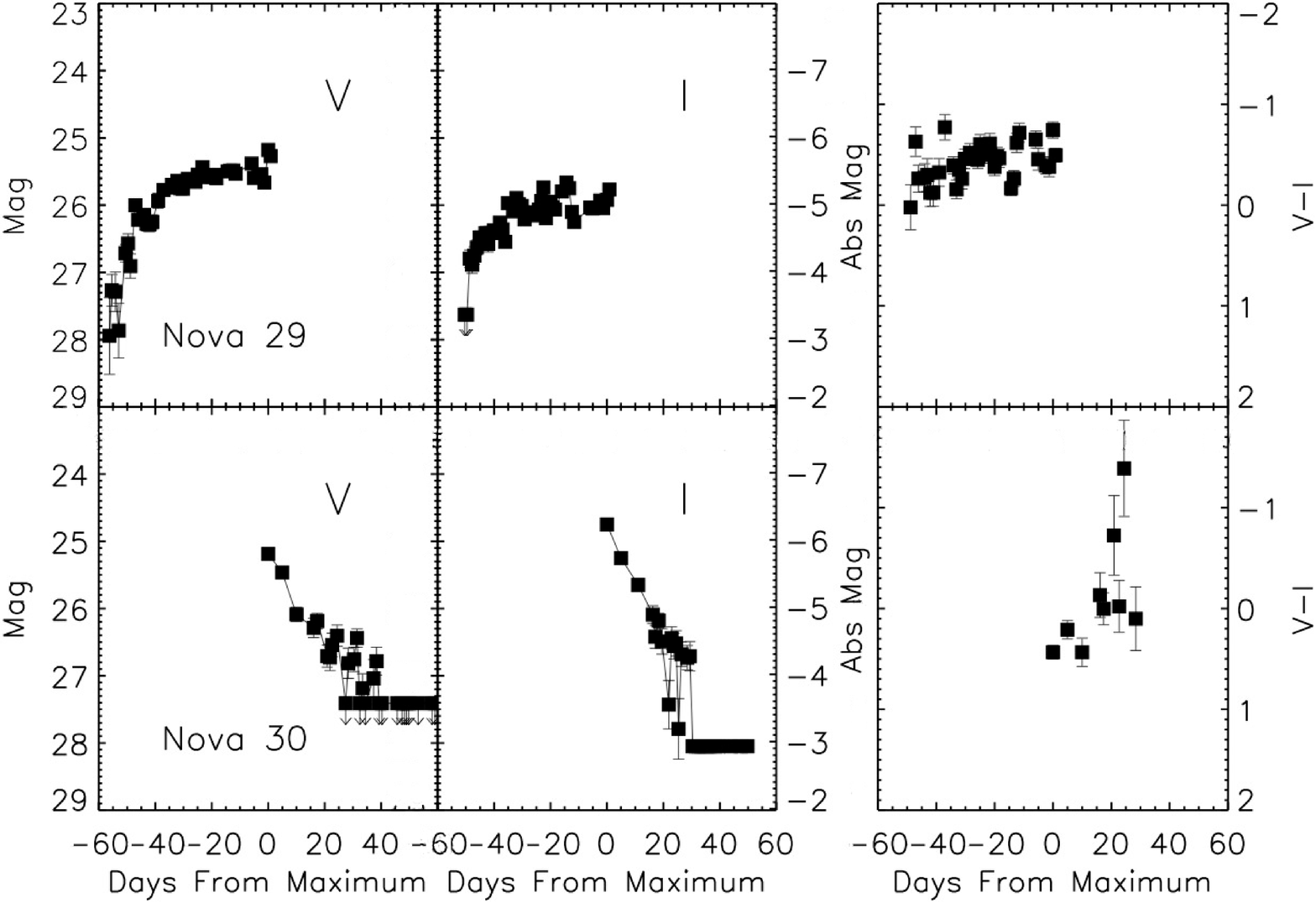}
\caption{Same as Figure 3a, except for novae 29 and 30.}
\end{figure}

\clearpage

\begin{figure}
\centering
\figurenum{3p}
\epsscale{1.0}
\includegraphics[width=180mm]{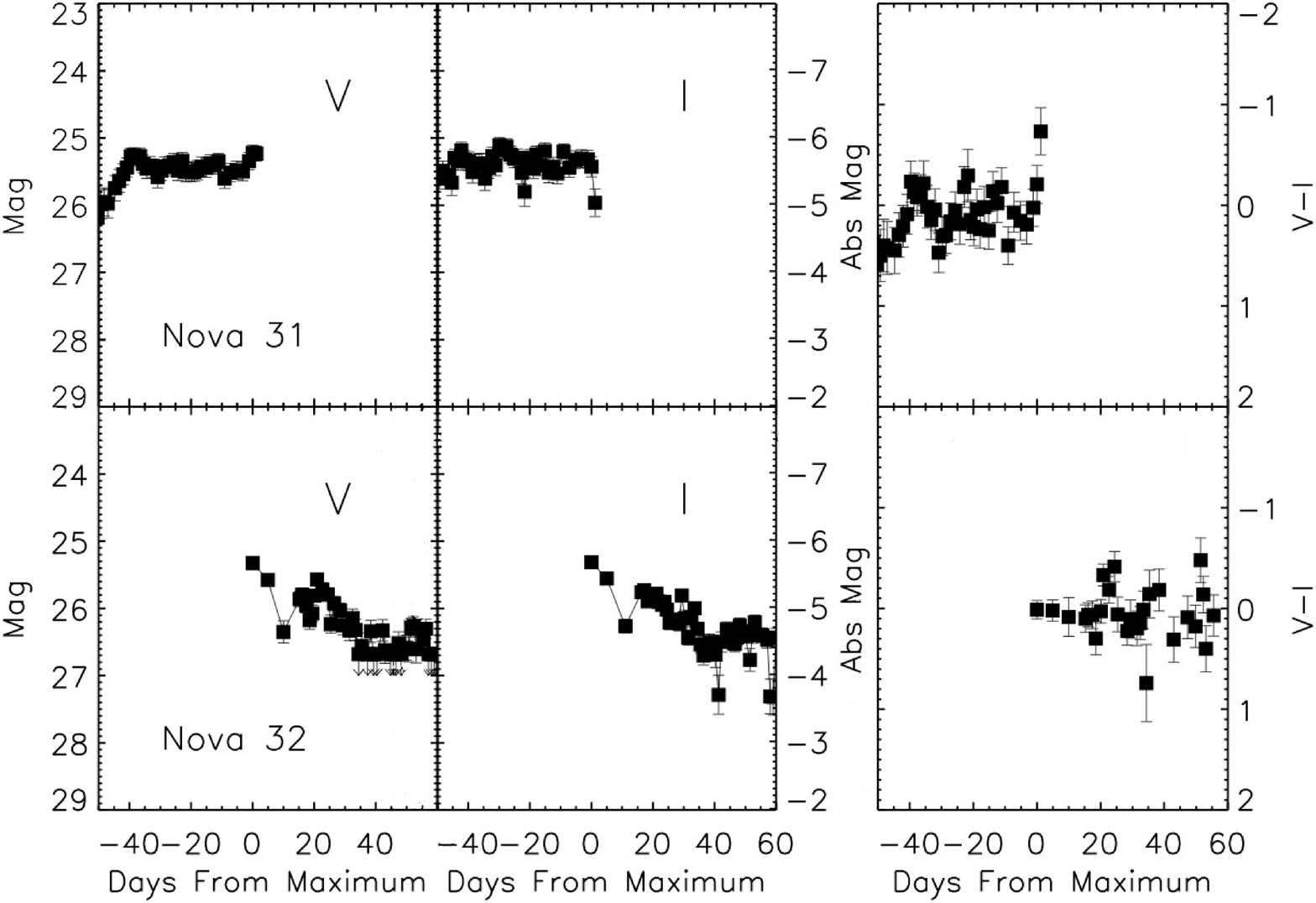}
\caption{Same as Figure 3a, except for novae 31 and 32.}
\end{figure}

\clearpage

\begin{figure}
\centering
\figurenum{3q}
\epsscale{1.0}
\includegraphics[width=180mm]{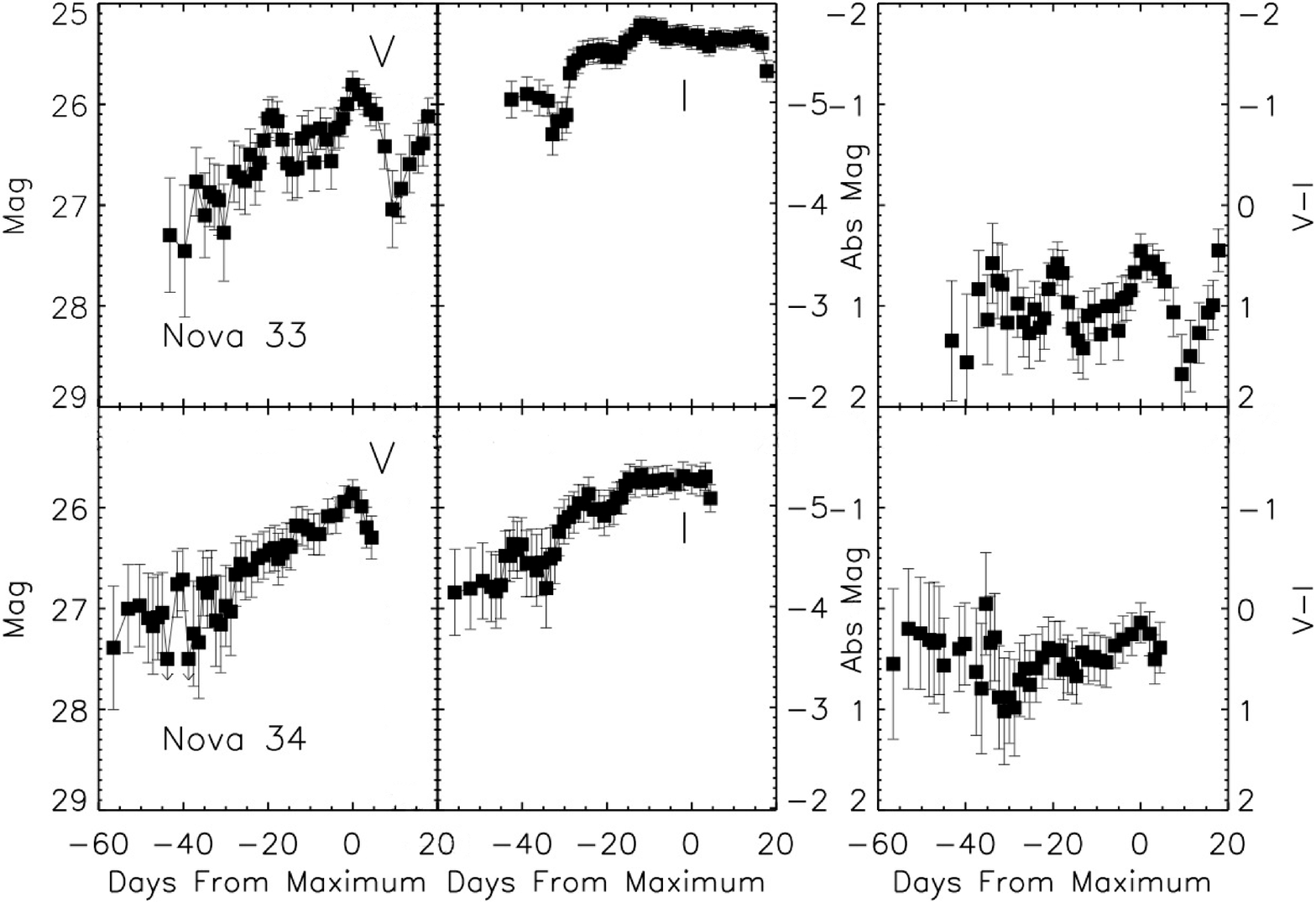}
\caption{Same as Figure 3a, except for novae 33 and 34.}
\end{figure}

\clearpage

\begin{figure}
\centering
\figurenum{3r}
\epsscale{1.0}
\includegraphics[width=180mm]{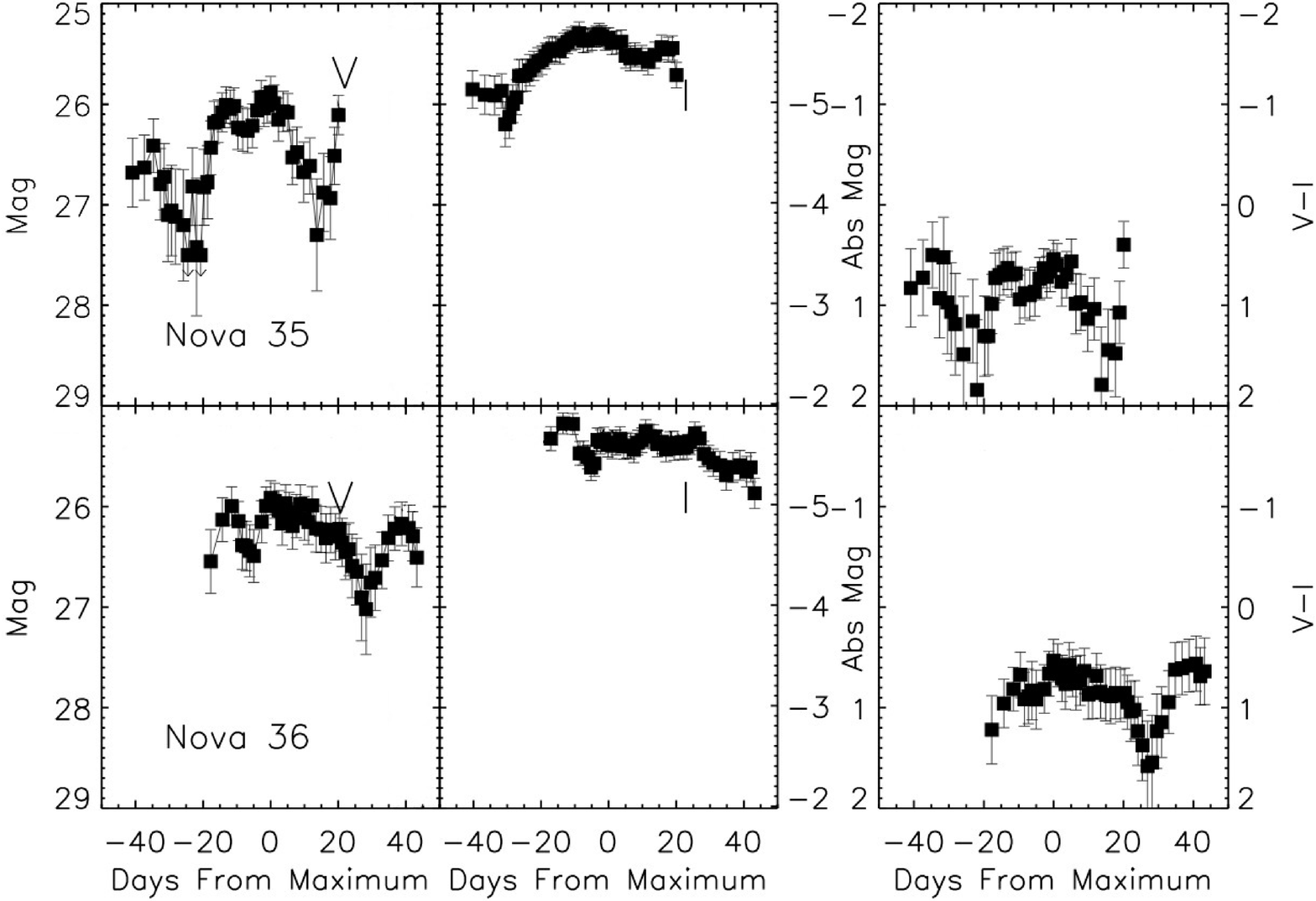}
\caption{Same as Figure 3a, except for novae 35 and 36.}
\end{figure}
\clearpage

\begin{figure}
\centering
\figurenum{3s}
\epsscale{1.0}
\includegraphics[width=180mm]{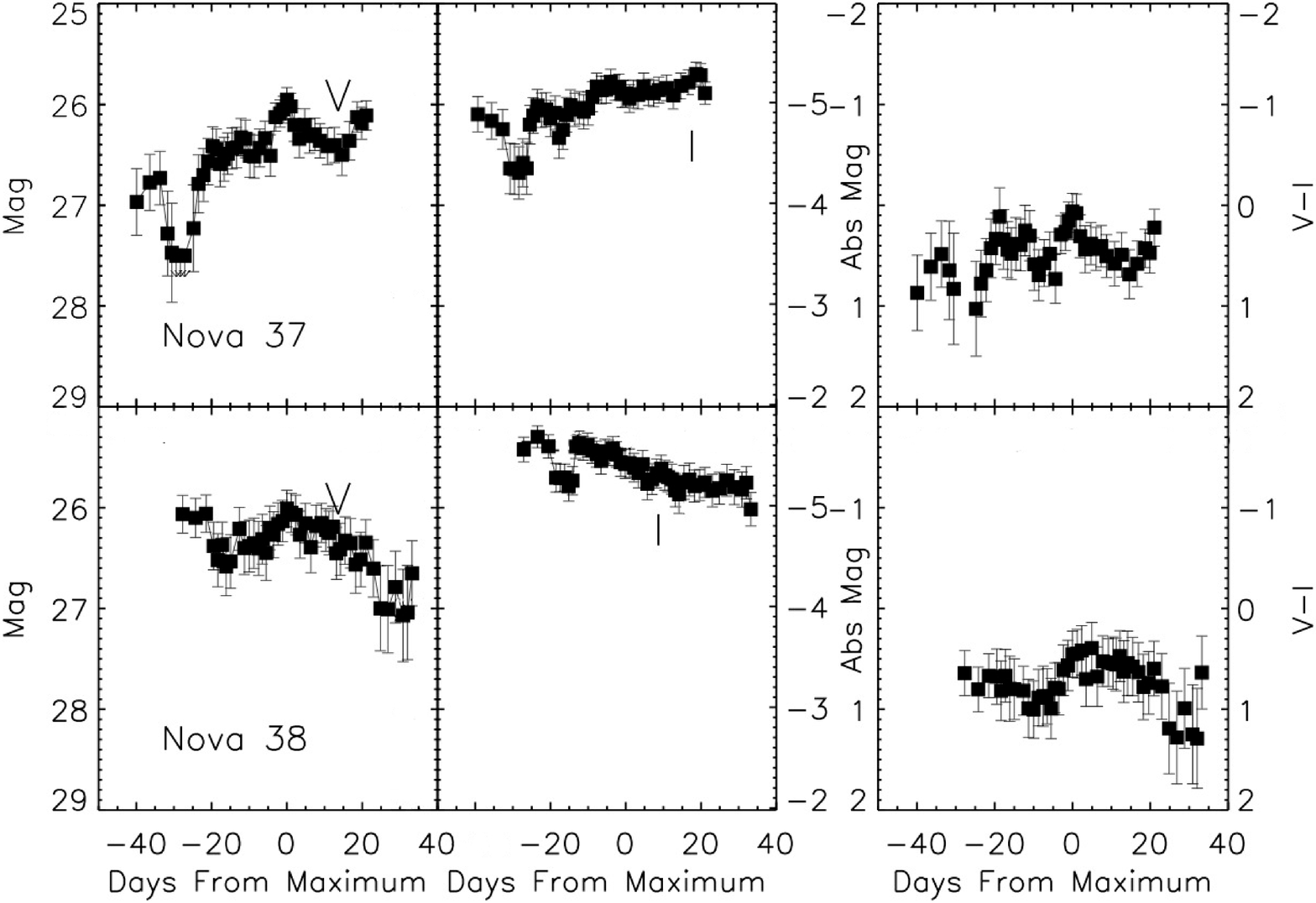}
\caption{Same as Figure 3a, except for novae 37 and 38.}
\end{figure}
\clearpage

\begin{figure}
\centering
\figurenum{3t}
\epsscale{1.0}
\includegraphics[width=180mm]{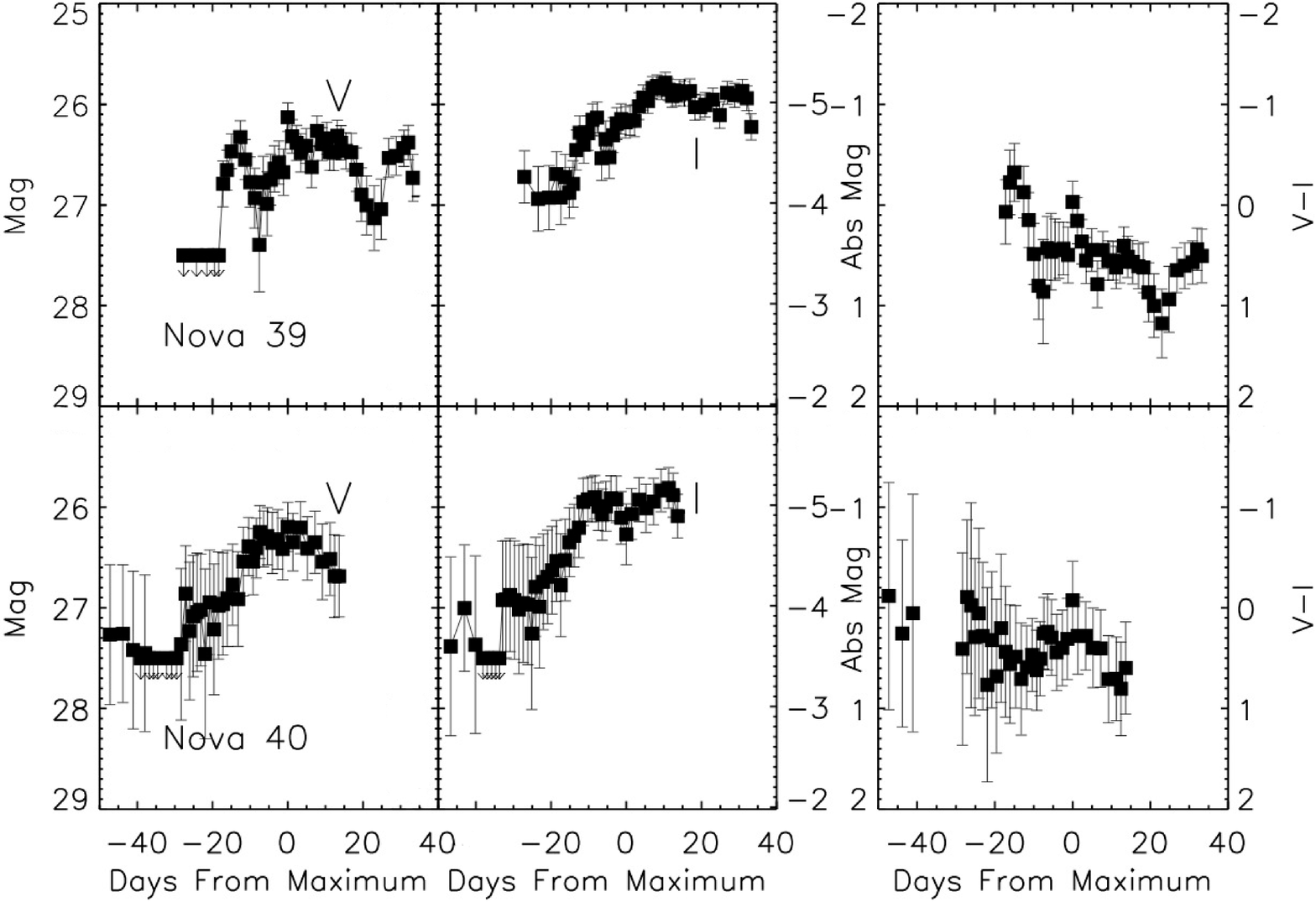}
\caption{Same as Figure 3a, except for novae 39 and 40.}
\end{figure}
\clearpage

\begin{figure}
\centering
\figurenum{3u}
\epsscale{1.0}
\includegraphics[width=180mm]{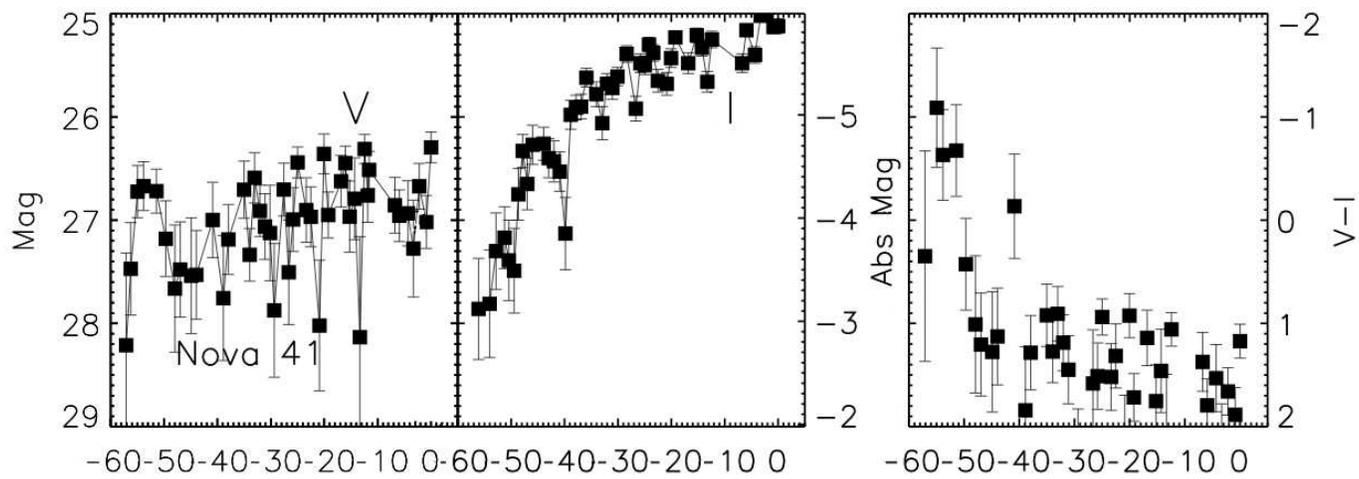}
\caption{Same as Figure 3a, except for nova 41.}
\end{figure}
\clearpage


\begin{figure}
\figurenum{4a}
\epsscale{1.0}
\includegraphics[width=180mm]{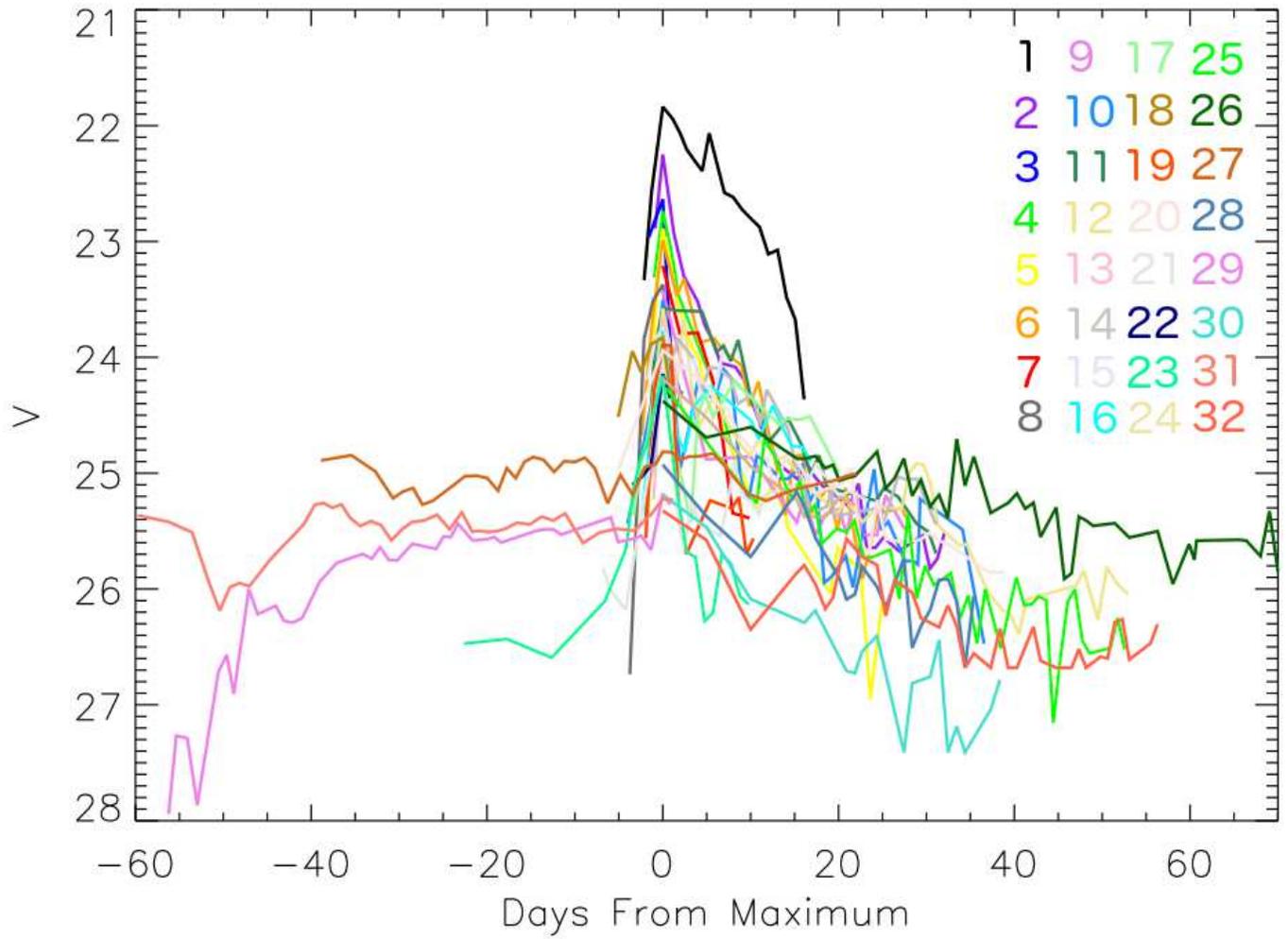}
\caption{32 M87 nova $V$ light curves overplotted as a function of days from maximum light. The 32 colored integers 
identify the 32 classical nova light curves with the same 32 novae displayed in Figures 1, 2 and 3.}
\end{figure}

\begin{figure}
\figurenum{4b}
\epsscale{1.0}
\includegraphics[width=180mm]{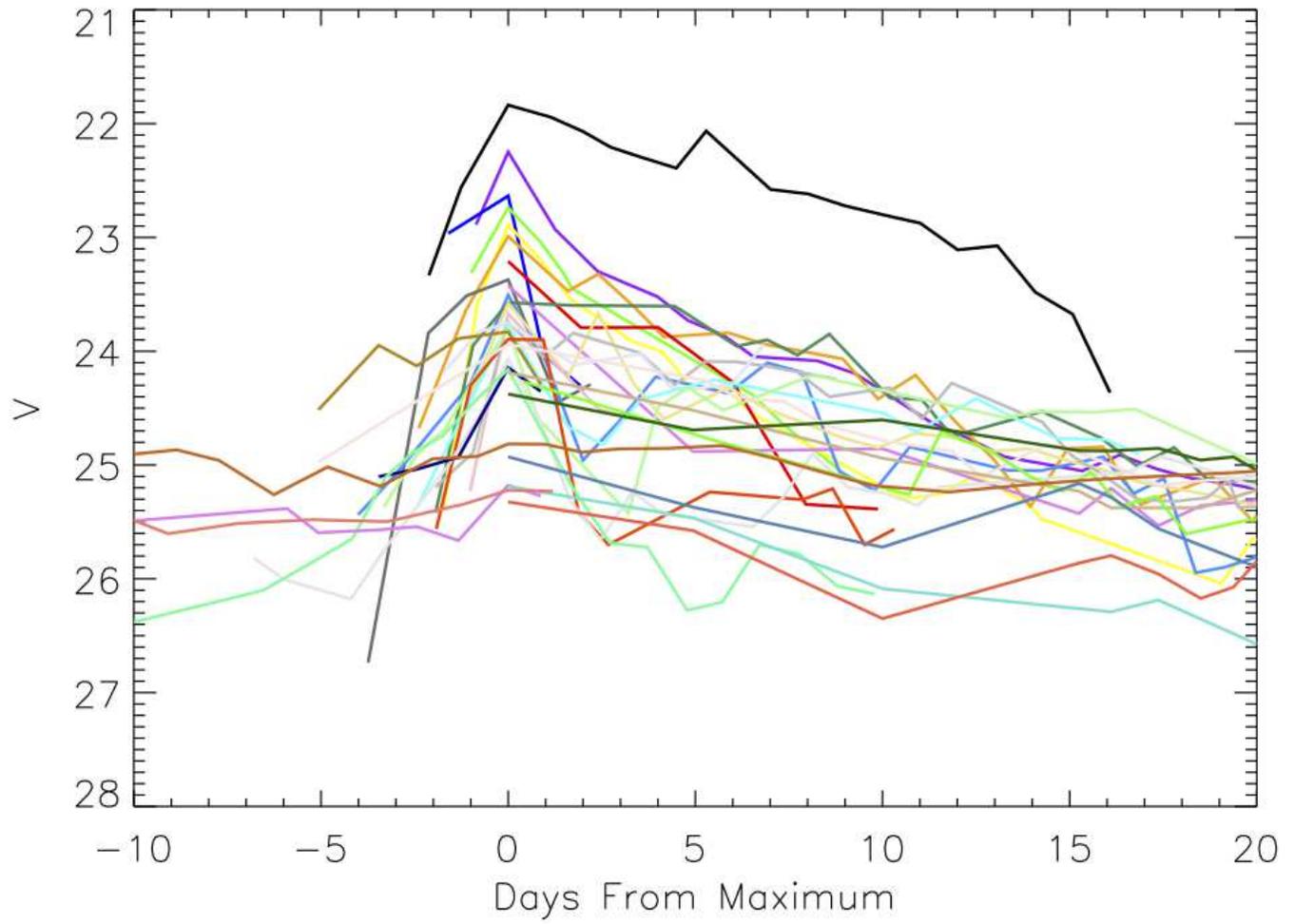}
\caption{Figure 4a zoomed in for more detail around the peak magnitude. The same color-coding scheme is used here, and in Figures 5 through 8, inclusive.}
\end{figure}

\begin{figure}
\figurenum{5}
\epsscale{1.0}
\includegraphics[width=180mm]{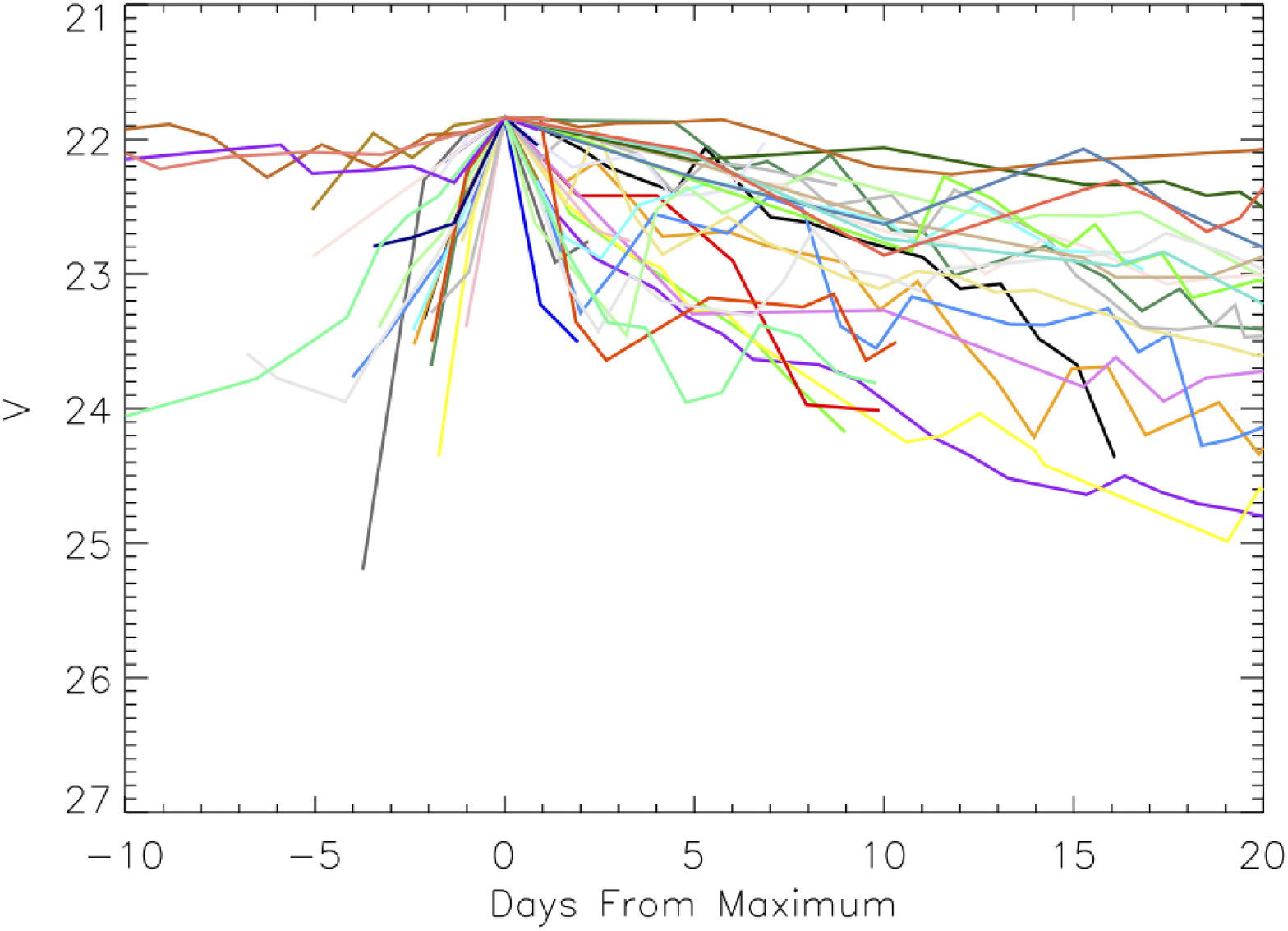}
\caption{32 M87 classical nova $V$ light curves overplotted to the same (arbitrary) maximum V magnitude, zoomed in for detail.}
\end{figure}

\clearpage

\begin{figure}
\figurenum{6a}
\epsscale{1.0}
\includegraphics[width=180mm]{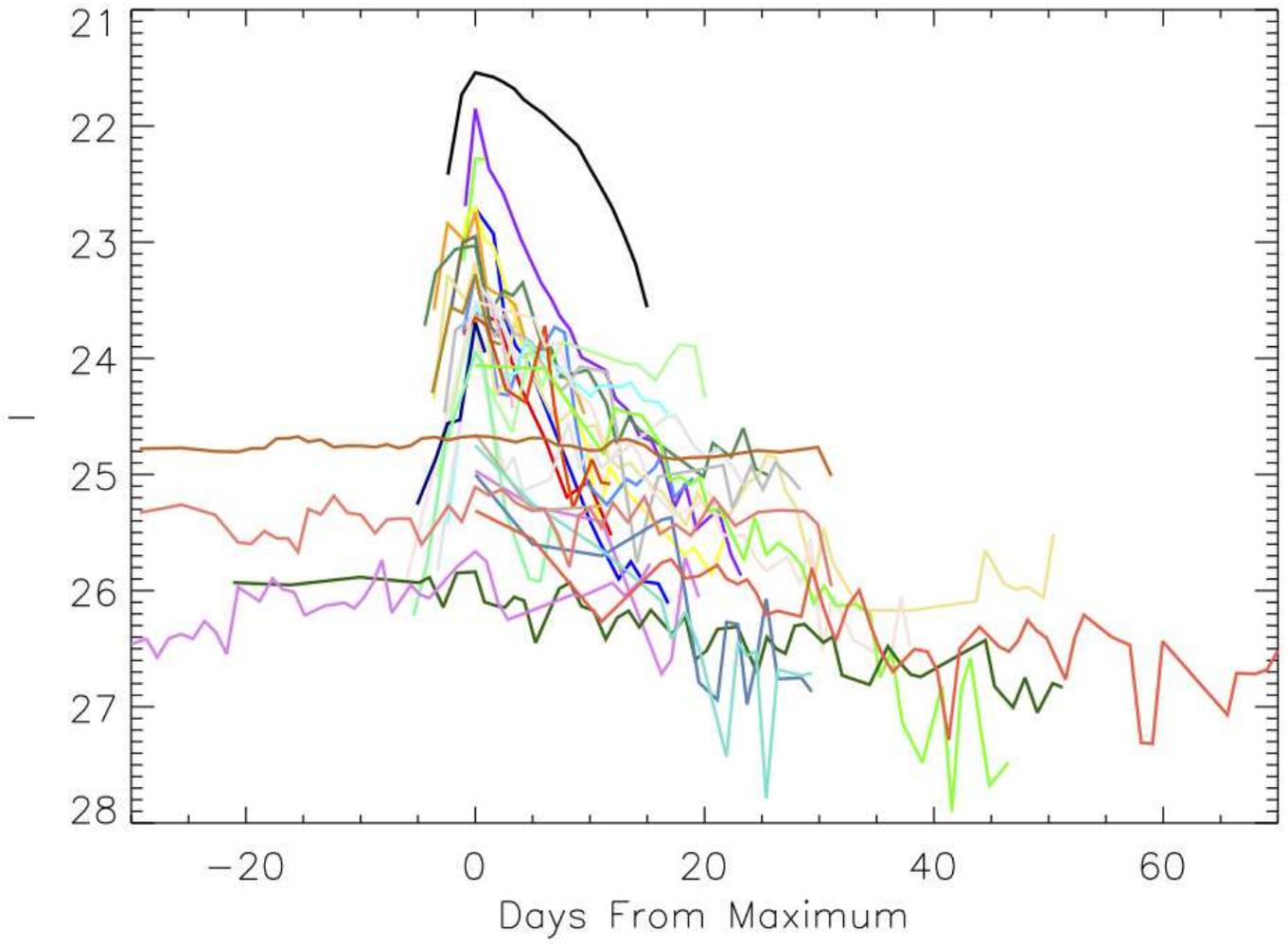}
\caption{32 M87 classical nova $I$ light curves overplotted as a function of days from maximum light.}
\end{figure}

\begin{figure}
\figurenum{6b}
\epsscale{1.0}
\includegraphics[width=180mm]{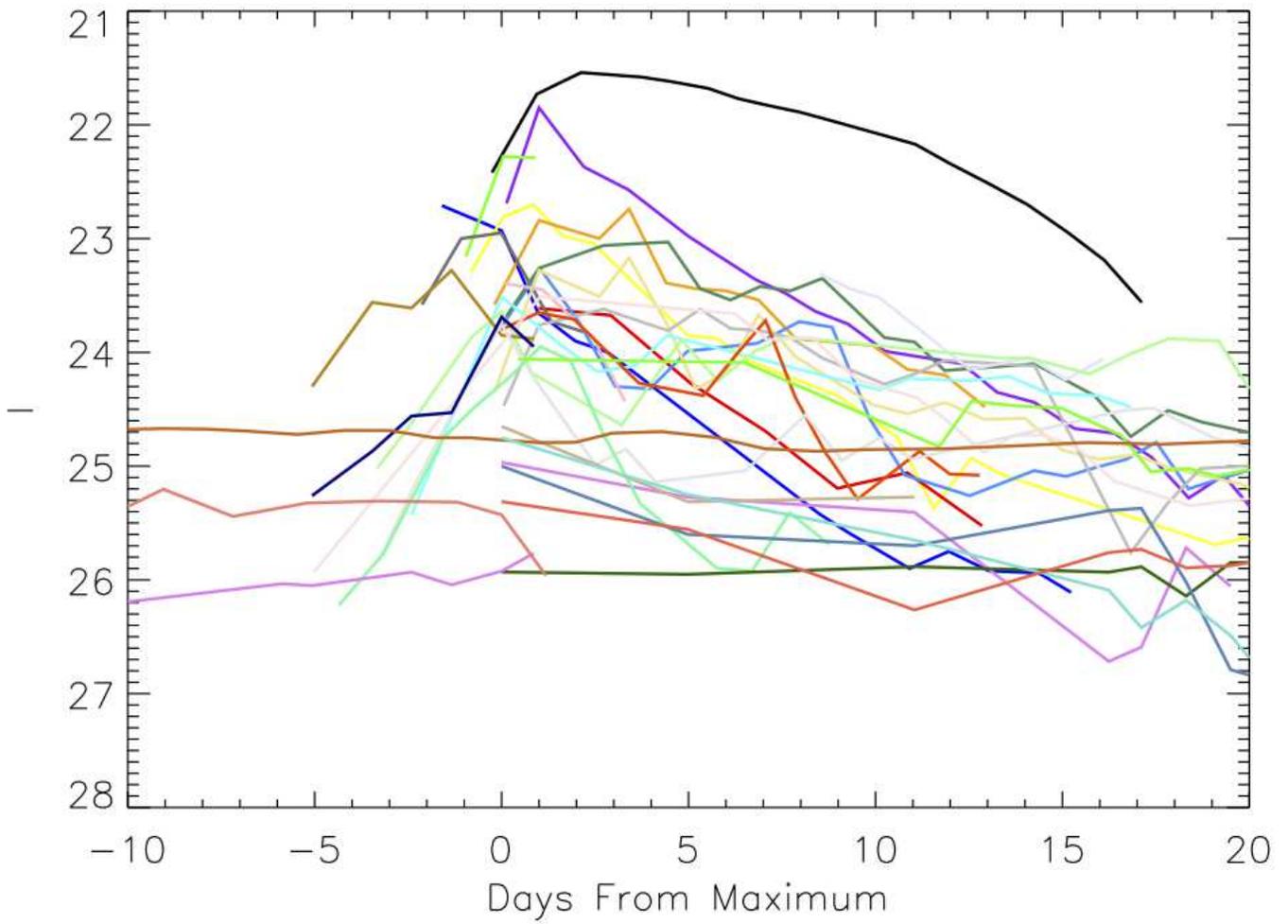}
\caption{Figure 6a zoomed in for more detail around the peak magnitude.}
\end{figure}

\begin{figure}
\figurenum{7}
\epsscale{1.0}
\includegraphics[width=180mm]{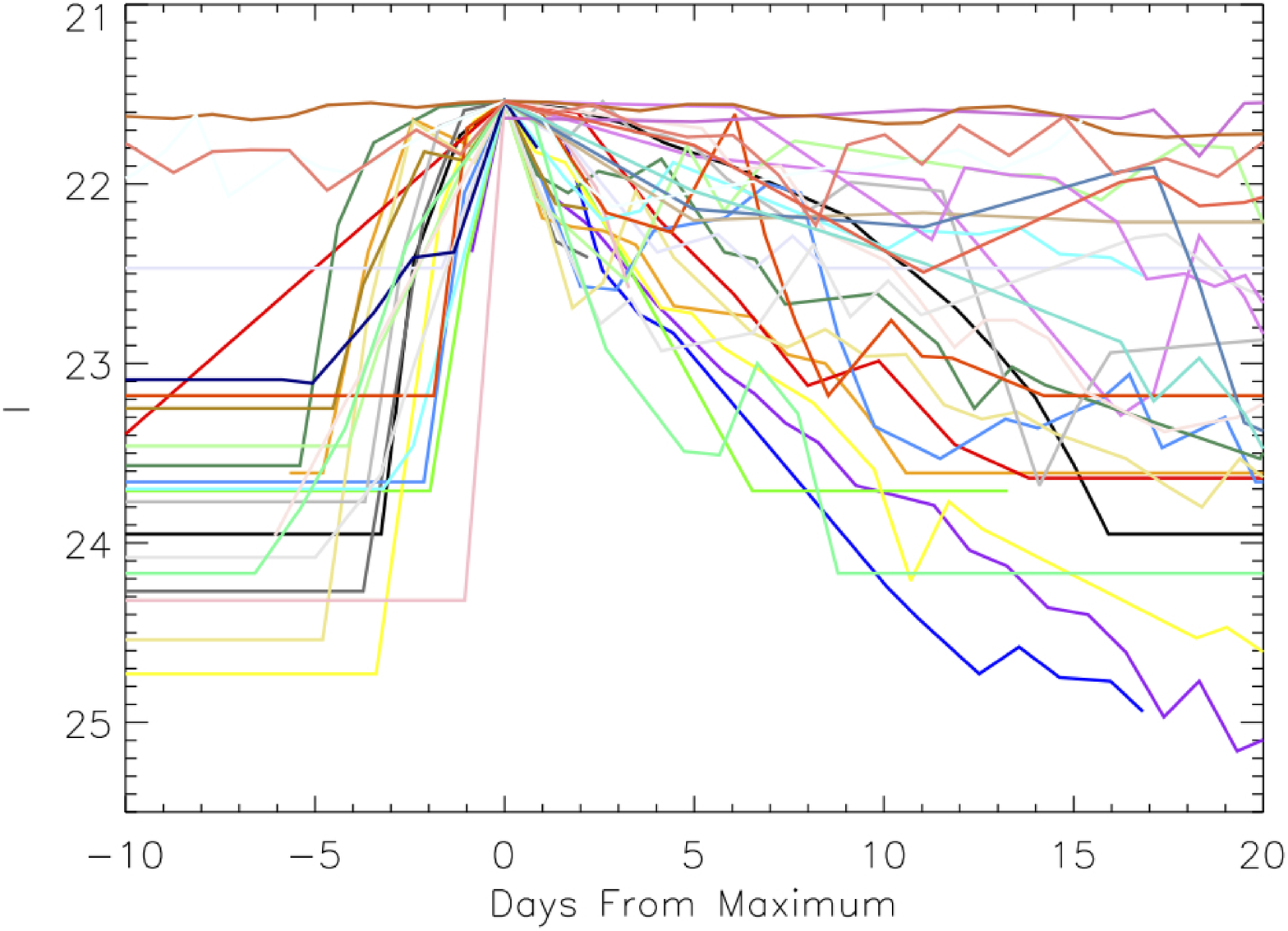}
\caption{32 M87 classical nova $I$ light curves overplotted to the same (arbitrary) maximum $I$ magnitude, zoomed in for detail.}
\end{figure}

\clearpage

\begin{figure}
\figurenum{8a}
\epsscale{1.0}
\includegraphics[width=180mm]{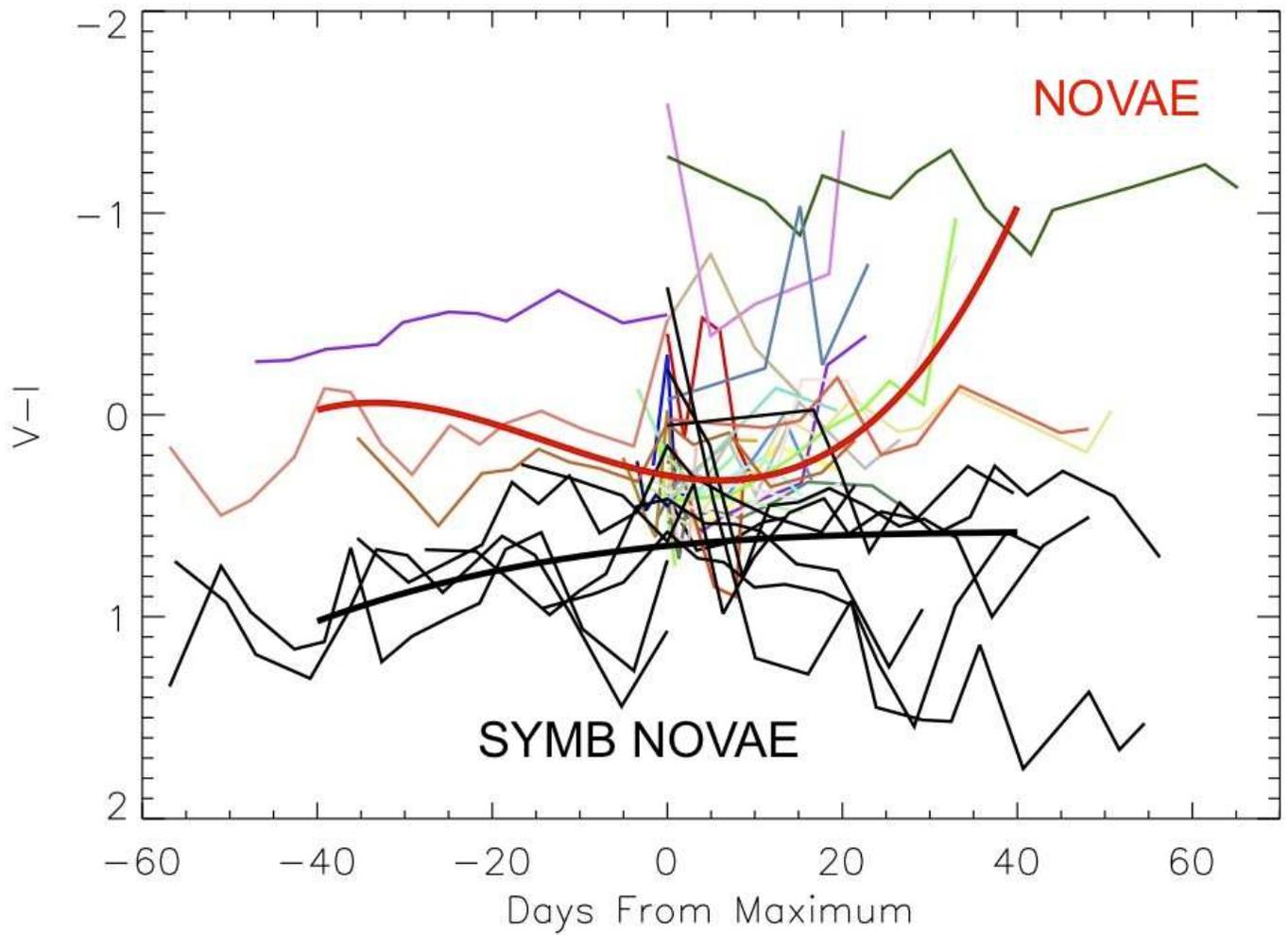}
\caption{Nova $(V-I)$ color curves overplotted as a function of days from maximum light. The thick red line is the median of 32 classical nova color curves,
while the thick black line labelled ``SYMB NOVAE" is the median of the brightest nine variables which are probably very slow and/or symbiotic novae.}
\end{figure}

\begin{figure}
\figurenum{8b}
\epsscale{1.0}
\includegraphics[width=180mm]{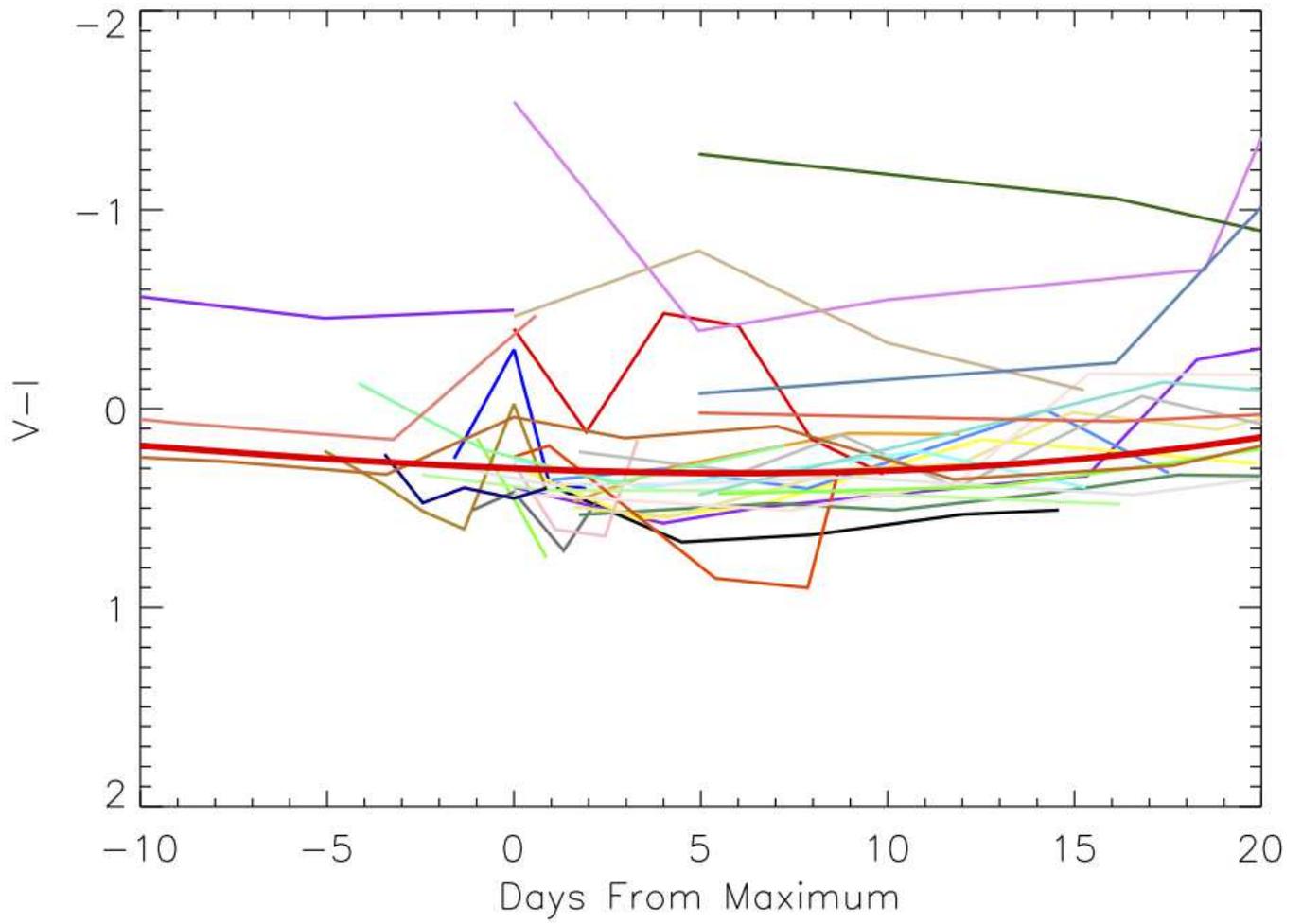}
\caption{Figure 8a zoomed in for more detail around the peak magnitude. The thick red line is the median of all nova color curves.}
\end{figure}

\clearpage


\begin{figure}
\figurenum{9}
\epsscale{1.0}
\includegraphics[width=180mm]{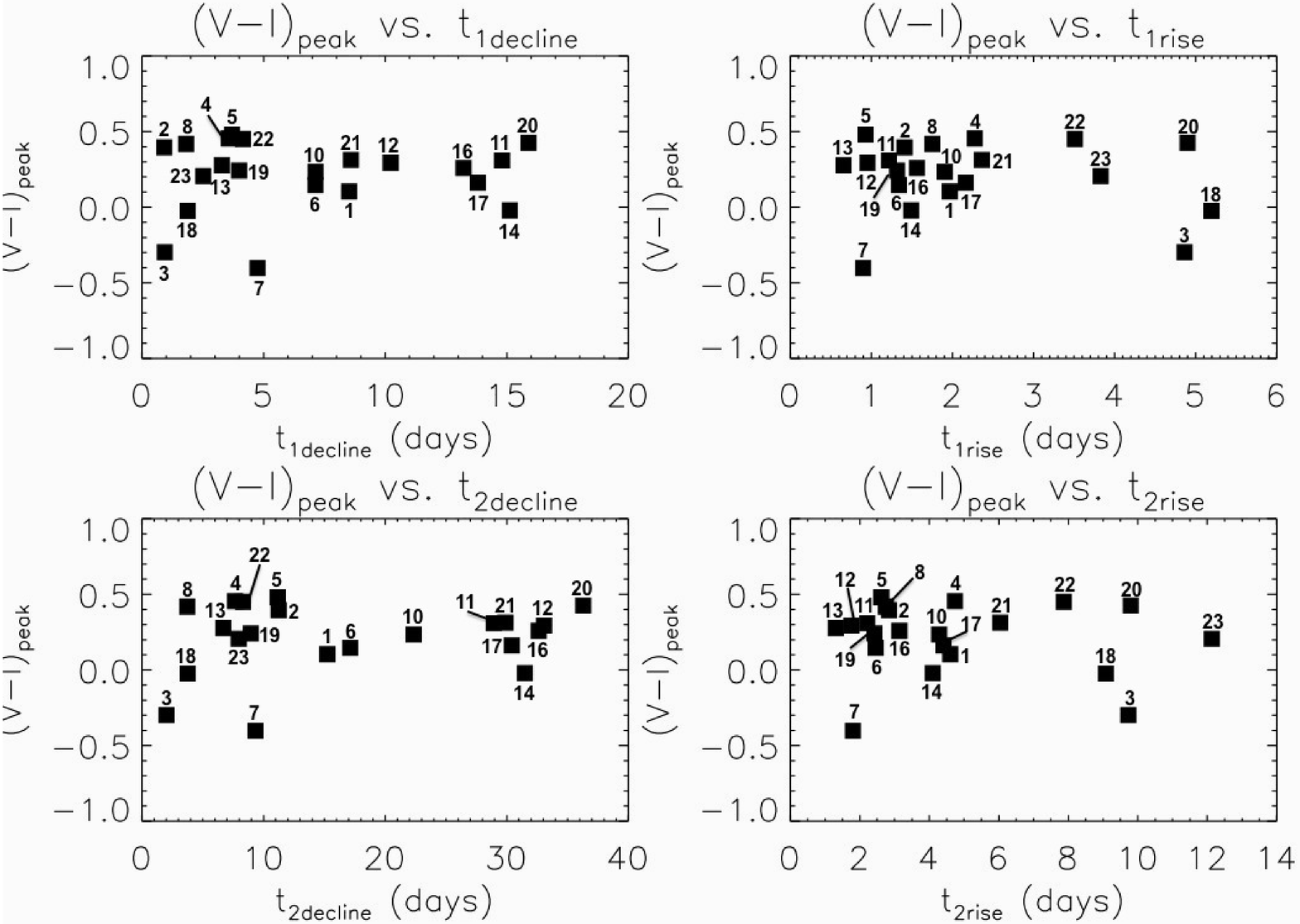}
\caption{$(V-I)_{peak}$ color of M87 novae versus times required to decline or rise by 1 or 2 magnitudes, respectively.  Only those novae which clearly show a rise, a peak, and a decline are used in figures 9 through 13, inclusive.  Rise and decline times are in days, and the nova number corresponds to those in Figures 1, 2 and 3.}
\end{figure}

\clearpage

\begin{figure}
\figurenum{10}
\epsscale{1.0}
\includegraphics[width=180mm]{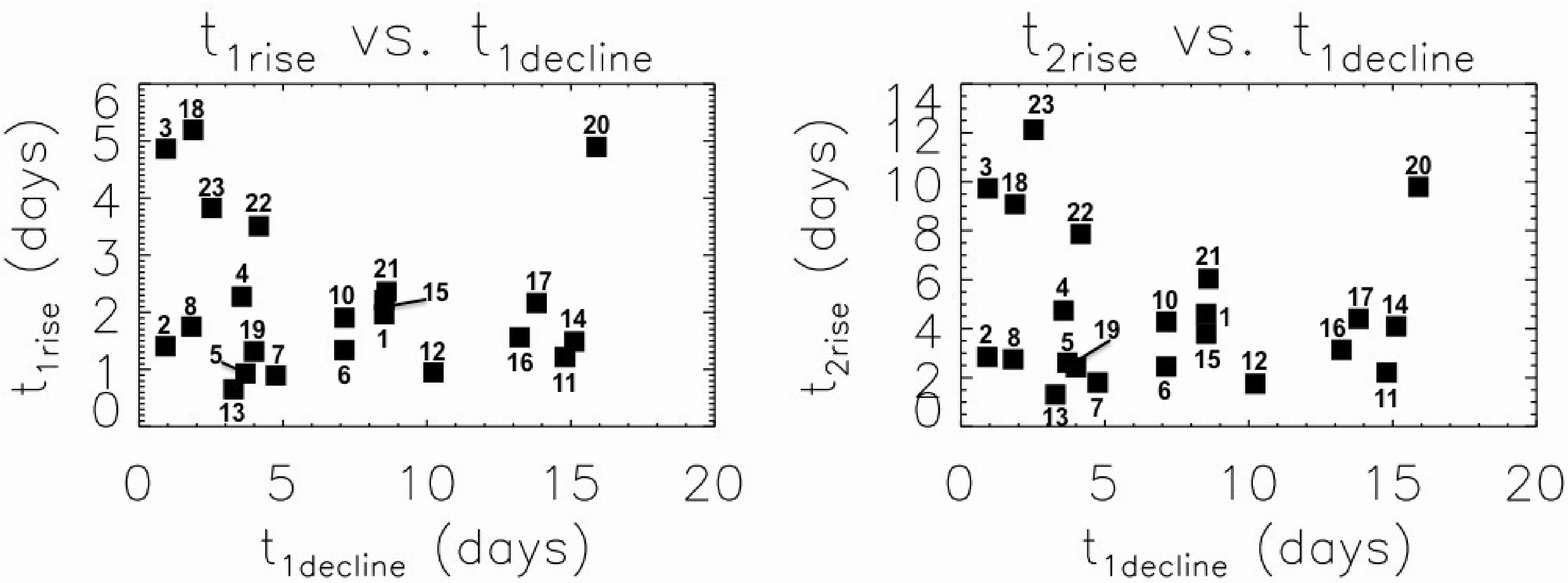}
\caption{Rise times of 1 and 2 magnitudes plotted as a function of decline time of 1 magnitude.}
\end{figure}

\clearpage

\begin{figure}
\figurenum{11}
\epsscale{1.0}
\includegraphics[width=160mm]{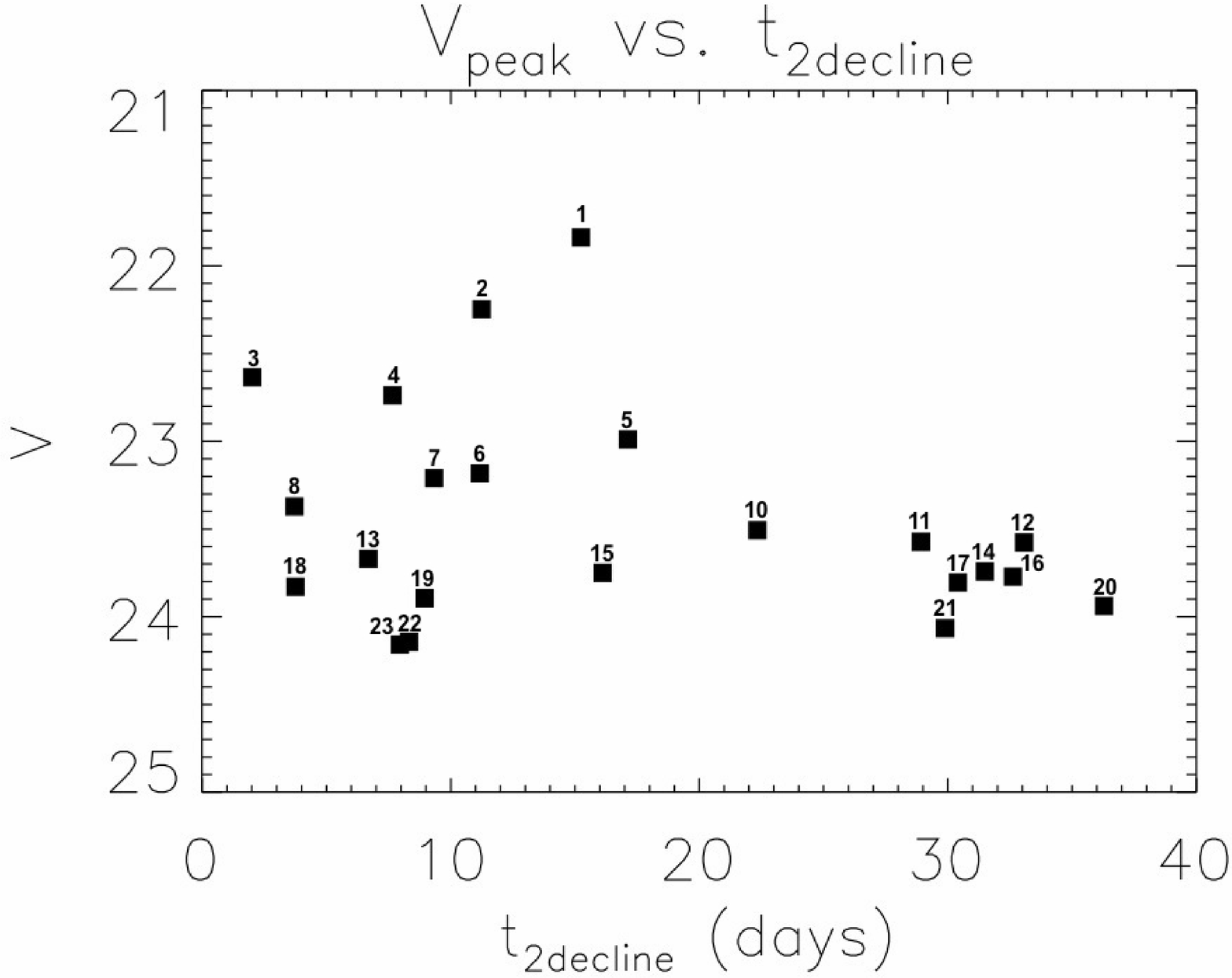}
\caption{Peak F606W ($V$) magnitude vs t$_{2decline}$ - the MMRD relation.}
\end{figure}


\begin{figure}
\figurenum{12}
\epsscale{1.0}
\includegraphics[width=180mm]{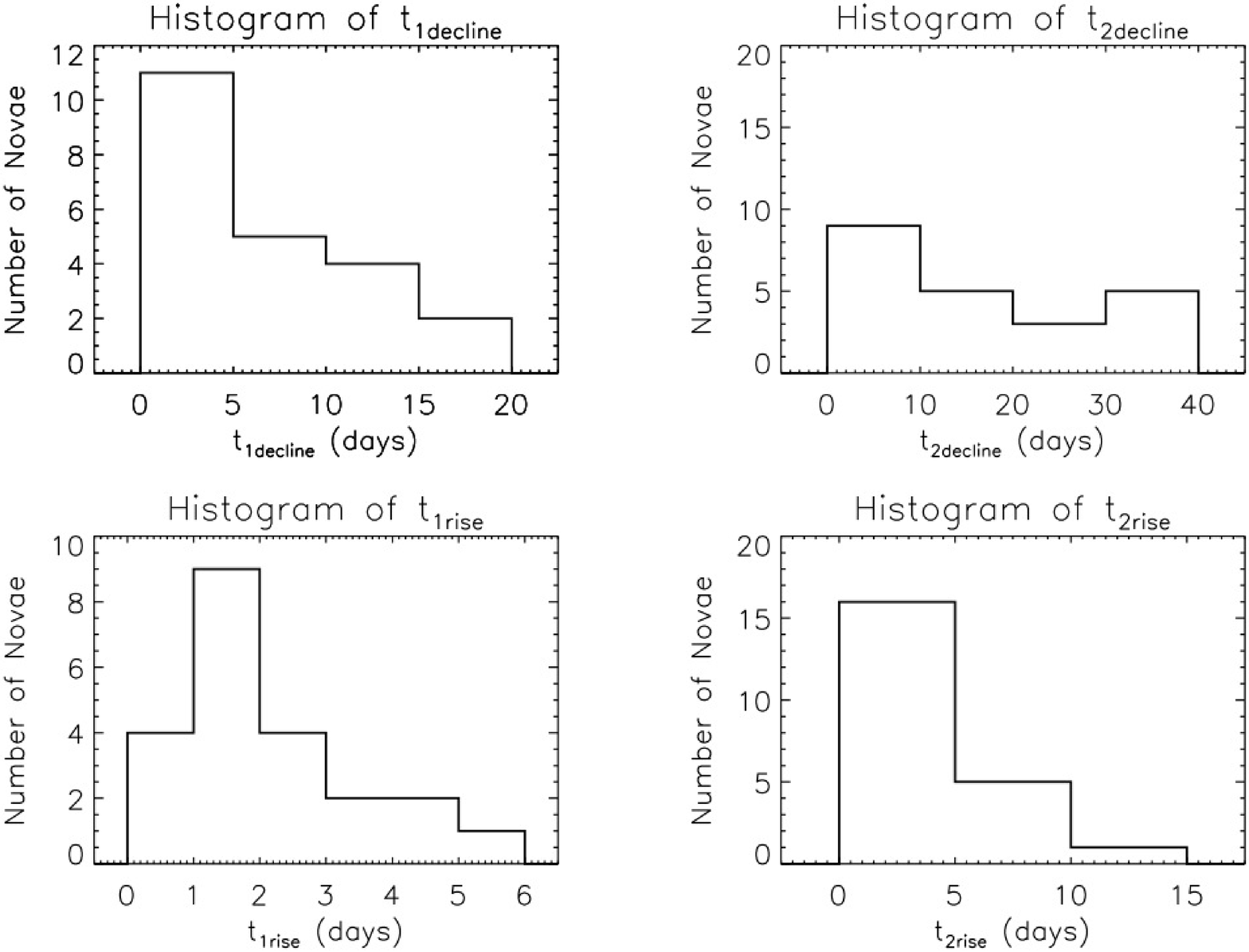}
\caption{Histograms of decline and rise times for M87 novae.}
\end{figure}

\clearpage


\begin{figure}
\figurenum{13}
\epsscale{1.0}
\includegraphics[width=180mm]{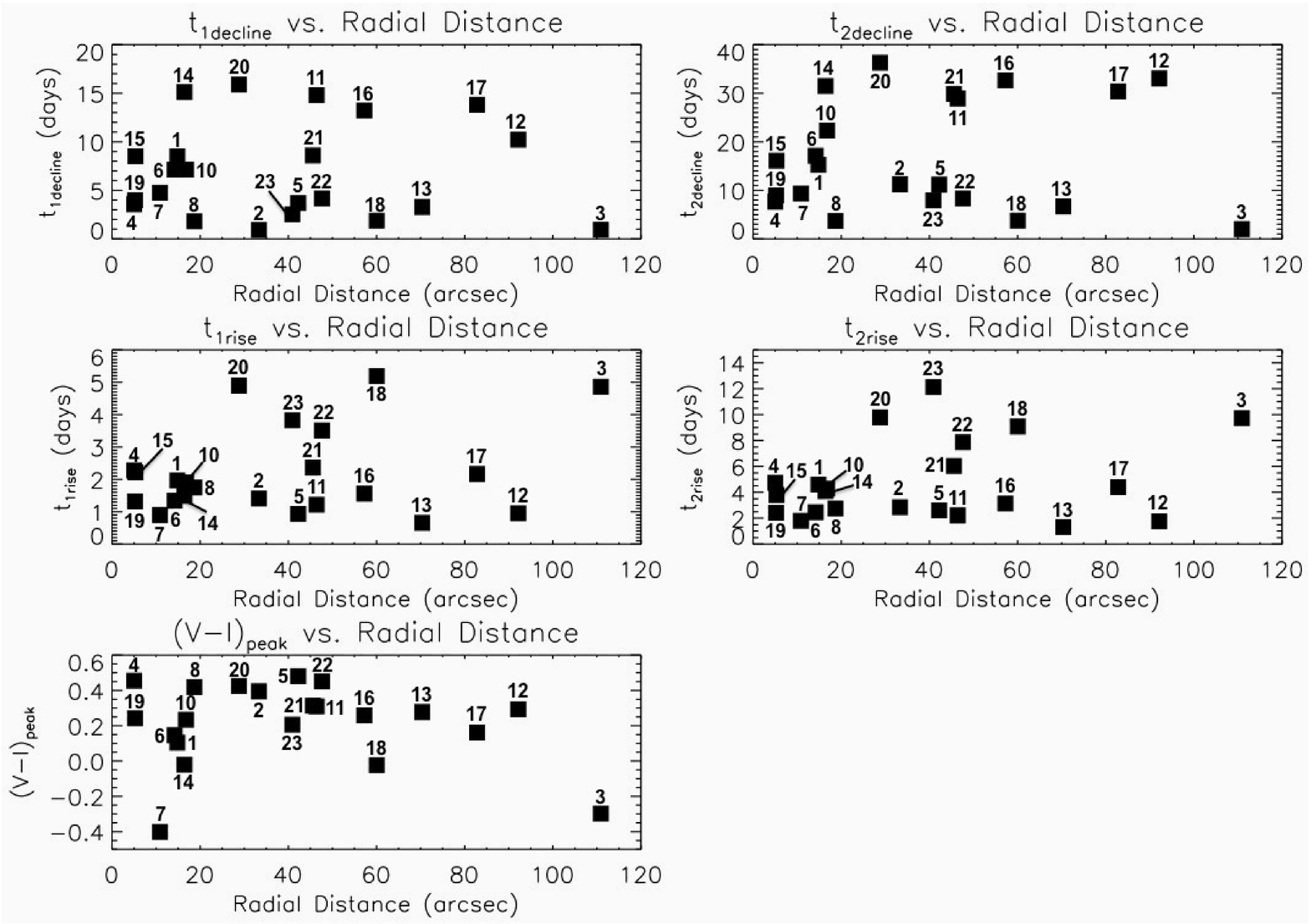}
\caption{Rise and decline times and $(V-I)_{peak}$ colors of classical novae as a function of distance from the center of M87.}
\end{figure}

\clearpage


\begin{figure}
\figurenum{14}
\epsscale{1.0}
\includegraphics[width=180mm]{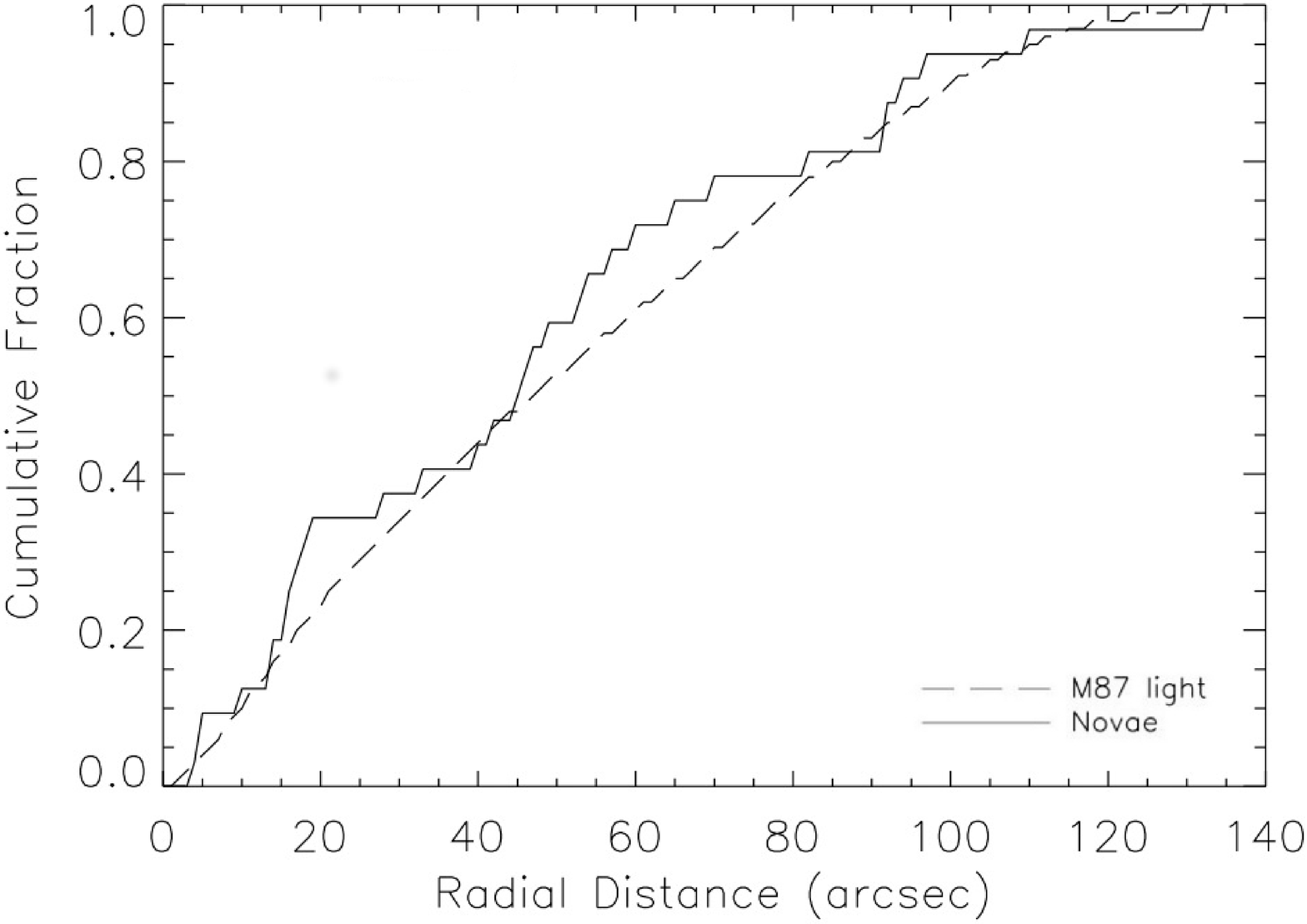}
\caption{The cumulative fraction of classical novae and light of M87 as a function of distance from the center of the galaxy.}
\end{figure}

\clearpage


\begin{figure}
\figurenum{15}
\epsscale{1.0}
\includegraphics[width=180mm]{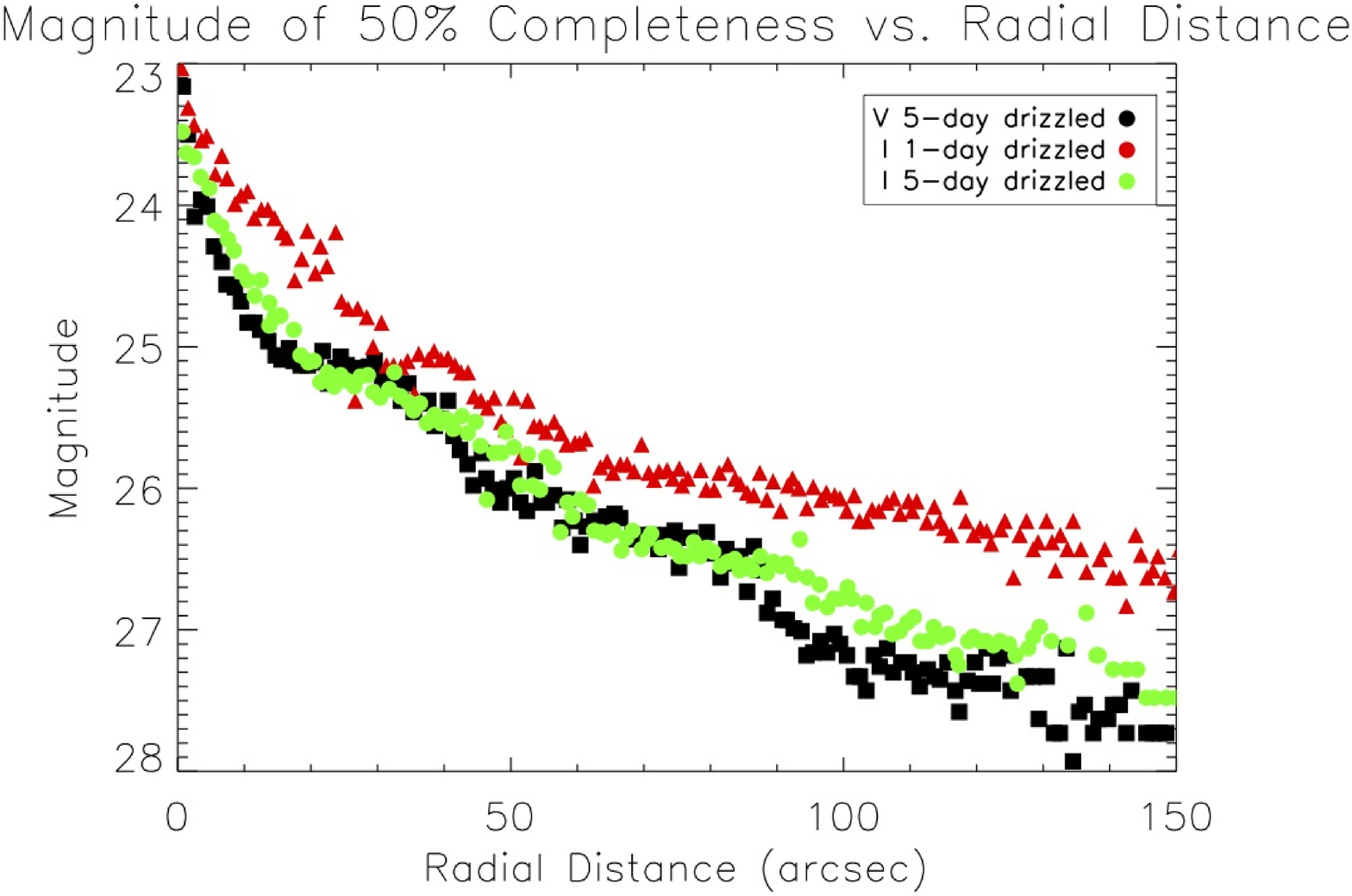}
\caption{Magnitude at 50\% recovery in each radial bin for 5-day drizzled $V$ images (black squares), single-day $I$ images (red triangles), and 5-day drizzled $I$ images (green circles).}
\end{figure}

\clearpage

\begin{figure}
\figurenum{16}
\epsscale{1.0}
\includegraphics[width=180mm]{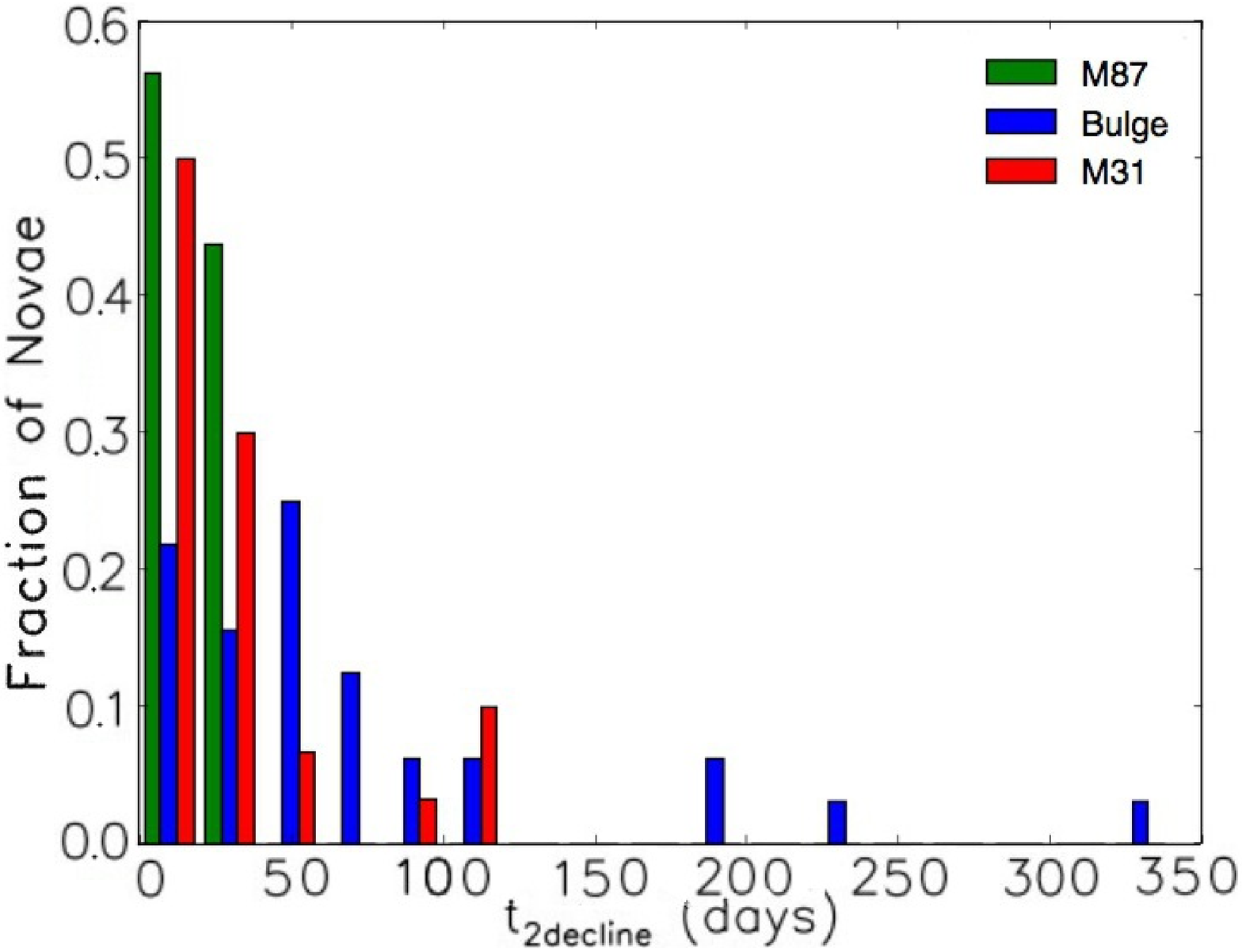}
\caption{Histograms of the distributions of times to decay by 2 magnitudes for the most complete samples of novae in the Galactic Bulge, M31 and M87 (the present work). The samples have time baselines of about 15 years (the Bulge), 1.5 years (M31), and 72 days (M87). Longer baselines lead to the detections of more slowly decaying, intrinsically fainter novae, important ingredients in correctly deriving the rate of nova eruptions in a galaxy. See text for details.}
\end{figure}

\clearpage

\begin{figure}
\figurenum{17}
\epsscale{0.8}
\plotone{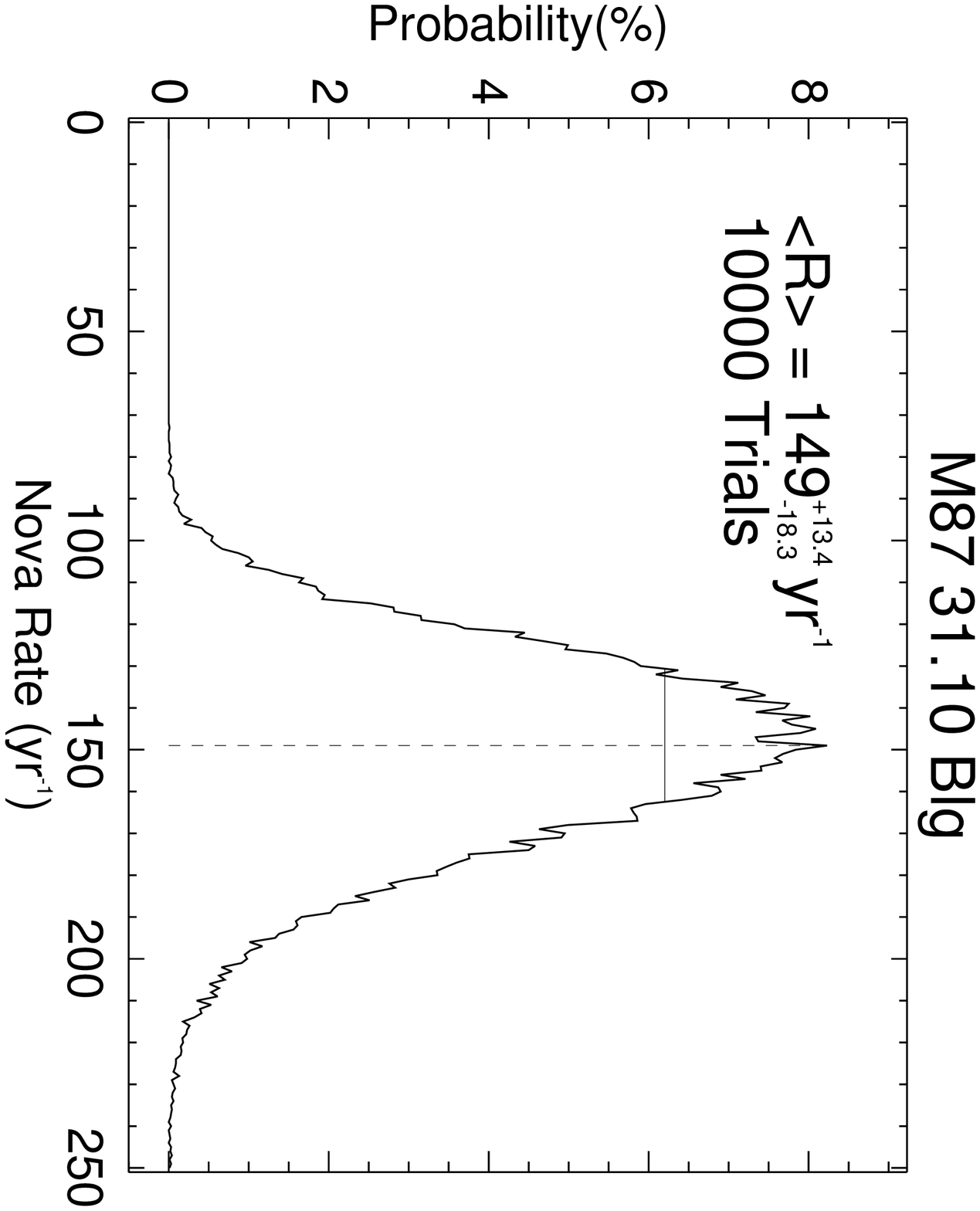}
\caption{The distribution of probabilities of varying annual classical nova rates in the area of M87 imaged by HST. The most probable rate in the 202" x 202" HST FOV is $R = 149_{-18.3}^{+13.4}$ classical novae/year, which corresponds to a global M87 classical nova rate of $363_{-45}^{+33}$ novae/year. That conservative rate ONLY assumes the brightest 32 novae in our sample to be true novae. The nova rate is significantly higher if some or all of the slow/symbiotic nova candidates (numbered 33-41, inclusive) are, in fact, novae.}
\end{figure}

\clearpage

\end{document}